\documentclass{aa}  
\usepackage{graphicx}
\usepackage{txfonts}

\renewcommand{\deg}{\ensuremath{^{\circ}}}

\newcommand{\mum}{\ensuremath{ \mu\mathrm{m}}}

\newcommand{\kms}{\ensuremath{\mathrm{km\,s^{-1}}}}

\begin{document}

   \title{Heliospheric modulation of the interstellar dust flow on to Earth}


   \author{Peter Strub\thanks{Contact: P.Strub@gmail.com, info@veerlesterken.ch }
          \inst{1}
          \and
          Veerle J. Sterken\inst{2}
          \and
          Rachel Soja\inst{3}
          \and
          Harald Kr\"{u}ger\inst{1}
          \and
          Eberhard Gr\"{u}n\inst{4,5}
          \and
          Ralf Srama\inst{3}
          \fnmsep
          }

   \institute{Max-Planck-Institut f\"{u}r Sonnensystemforschung,
             Justus-von-Liebig-Weg 3, 37077 G\"{o}ttingen, Germany
         \and
             Astronomical Institute University of Bern, Sidlerstrasse 5, 3012 Bern, Switzerland
         \and
             Institut f\"{u}r Raumfahrtsysteme, Universit\"{a}t Stuttgart, Pfaffenwaldring 29, 70569 Stuttgart, Germany
        \and
             Max-Planck-Institut f\"{u}r Kernphysik, Saupfercheckweg 1, 69117 Heidelberg, Germany
                \and
             Laboratory for Atmospheric and Space Physics, University of Colorado, Boulder, CO 80303, USA
             }

   \date{}


  \abstract
   {}
   {Based on  measurements by the Ulysses spacecraft and high-resolution modelling of the motion of interstellar dust (ISD) through the heliosphere we  predict the ISD flow in the inner planetary system and on to the Earth. This is the third paper in a series of three about the flow and filtering of the ISD.}
  {Micrometer- and sub-micrometer-sized dust particles are subject to solar gravity and radiation pressure as well as to interactions 
with the interplanetary magnetic field
 that result in a complex size-dependent flow pattern of ISD in the planetary system. With high-resolution dynamical modelling we study the time-resolved flux and mass distribution of ISD and assess the necessary requirements for detection of ISD near the Earth. } 
  {Along the Earth orbit the density, speed, and flow direction of ISD depend strongly on the Earth's position and the size of the interstellar grains. A broad maximum of the ISD flux ($\sim2 \times 10^{-4} \mathrm{m}^{-2} \mathrm{s}^{-1}$ of particles with radii $\gtrsim 0.3\,\mum$) occurs in March when the Earth moves against the ISD flow. During this time period the relative speed with respect to the Earth is highest ($\sim60\,
   \mathrm{km\,s^{-1}}$), whereas in September when the Earth moves with the ISD flow, both the flux and the speed are lowest ($\lesssim 10\, \mathrm{km\,s^{-1}}$). The mean ISD mass flow on to the Earth is approximately $ 100\,\mathrm{kg\,year^{-1}}$ 
   with the highest flux of $\sim3.5\,\mathrm{kg\,day^{-1}}$ occurring for about 2 weeks close to the end of the year when the Earth passes near the narrow gravitational focus region of the incoming
ISD flow, downstream from the Sun. The phase of the 22-year solar wind cycle has a strong effect on the number density and flow of sub-micrometer-sized ISD particles. During the years of maximum electromagnetic focussing (year 2031 +/- 3) there is a chance that  ISD particles with sizes even below $0.1\,\mum$ can reach the Earth. }
   {We demonstrate that ISD can be effectively detected, analysed, and even collected by space probes at 1\,AU distance from the Sun.}

   \keywords{interstellar dust --
                dust --
                heliosphere
               }

   \maketitle
%

\section{Introduction}

Interstellar dust (ISD) particles are messengers from the remote sites where they formed and from the environment that they traversed during their journey through space and time. They are born as stardust and take their initial elemental and isotopic signatures from the cool atmospheres of giant stars or from stellar explosions. Ultraviolet (UV) irradiation, interstellar shock waves, and mutual collisions modify and deplete particles in the interstellar medium (ISM). In dense molecular clouds particles grow by agglomeration and accretion.

The Sun and the heliosphere are surrounded by a local dense warm cloud of gas and dust, the Local Interstellar Cloud (LIC). About 1\% of the mass of this cloud is ISD. The motion of the heliosphere with respect to this cloud causes an inflow of the ISD into the heliosphere from a direction of 259\deg\ ecliptic longitude, and +8\deg\ ecliptic latitude \citep{landgraf1998a,frisch1999a,strub2015a}. The relative velocity of the ISD in the heliosphere is $26\,\mathrm{km\,s^{-1}}$ \citep{gruen1994a,strub2015a}. 

Most of our knowledge of ISD comes from astronomical observations: modelling combined with observations of the dependence of starlight extinction on wavelength reveals some material properties and allows an ISD size distribution to be derived \citep{mathis1977,draine1984,weingartner2001,zubko2004}.

Interstellar dust was first positively identified inside the solar system more than two decades ago. After its fly-by of Jupiter, the dust detector onboard the Ulysses spacecraft detected impacts of micron and submicron-sized particles ($10^{-17}$ to $10^{-14}$\,kg) predominantly from a direction  opposite to the expected impact direction of interplanetary dust particles \citep{gruen1993a}. 
On average, the measured impact velocities exceeded the local solar system escape velocity \citep{gruen1994a}. 
Subsequent analysis showed that the motion of the ISD particles through the solar system was parallel to the flow of neutral interstellar hydrogen and helium gas \citep{frisch1999a}. While the ISD flow persisted at higher latitudes above the ecliptic plane, even over the poles of the Sun, the interplanetary dust is strongly depleted away from the ecliptic plane. 

The flow patterns of ISD through the heliosphere were discussed in detail by \citet{sterken2012a}. \citet{sterken2013a} described the modulation of the ISD size distribution at Saturn, Jupiter, and asteroid distances from the Sun (3 AU). 
The observed variation of the ISD flow is due to the modulation of the dust stream by the Lorentz force, the radiation pressure force, and gravity \citep{landgraf:00a}.
Particles with a radius $a_d\gtrsim 100$\,nm pass the heliospheric bow shock and enter the heliosphere \citep{linde2000a}. As a consequence, interstellar particles constitute the dominant known particulate component in the outer solar system (in number flux, not in mass flux). 

Although a large portion of the ISD measurements by Ulysses were obtained far from the ecliptic plane and outside the inner solar system, it is also known that ISD particles of various sizes can reach the Earth's orbit \citep{altobelli2003,altobelli2005a,altobelli2006}. This opens the possibility for Earth-orbiting spacecraft to detect and study ISD. Previous models, while accurately describing the ISD environment at larger heliocentric distances, did not have the resolution to enable a good time-resolved understanding of the dust environment at the Earth \citep{gruen1994a,landgraf2000c,sterken2012a}. In this paper we describe a model of sufficiently high spatial resolution to allow the study of the characteristics of the ISD flow at the Earth. We study the time-resolved flux and mass distribution and assess the necessary requirements for detection of ISD near the Earth. 

Section \ref{sec:modeling} describes the physical modelling of the ISD dynamics in the solar system, including initial conditions, the three modelled forces, and the material assumptions for the modelling. Section \ref{secISDat1AU} presents the Monte Carlo computer simulation results of the ISD flow at Earth orbit. Section \ref{sec:discussion} discusses in detail the temporal evolution of ISD observing conditions (density, speed, impact direction and size distribution) relative to Earth and the total mass influx on to Earth. Section \ref{isdmissions} concludes this study with an outlook for ISD missions at 1\,AU.

\section{A time-resolved model of the ISD flow through the solar system}\label{sec:modeling}

The ISD model is a three-dimensional (3D), time-resolved representation of the flow properties of ISD particles in the inner solar system created using a Monte Carlo simulation. It is generated by integrating the equation of motion of ISD particles as they move through the solar system in order to obtain particle trajectories. These trajectories are used to produce data cubes that contain the average density, velocity components, and velocity dispersion of particles in each region of space (within $10$\,AU of the Sun) and time (within the solar cycle) for a given particle size.

\citet[][Fig.~3-4]{sterken2015a} described the effect of different interplanetary magnetic field (IMF) models on the simulation results and used a constant rate of change of the solar magnetic dipole for describing the ISD flux near Saturn, Jupiter, and the asteroid belt.
The model is based on the work of \citet{landgraf:00a} and \citet{sterken2012a,sterken2013a}. The main goal of our model is to improve both the spatial and temporal resolution by a factor of six over the model of \citet{sterken2012a,sterken2013a}. 
This leads to a spatial mesh cell size of 0.25\,AU, with a temporal resolution 
of approximately 12.2~days, as opposed to the resolution of 1.5\,AU and 73\,days in the earlier model. 
Also, the variable rotation rate of the solar magnetic dipole from WSO Observations is used for discussing the ISD flow and flux near the Earth, approximated by a piecewise constant rotation rate. 

The initial conditions, relevant forces, and general flow properties are briefly introduced in Sections \ref{subsecICs} and \ref{subsecForces}. A more detailed description of the latter is given in \citet[][model and flow pattern]{sterken2012a} and \citet[][filtering]{sterken2013a}.


\subsection{Initial conditions} \label{subsecICs}
The model uses the same initial conditions as \citet{landgraf:00a} and~\citet{sterken2012a,sterken2013a}: the simulated ISD particles enter the solar system at a uniform direction and velocity, with their positions randomly distributed on a plane $50$\,AU upstream of the Sun and perpendicular to the velocity vector. We use an initial velocity of $v_\infty=26$\,kms$^{-1}$ and an inflow direction from an ecliptic longitude\footnote{We follow the convention to use $l, b$ to denote ecliptic coordinates given in the heliocentric frame of reference, and $\lambda, \beta$ in the geocentric frame of reference.} $l_{\mathrm{ecl}}= 259\deg$ and ecliptic latitude $b_{\mathrm{ecl}}=8\deg$. This is the same as the direction of the neutral gas flow inside the solar system~\citep{witte:93a,lallement2014a,wood2015a}, and it is compatible with the Ulysses measurements of the ISD flow~\citep{frisch:99a,strub2015a}.  This is equivalent to the ISD particles being at rest with respect to the LIC. 

Our model does not simulate the effect of the heliospheric boundary on the ISD trajectories, which results in a diminution of the small particle contribution \citep{linde2000a}. However, as the normalisation of the simulated dust densities is based on Ulysses dust fluxes measured inside the heliosphere, the time-averaged effects of the filtering at the heliospheric transition region have been accounted for. A possible time-dependence of the filtering as suggested by \citet{sterken2015a} is not reflected in the present model.

\subsection{Relevant forces}\label{subsecForces}
The trajectories of charged ISD particles were integrated under solar gravity, the solar radiation pressure force, and the Lorentz force from the charged particles moving through the IMF. The solar radiation pressure force and gravity are not considered variable in time and therefore produce a stationary flow pattern. The Lorentz force depends on the solar cycle, and therefore leads to a temporal variability in the flow of ISD \citep{morfill1979a,landgraf2000c,sterken2012a}. 


The solar radiation pressure force and gravity both decline with the square distance to the Sun. We can therefore use the solar radiation pressure force divided by the gravity as a dimensionless constant parameter ($\beta$) that depends on the particle material properties such as particle size, morphology, and the reflectivity and absorptivity of the particle integrated over the solar spectral energy distribution (SED) \citep{burns1979a}.

We use the same $\beta$-curve as \citet{sterken2013a}, which was adapted from the curve for astronomical silicates given in \citet{gustafson1994a} by scaling it to match the maximum value $\beta_{max}\simeq1.4 \-- 1.8$ (average 1.6) measured by Ulysses \citep{landgraf:99a}. \citet{sterken2013a} discuss the filtering of ISD at different places in the solar system (as close as the asteroid belt) for this $\beta$-curve, as well as independently of any $\beta$-curve. \citet{landgraf2000c} used the astronomical silicates $\beta$-curve with a maximum of 1.4 \citep{gustafson:94a,draine1984}.


The Lorentz force is important for the dynamics of ISD particles with radius  $a_d \lesssim 0.4\,\mum$. The particles become charged as they move through the heliosphere, as a result of (1) predominantly the loss of electrons through photoionisation by solar UV light,  (2) the collection of ions and electrons from the ambient solar wind plasma, and (3) secondary electron emission. Similar to previous studies, we assume a constant equilibrium potential of +5\,V, which is compatible with Cassini CDA measurements of charged particles~\citep{kempf:04a} and theoretical charging models \citep{mukai1981a,horanyi1996b}. A constant charge of +5\,V is a good approximation for particles moving through the solar wind plasma between 50 and 1\,AU from the Sun \citep{slavin2012a,kimura1998a}. The resulting Lorentz force depends on the particle's charge to mass ratio ($Q/m$), its velocity relative to the solar wind velocity (${\bf \dot{r}}_{p,sw}$), and the strength of the IMF at the particle's location, ${\bf B}_{sw}$. $Q/m$ is defined using the $+5$\,V potential, along with an assumption of compact spherical particles with a constant bulk density of $2\,\mathrm{g\,cm^{-3}}$.

The interplanetary magnetic field is the continuation of the solar magnetic field in interplanetary space, frozen into the plasma of the solar wind. Due to the Sun's rotation, its outward motion leads to a spiral pattern of alternating magnetic field polarities called the Parker spiral~\citep{parker1958}.

The Parker model determines the direction and strength of the IMF, but the polarity of the field depends on the position with respect to the heliospheric current sheet (HCS) and on the phase of the solar cycle. We approximate the HCS as a plane that separates the regions of positive polarity from the regions of negative polarity. It is aligned with the solar equatorial plane during solar minimum, while at solar maximum it is tilted by 90$^\circ$, parallel to the solar rotation axis. As a result, the modelled HCS turns around a reference axis in the solar equatorial plane at a rate of 360$^\circ$ over 22 years. Additionally, the HCS follows the solar rotation. This leads to focussing and defocussing configurations of the
IMF for ISD particles; see Table~\ref{tab:solarcyclemodel}.

\begin{figure}
   \centering
   \includegraphics[width=\columnwidth]{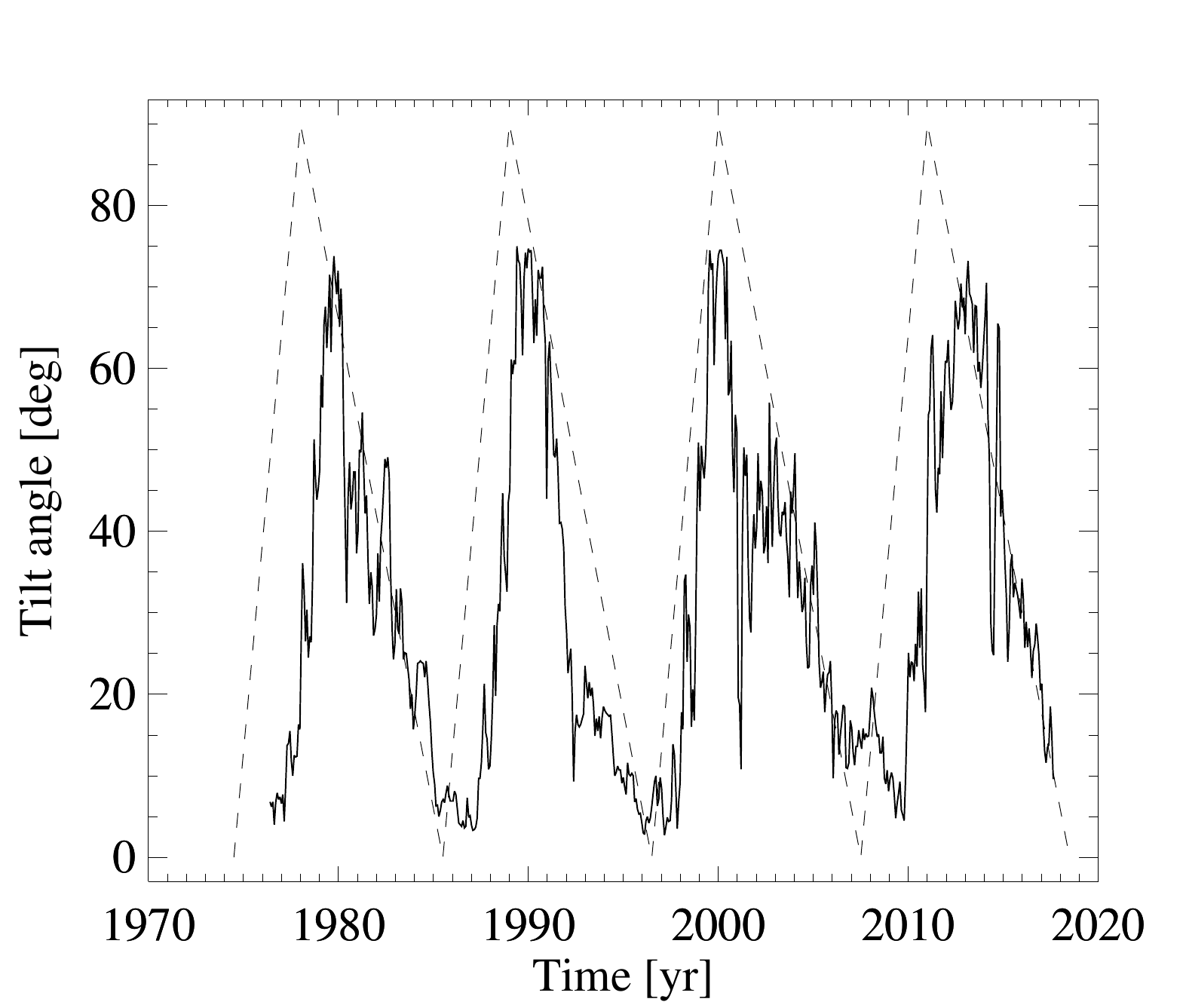}
      \caption{Tilt of the heliospheric current sheet, computed from measurements at the Wilcox Solar Observatory~\citep[solid line, ][]{WSO:18}, and the piecewise-linear tilt used in the simulation (dashed line).  }
   \label{fig:WSOsolarcycle}
   \end{figure}

In contrast to the approximations used in models 1 and 2 of \citet{sterken2012a}, the rotation rate of the HCS is not constant over time. Instead, the rate is piecewise constant with a faster rate on the two intervals from solar minimum to maximum, and a slower rate from solar maximum to minimum. This improves the representation of the observed rotation rates of the HCS dipole (Fig.~\ref{fig:WSOsolarcycle}). It is somewhat similar to model 3 of \citet{sterken2012a}, which utilises the measured inclination of the solar magnetic dipole, but here we approximate the inclination with a monotonous function instead of the noisy data from the observations to speed up the calculations, and to extrapolate into the future.

We assume one sidereal rotation every 25.38 days for both the poles and the equator of the Sun. This is the Carrington rotation rate of the Sun which corresponds to the sidereal rotation rate at 26\deg\ latitude. As we average the magnetic field over the solar rotation period, the only effect of the solar rotation rate on the simulation is as a scaling factor for the azimuthal component of the solar magnetic field \citep[][equation 28]{sterken2012a}. We adopt the IMF description given in \citet{parker1958}, use a magnetic field strength of $B_0=2300\,$nT at 10~solar~radii from the Sun \citep{cravens1997}, and assume a constant solar wind speed of $400$\,kms$^{-1}$.
As proposed by \citet{landgraf1998a}, the magnetic field strength is averaged over one solar rotation. This facilitates the use of larger integration time steps for faster computation. According to our test simulations with a magnetic field following the solar rotation \citep[similar to model 2 of][]{sterken2012a}, the error of this approximation is below a cell size for all particles except those that pass within 0.25\,AU of the Sun. As these particles constitute only a very small fraction in the inner solar system, we consider this a good trade-off. 

\begin{table}[htb] 
  \centering
  \caption{An overview of the modelled solar cycle.}
  \label{tab:solarcyclemodel}
   \begin{tabular}{llll} \hline\hline
   Year                 & Year          & Min / Max             & Cycle                 \\ 
   (ISD model)  & (WSO) &                               &                               \\ \hline
   1974.5               & 1976          & Min                   & defocus                                         \\ 
   1978         &                               & Max                   & defocus $\to$ focus               \\ 
   1985.5               & 1987          & Min                   & focus                                   \\ 
   1989         &                       & Max                   & focus $\to$ defocus         \\ 
   1996.5               &                       & Min                   & defocus                                 \\ 
   2000         & 2000          & Max                   & defocus $\to$ focus           \\
   2007.5               & 2009          & Min                   & focus                                 \\ 
   2011         & 2013          & Max                   & focus $\to$ defocus           \\ 
   2018.5               &                       & Min                   & defocus                                 \\ 
   2022         &                       & Max                   & defocus $\to$ focus             \\
   2029.5               &                       & Min                   & focus                                   \\
   2033         &                       & Max                   & focus $\to$ defocus         \\ 
   2040.5               &                       & Min                   & defocus                                 \\ \hline
\end{tabular}
\end{table}

\subsection{Resulting simulation output}\label{subsecDatabase}
The simulations produce a set of data cubes, one for each simulated particle size with specific $\beta$ and $Q/m$. The cubes contain the spatial densities, velocity components, and the velocity dispersion inside a box within $10$\,AU of the Sun (between $-10$\,AU and $+10$\,AU in each coordinate), at a spatial resolution of $0.25$\,AU, and a time resolution of 12.2 days.

Thirteen particle sizes between 0.05 and 4.9\,\mum\ were used for calculations.
(Table~\ref{tbl:simsizes}).
However, for quantitative predictions in this study, only seven sizes between 0.11 and 2.29\,$\mu$m are used: particle sizes below and above this interval were not present with a sufficient signal-to-noise ratio (S/N) in the Ulysses dataset used for normalisation. Furthermore, there is a gap between $0.1$ and $0.34$\,$\mu$m because those particles do not reach the Earth's orbit for the $\beta$-curve we assumed. The two smallest simulated particle sizes are used in Section~\ref{sec:discussion} to discuss the possibilities of detecting such particles at Earth orbit.

The normalisation of the simulated dust flux was achieved using the ISD mass distribution measured by Ulysses. Over several extended periods comprising a total of 13 years, the Ulysses dust detector identified particles of interstellar origin and measured their mass distribution \citep{krueger2015a}. The simulated dust flux was extracted for the same periods and normalised such that the integrated number of impacts over the whole measurement period matches the Ulysses impact rates for all simulated sizes.

The general ISD flow pattern and resulting flux filtering and enhancements are discussed in \citet{sterken2012a,sterken2013a}. In Appendix A we give an overview of the results of our simulations 
showing the spatial distribution of ISD grains, which varies markedly with the solar cycle. Nevertheless, there is some systematic structure that can be predicted. This will be looked at in the following sections.

In Section~3 we show the variations of four parameters
along the Earth's orbit over the whole solar cycle of 22 years: ISD density and velocity, the angular deflection for different particle sizes, and the widening of the flow direction due to dynamic effects. All these quantities are given in the heliocentric ecliptic reference frame (i.e. not accounting for the relative velocity of the Earth, in order to show the underlying modulation of the ISD flow). In Section 4 we show the ISD mass inflow rate on to Earth as well as the ISD count rate and impact velocity for a dust detector on an Earth-like orbit (e.g. at one of the Lagrangian points), taking into account the relative motion of the Earth. The variations of the dust flow registered at Earth are a superposition of the spatial and time variabilities of the ISD dust flow through the heliosphere and the sampling of these inhomogeneities by the Earth as it moves on its orbit. We note that the fluxes are calculated assuming a detector with a $4\pi$ field of view. However, as the ISD flow is collimated, the results are virtually the same for a more realistic detector geometry with an opening angle of $\lesssim2\pi$ pointing towards the dust apex direction.

\section{The ISD flow at 1\,AU from the Sun} \label{secISDat1AU}
Here we give an overview of the heliocentric ISD flow sampled at the positions of the Earth along its orbit. We discuss the time variations of ISD density, heliocentric speed, the heliocentric flow direction, and the spread of the flow direction for three different dust particle sizes (Table~\ref{tbl:simsizes}) representative for three different size regimes, each dominated by a different force: Electromagnetic force ($a_d=0.072\,\mum$), radiation pressure (0.34\,\mum), and gravity (0.49\,\mum). ISD particles with $\beta>1.4$ cannot be observed at Earth orbit because the Sun's radiation pressure is too high. For the $\beta$-curve used in our simulations, this applies to particle sizes of $0.15\,\mum<a_d<0.3\,\mum$ (Appendix.~\ref{sec:general_flow}).

\begin{figure*}
\centering
\begin{tabular}{ccc}
\includegraphics[width=0.3\textwidth]{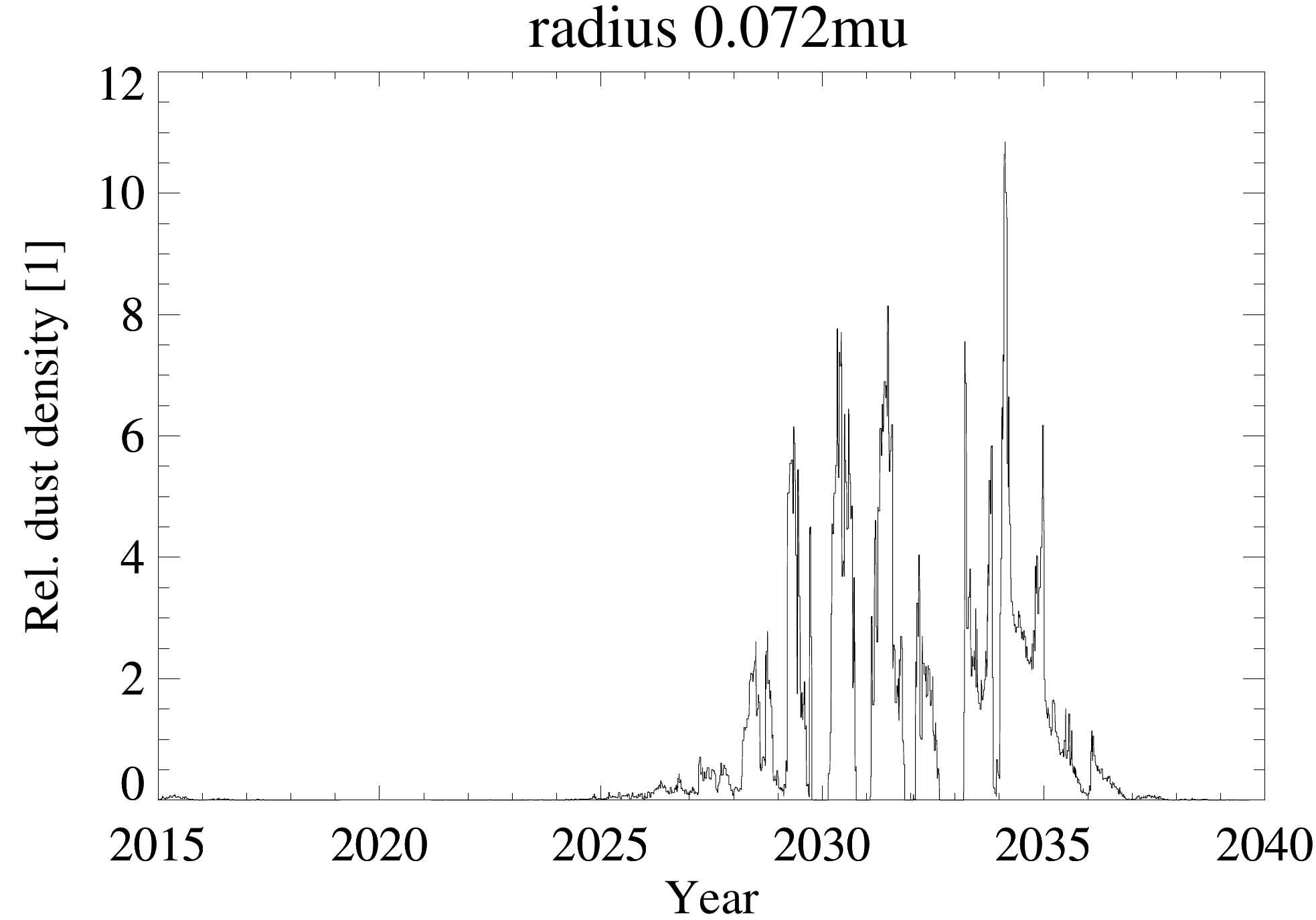} &
\includegraphics[width=0.3\textwidth]{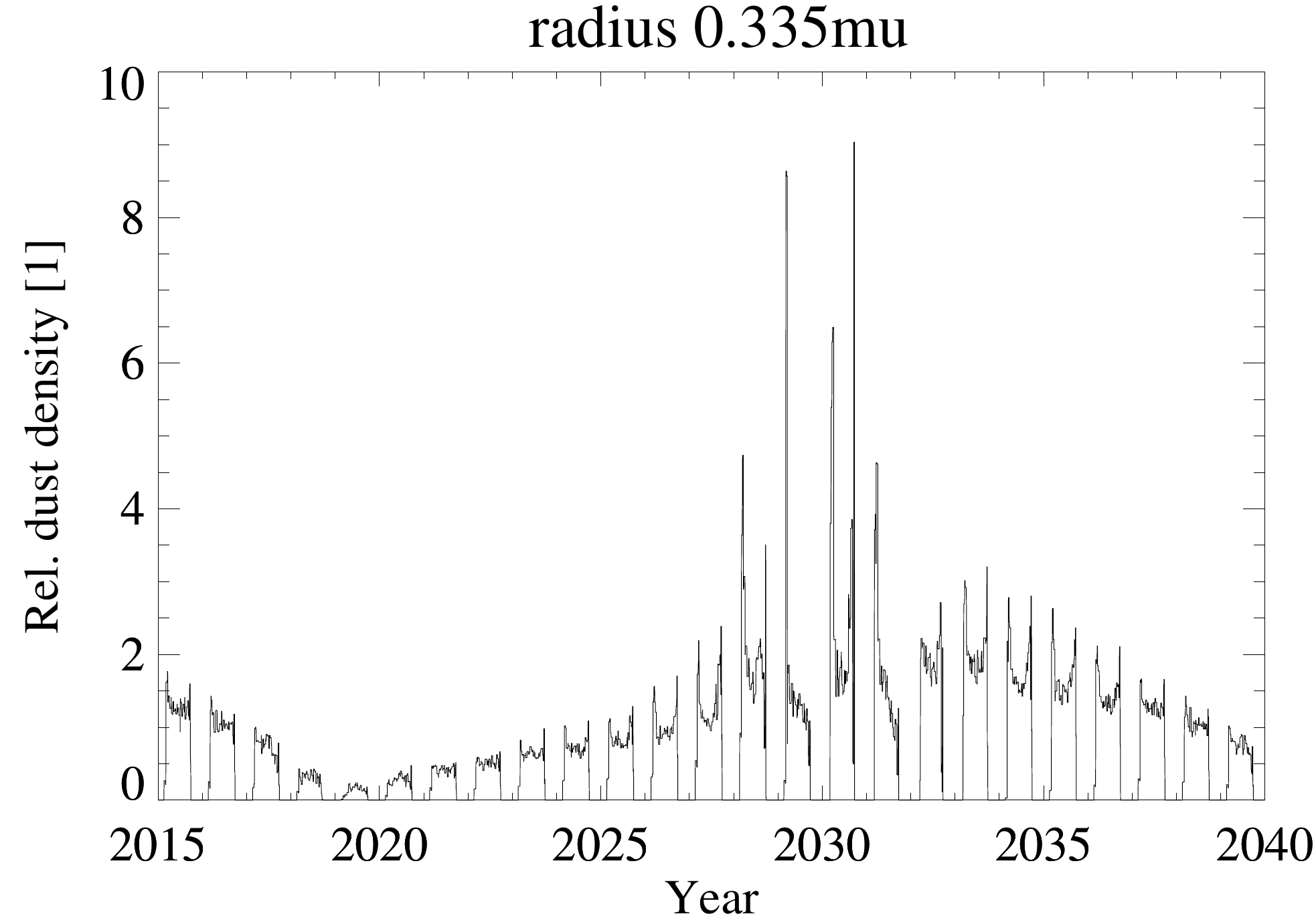} &
\includegraphics[width=0.3\textwidth]{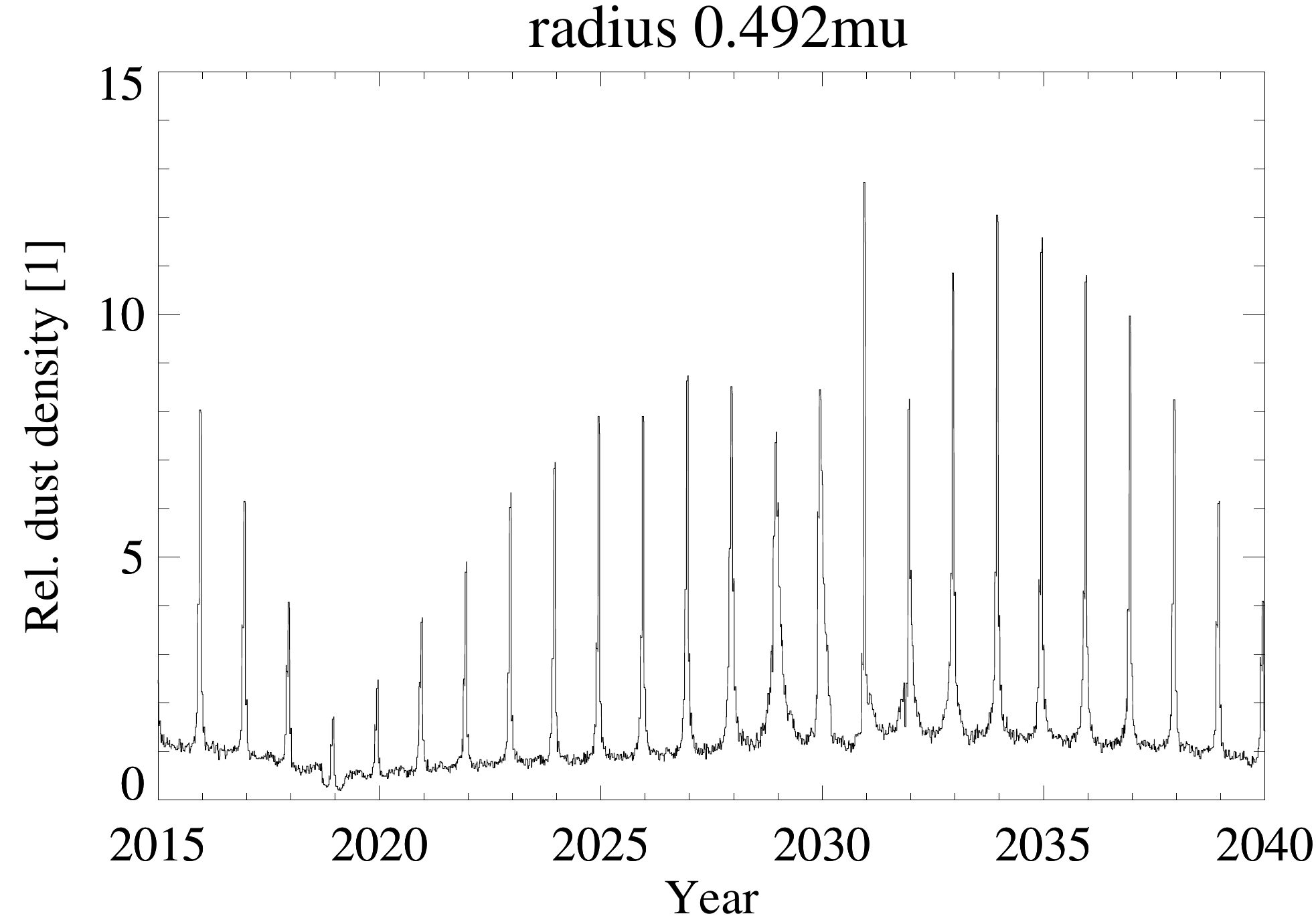} \\
\includegraphics[width=0.3\textwidth]{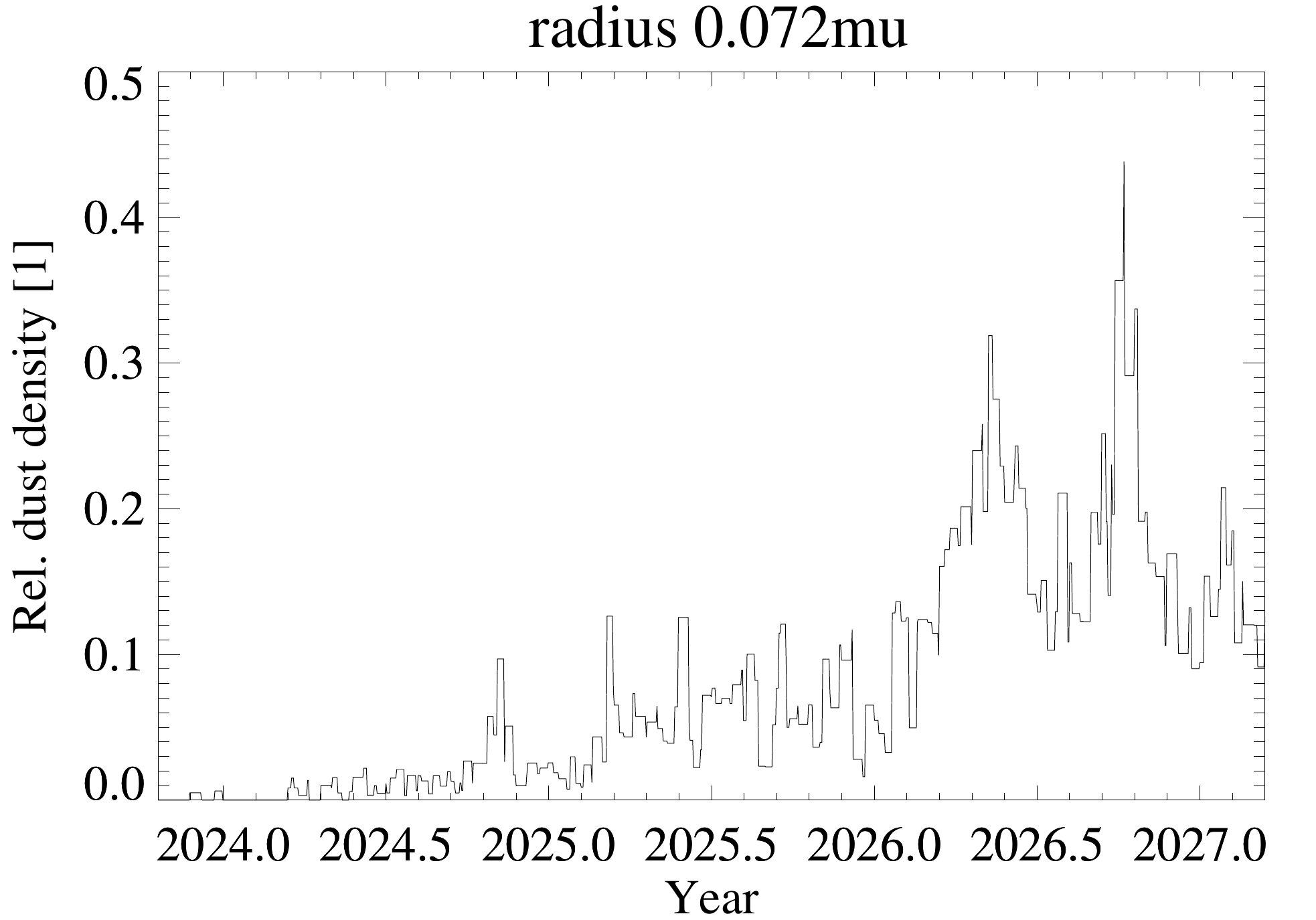} &
\includegraphics[width=0.3\textwidth]{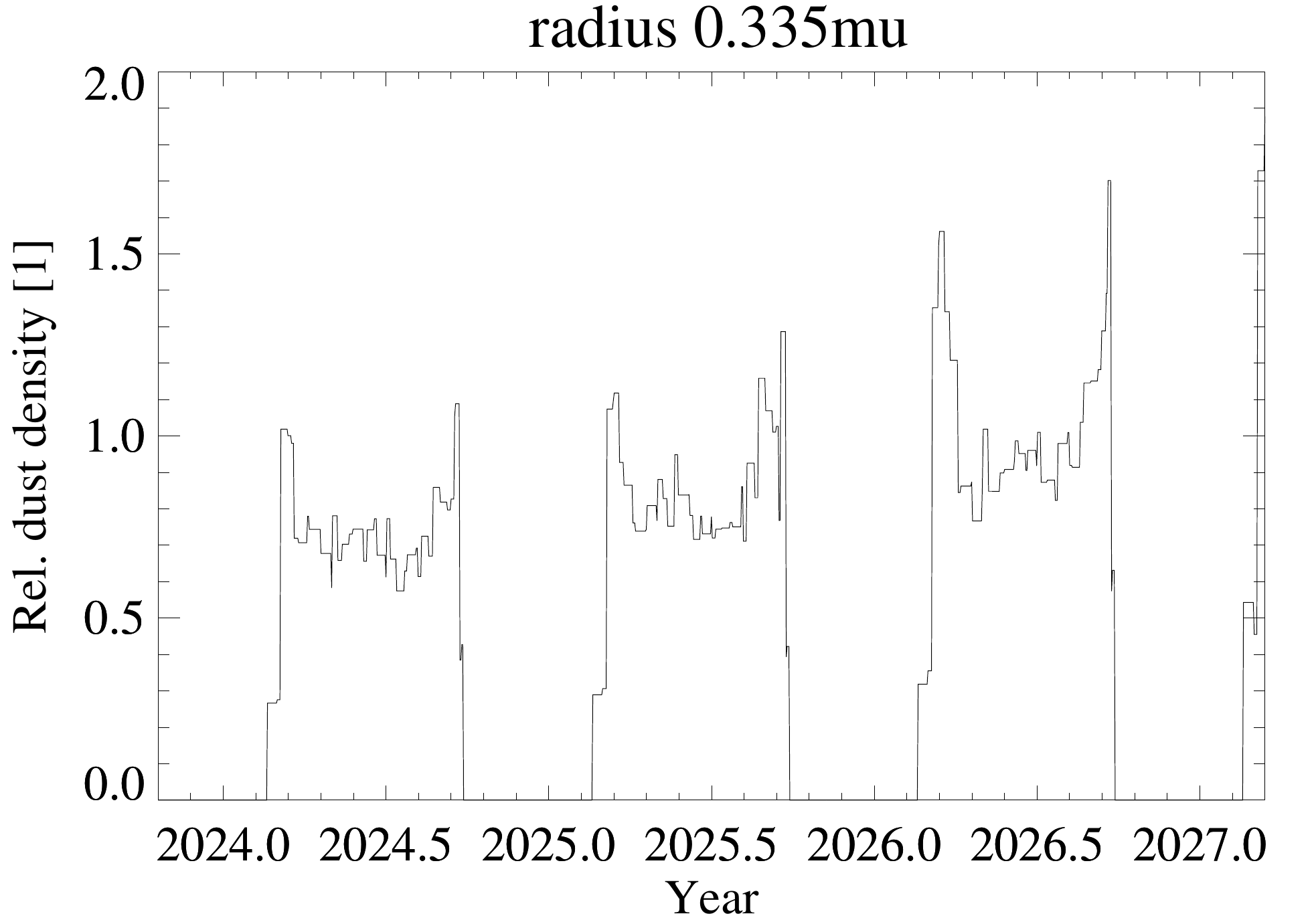} &
\includegraphics[width=0.3\textwidth]{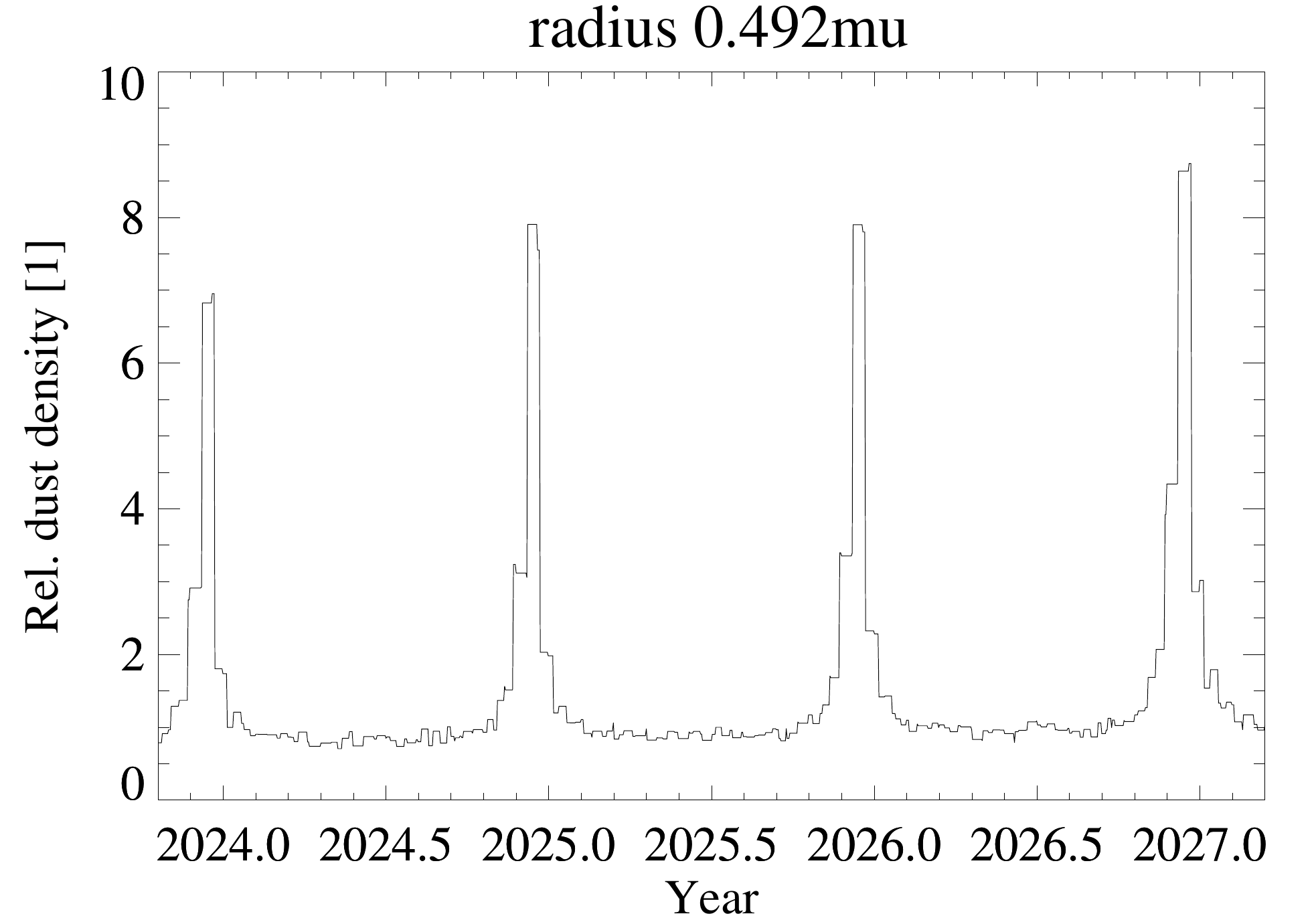} \\
\end{tabular}
\caption{Simulated relative ISD density (relative to the density of the undisturbed ISD flow) at Earth's position ({\em top row}), and the same figures zoomed in to a range of 3 years ({\em bottom row}).
}
\label{fig:earthorbit_first}
\label{fig:earthorbit_dens}
\label{fig:earthorbit_dens_zoomed}
\end{figure*}

\begin{figure*}
\centering
\begin{tabular}{ccc}
\includegraphics[width=0.3\textwidth]{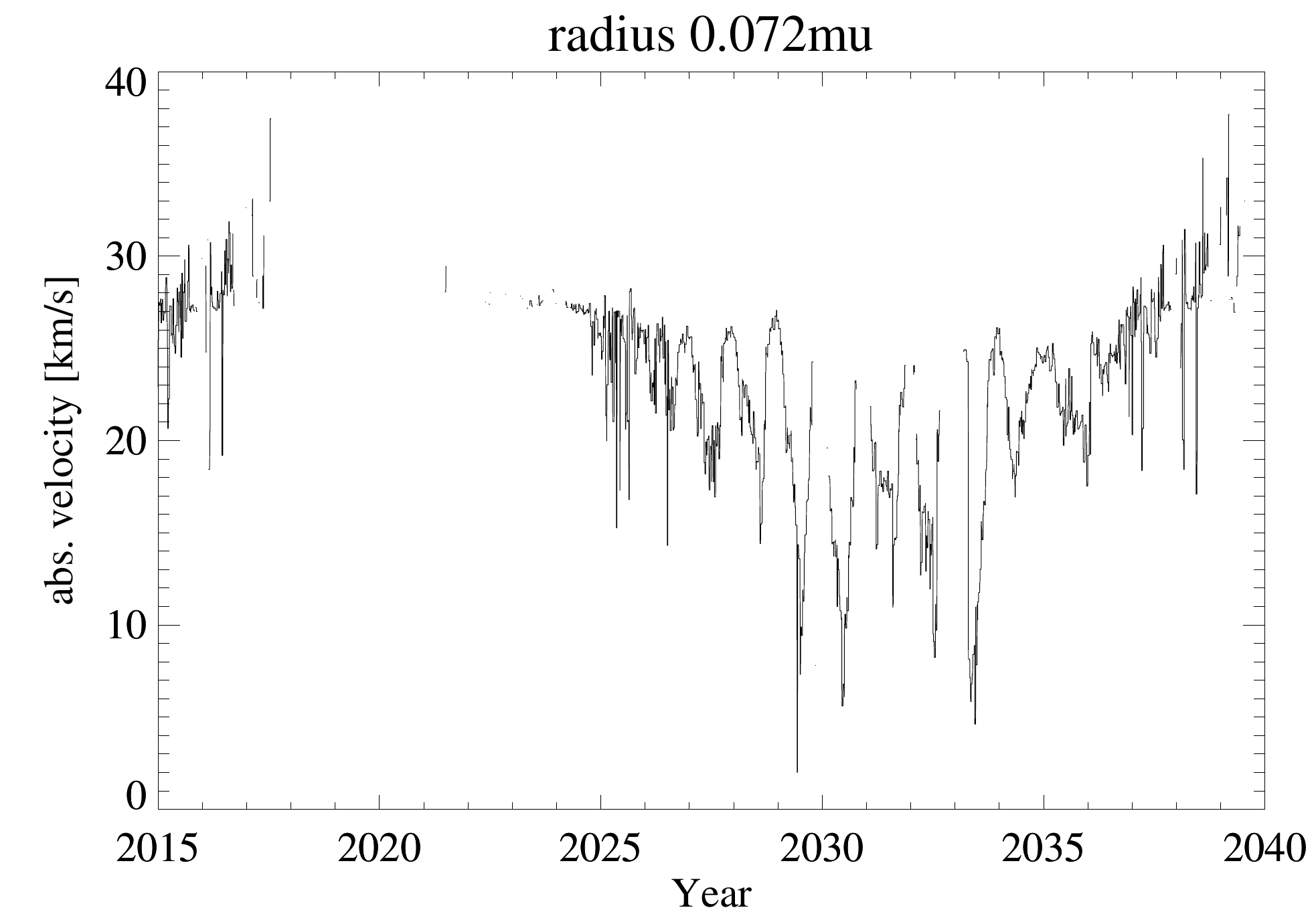} &
\includegraphics[width=0.3\textwidth]{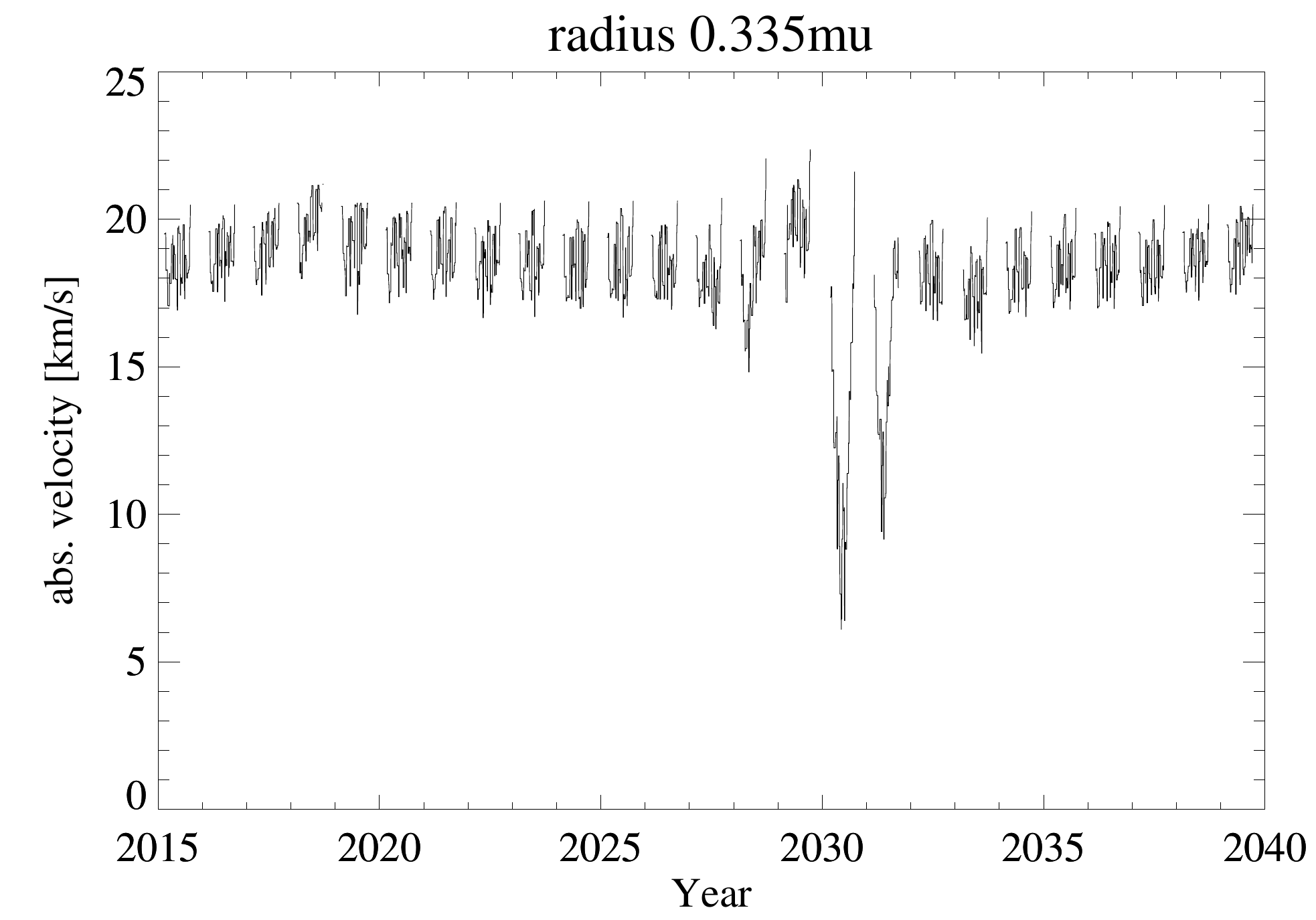} &
\includegraphics[width=0.3\textwidth]{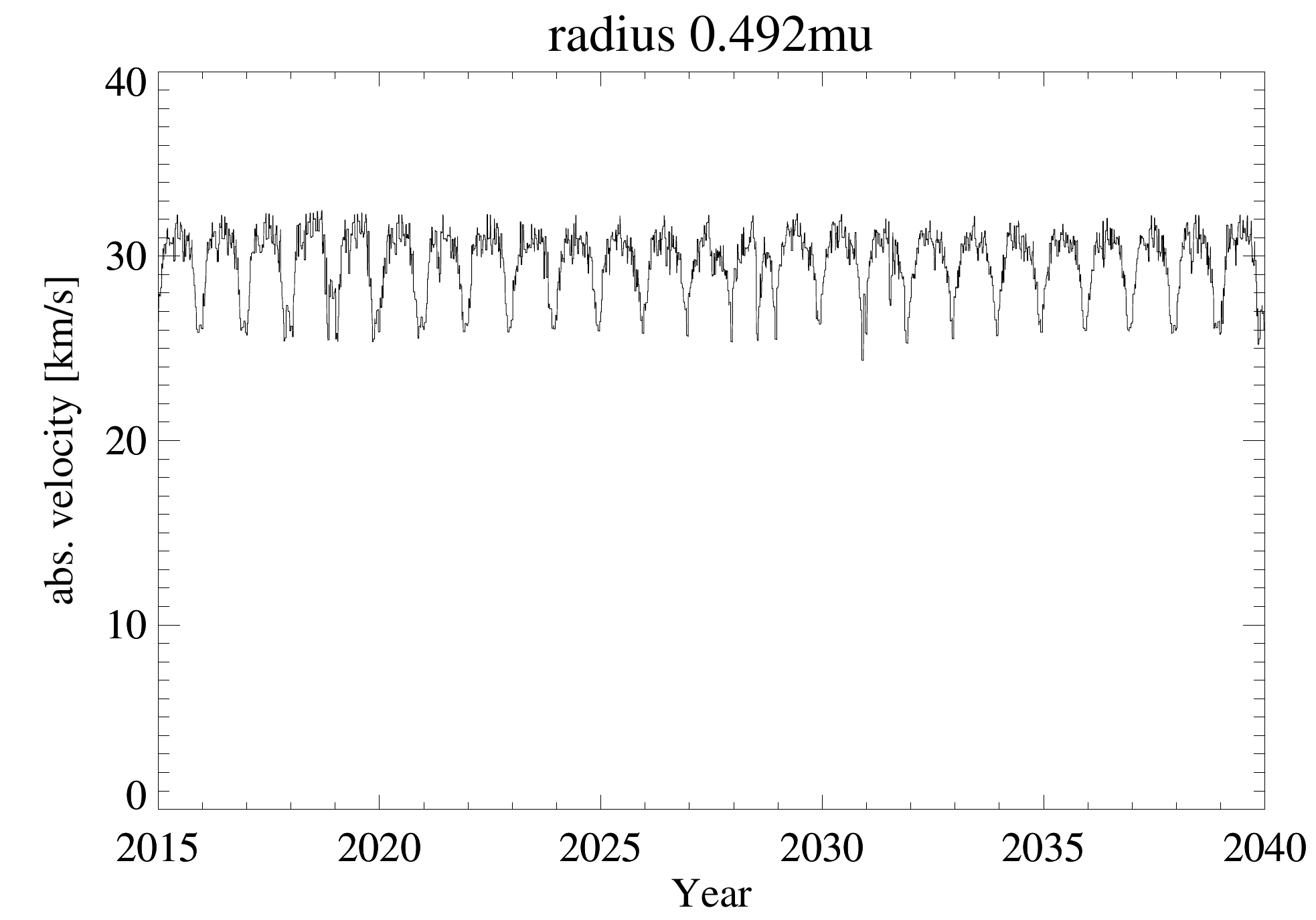} \\
\end{tabular}
\caption{Simulated heliocentric dust speed at Earth's position in the ecliptic reference frame. We note the different y-axis scalings.}
\label{fig:earthorbit_vel}
\end{figure*}

\begin{figure*}
\centering
\begin{tabular}{cc}
\includegraphics[width=0.4\textwidth]{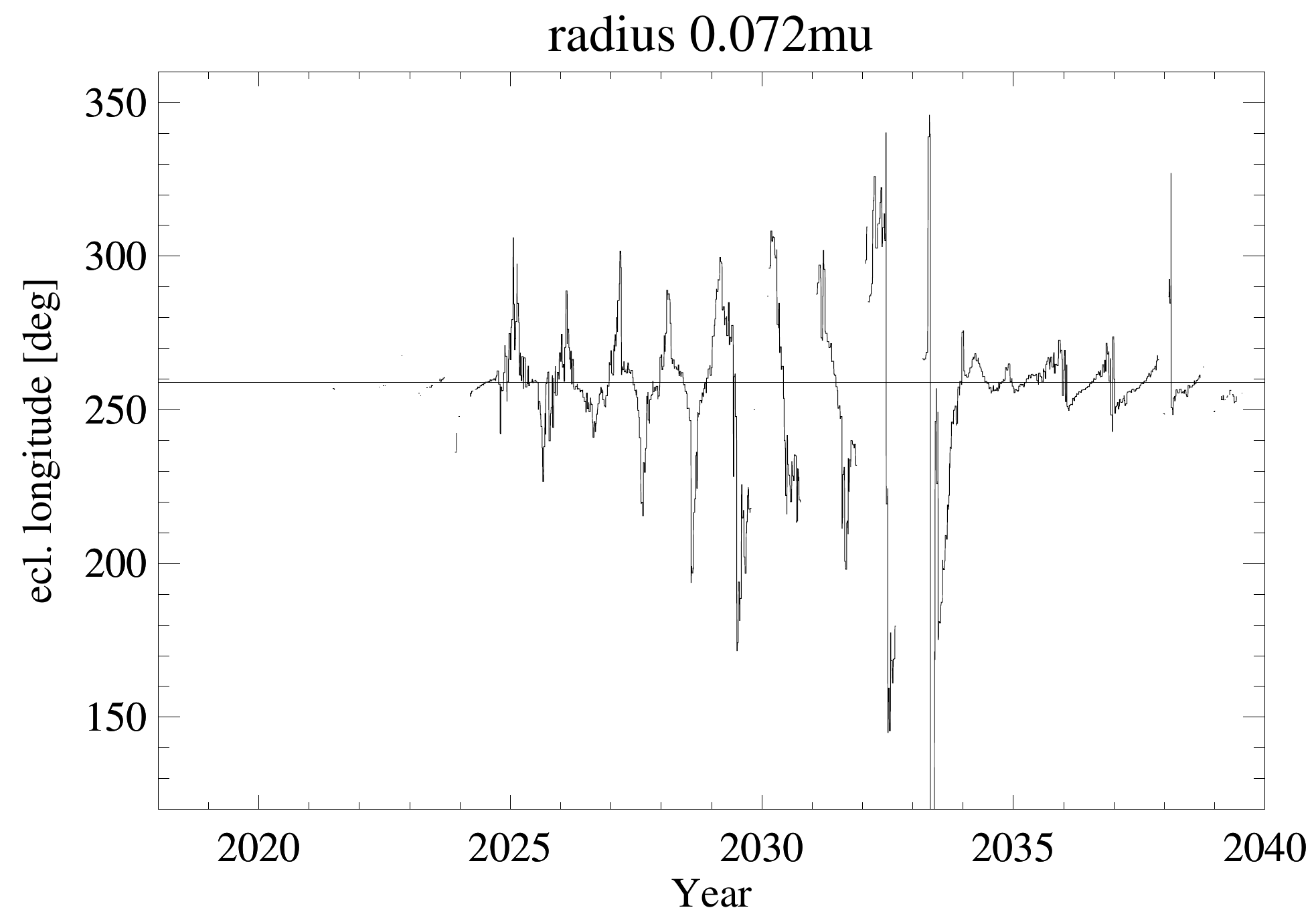} &
\includegraphics[width=0.4\textwidth]{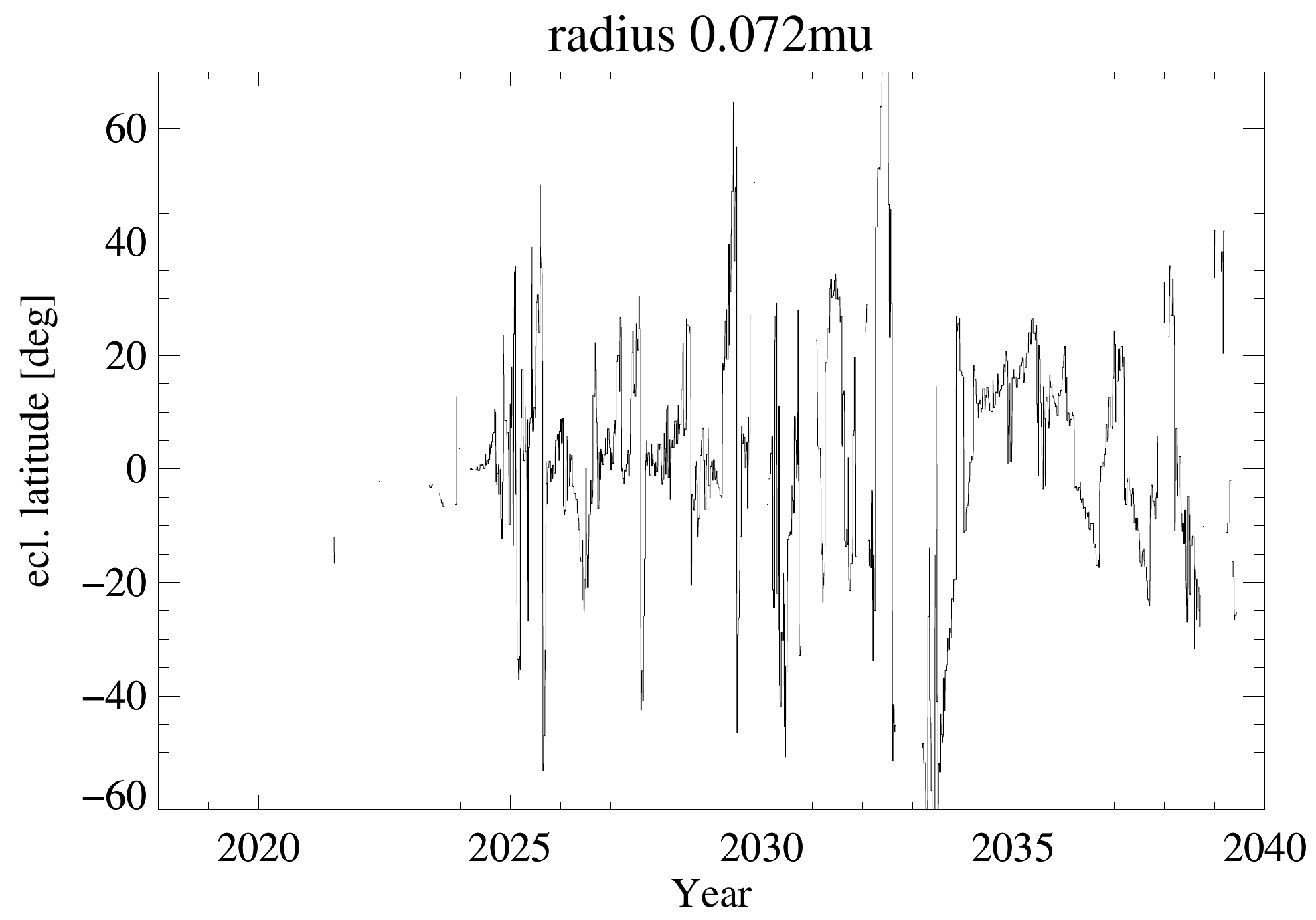} \\
\includegraphics[width=0.4\textwidth]{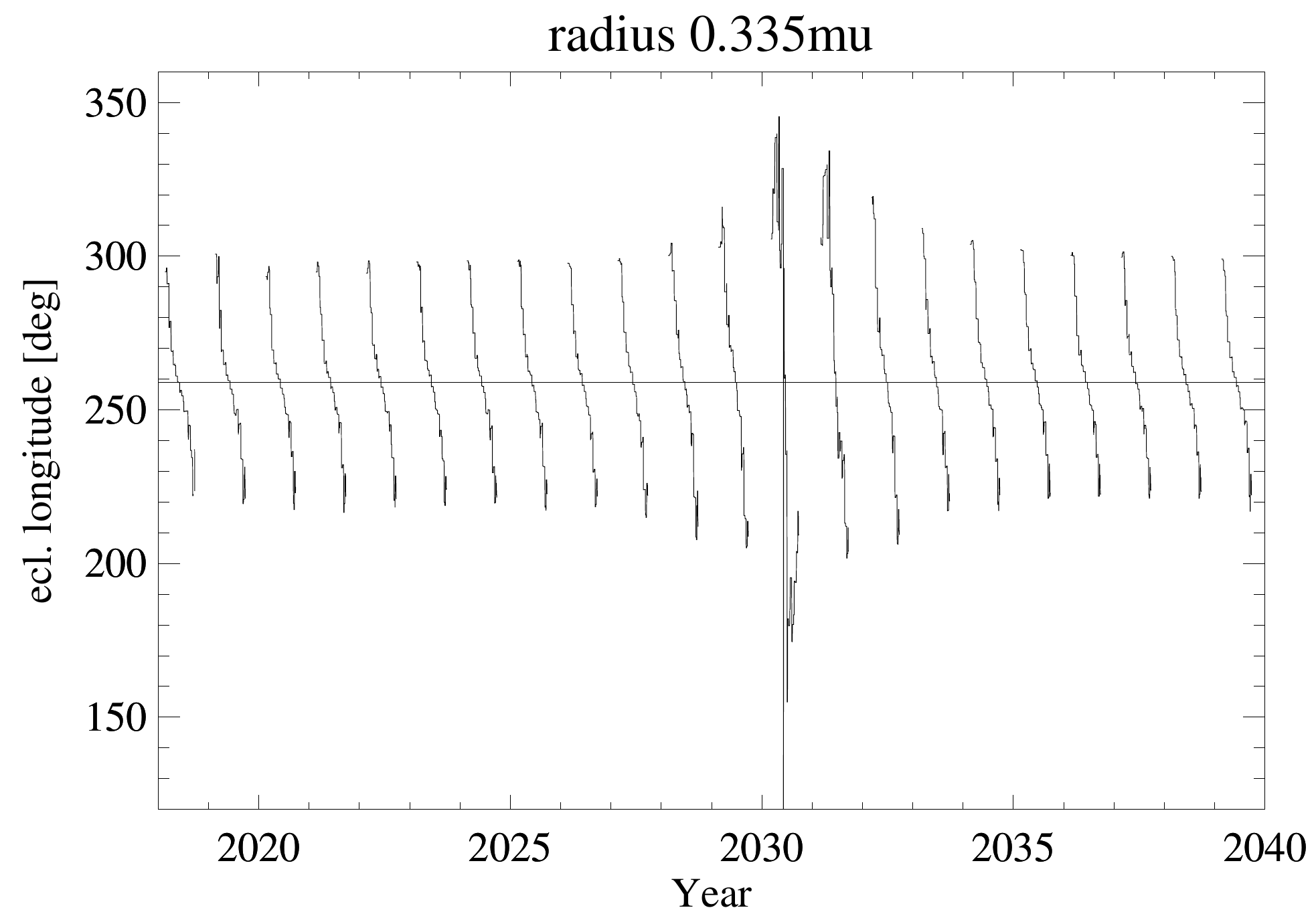} &
\includegraphics[width=0.4\textwidth]{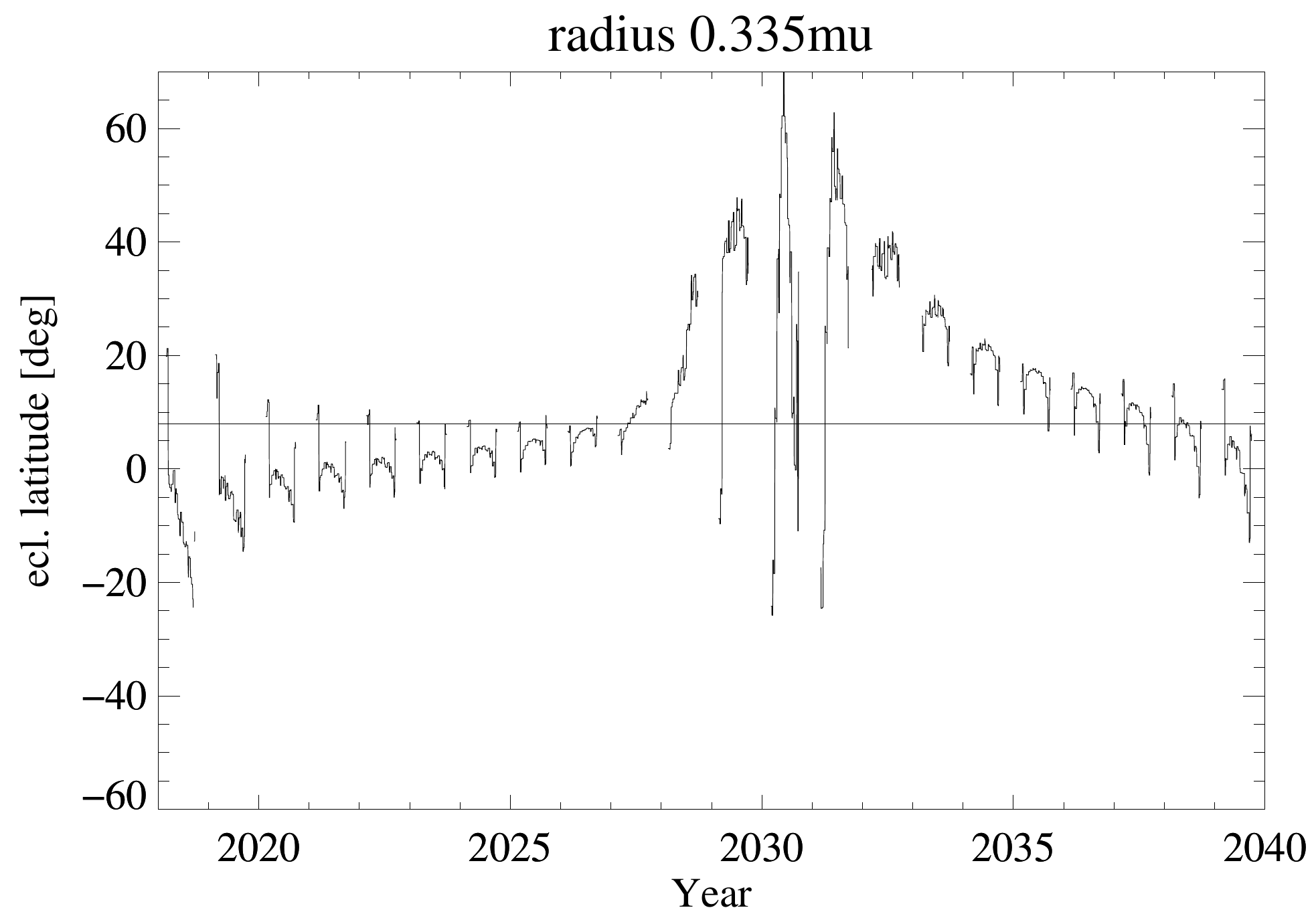}  \\
\includegraphics[width=0.4\textwidth]{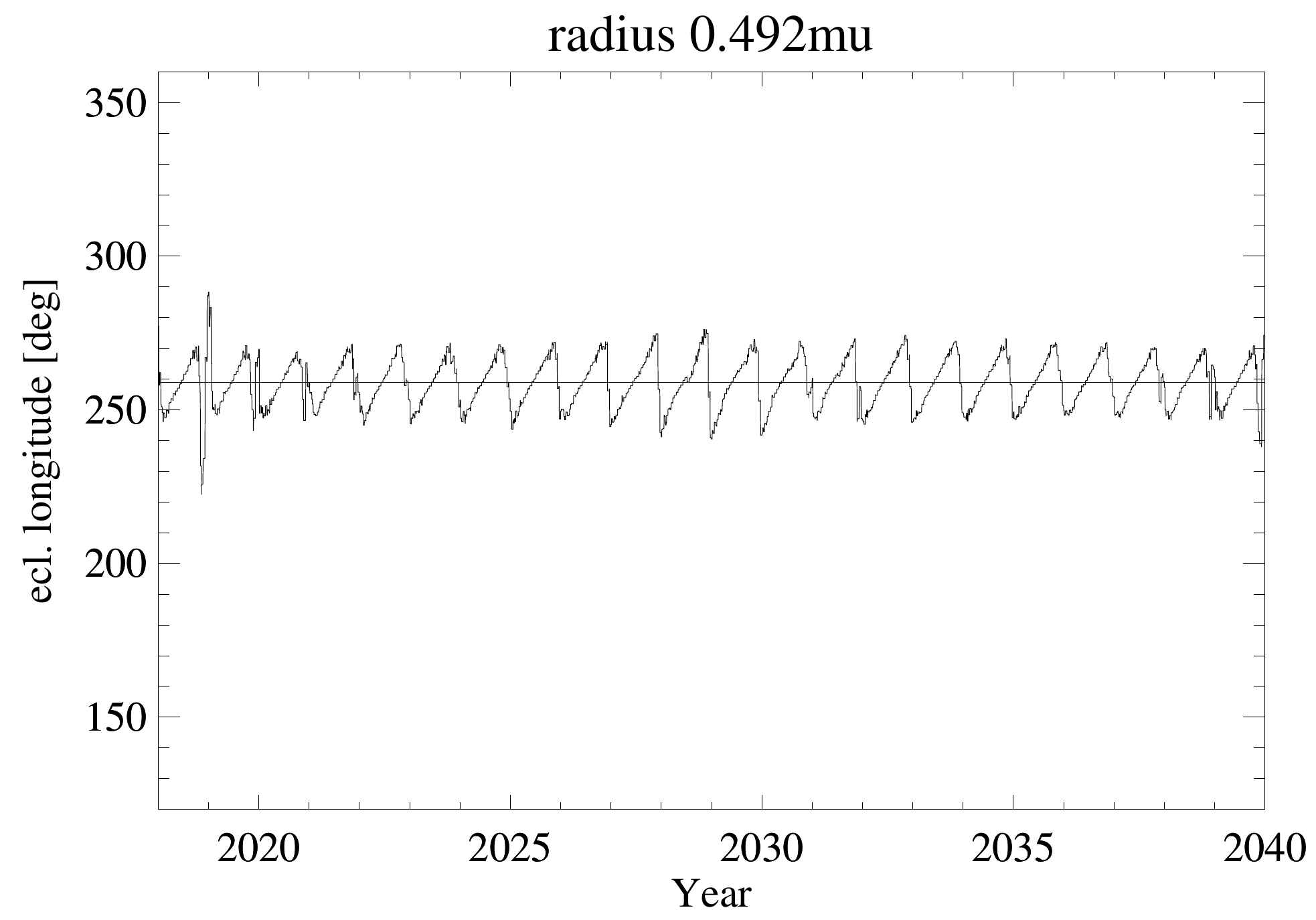} &
\includegraphics[width=0.4\textwidth]{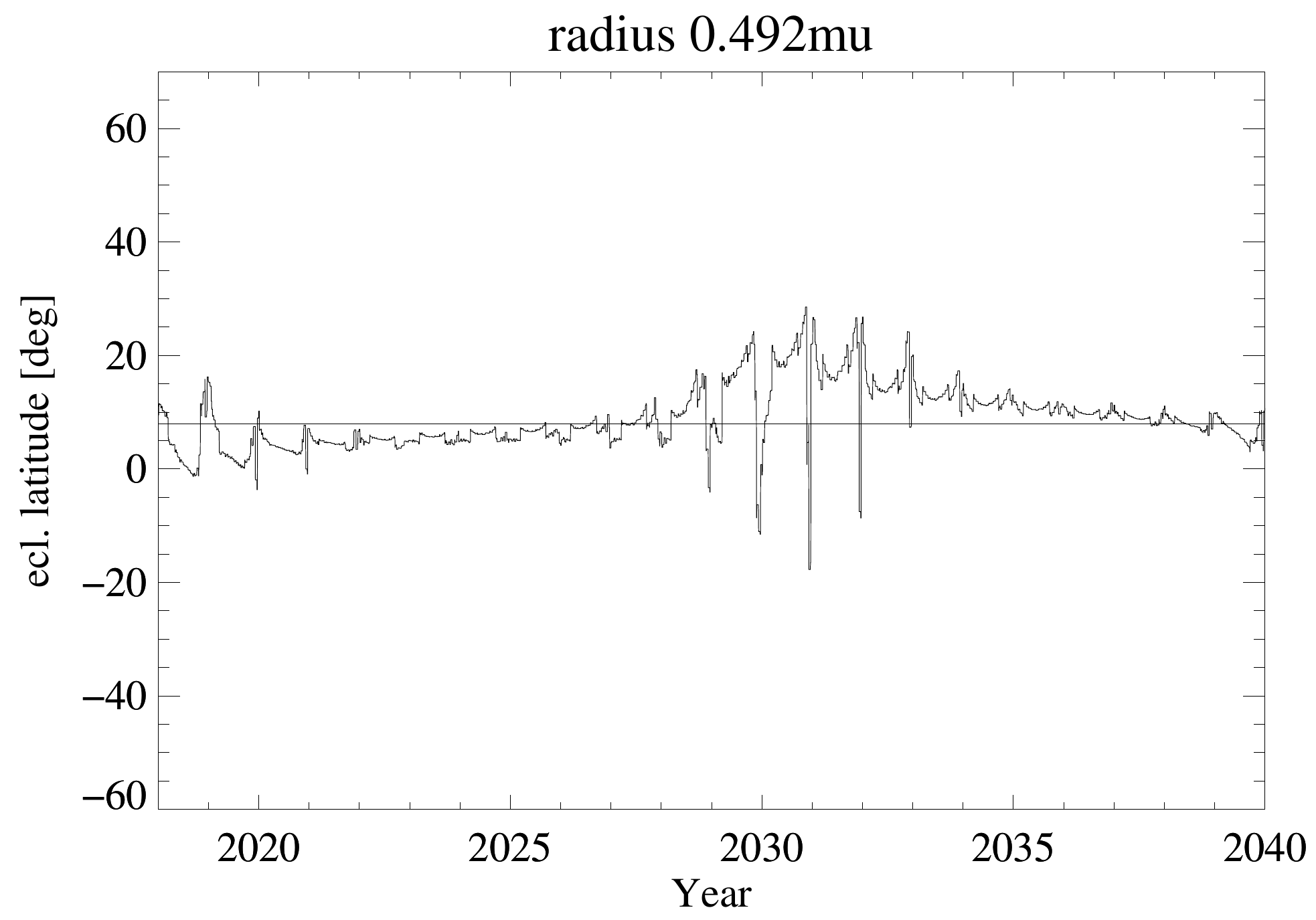}\\
\includegraphics[width=0.4\textwidth]{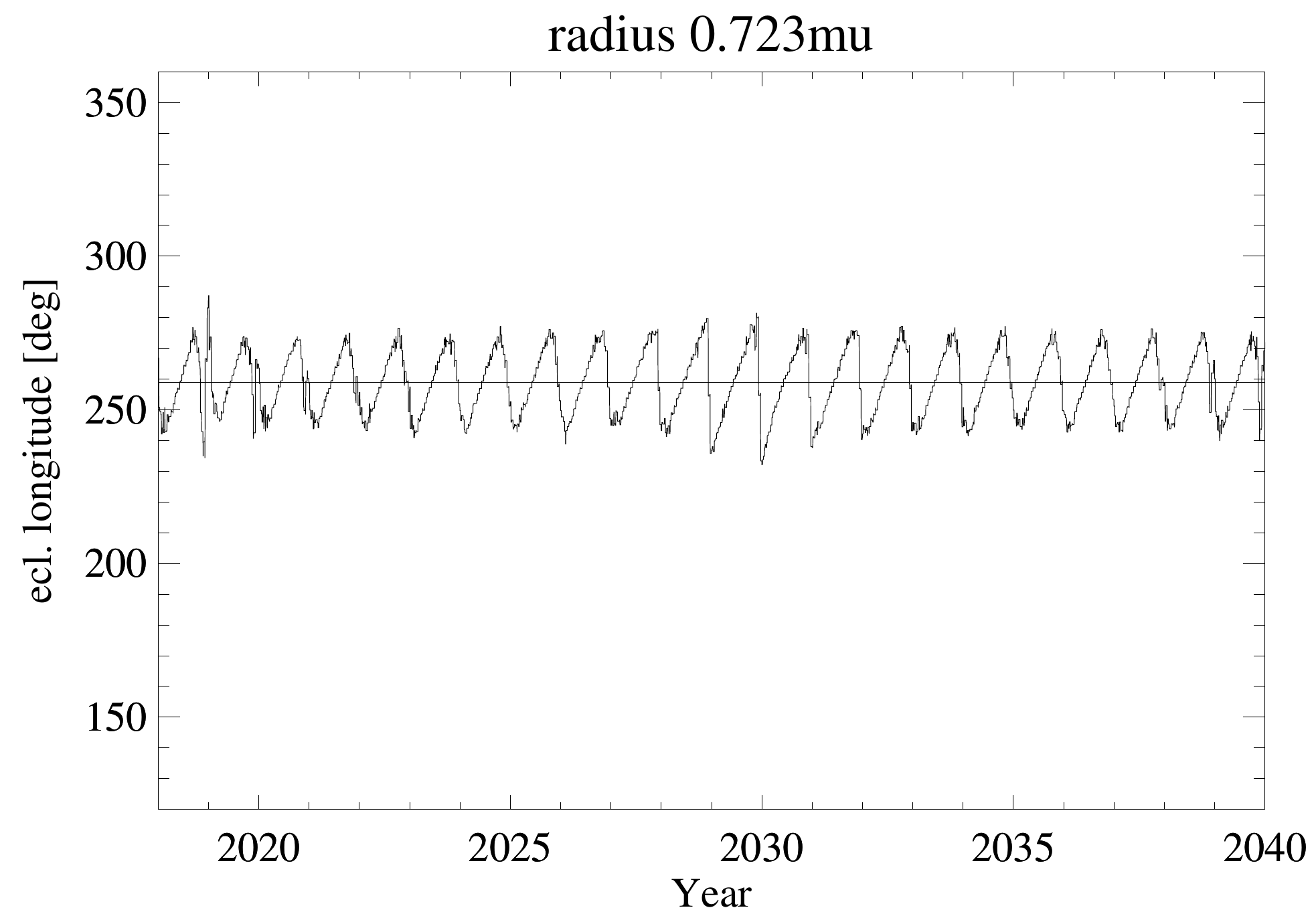} &
\includegraphics[width=0.4\textwidth]{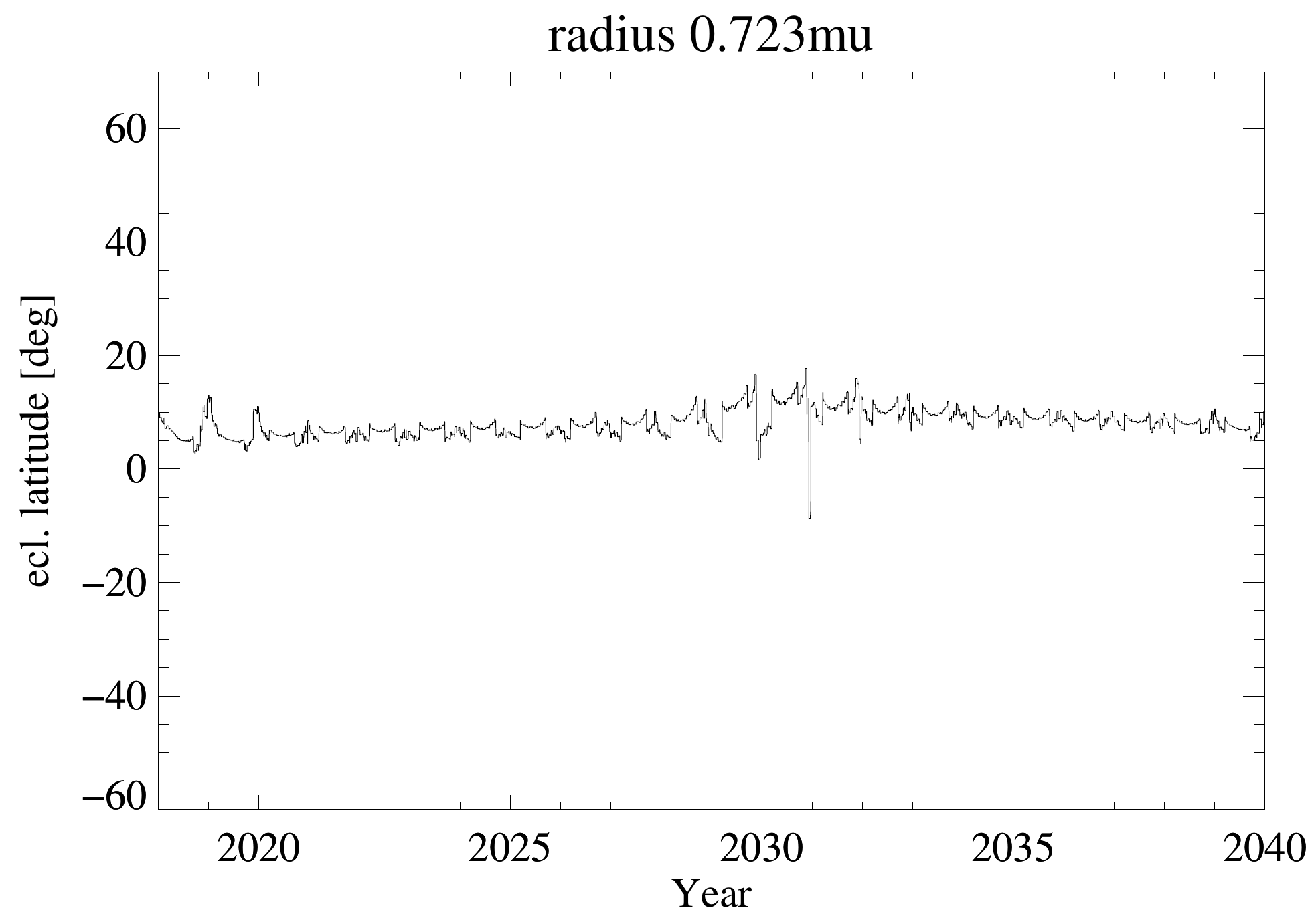}\\
\end{tabular}
\caption{Flow direction in ecliptic coordinates $(l, b)$ of ISD particles at Earth's position in the heliocentric ecliptic reference frame for particle sizes 0.07\,\mum, 0.34\,\mum, 0.49\,\mum, and 0.72\,\mum. The solid horizontal lines correspond to the incoming flow direction.}
\label{fig:lambet1}
\end{figure*}

\begin{figure*}
\centering
\begin{tabular}{ccc}
\includegraphics[width=0.3\textwidth]{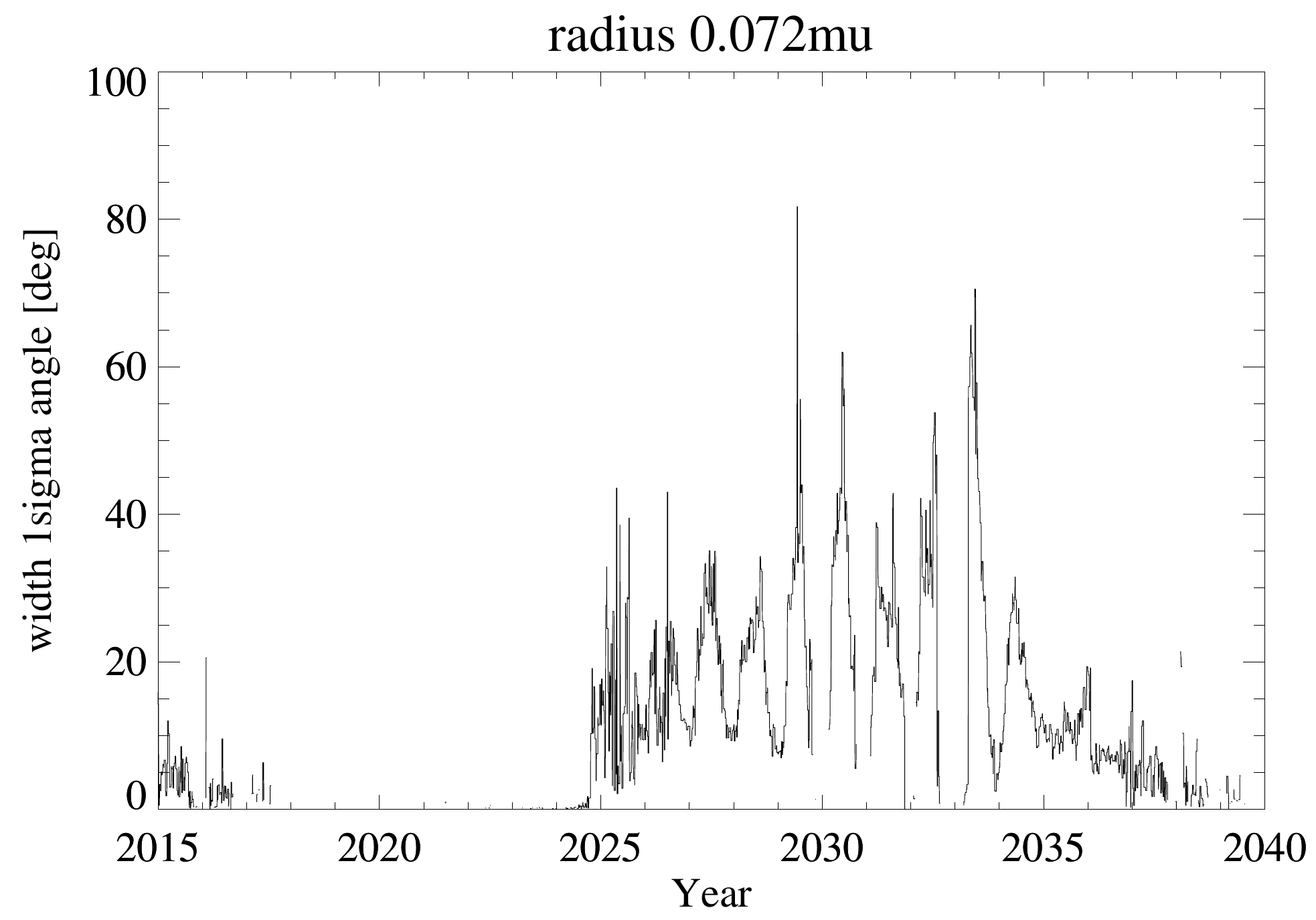} &
\includegraphics[width=0.3\textwidth]{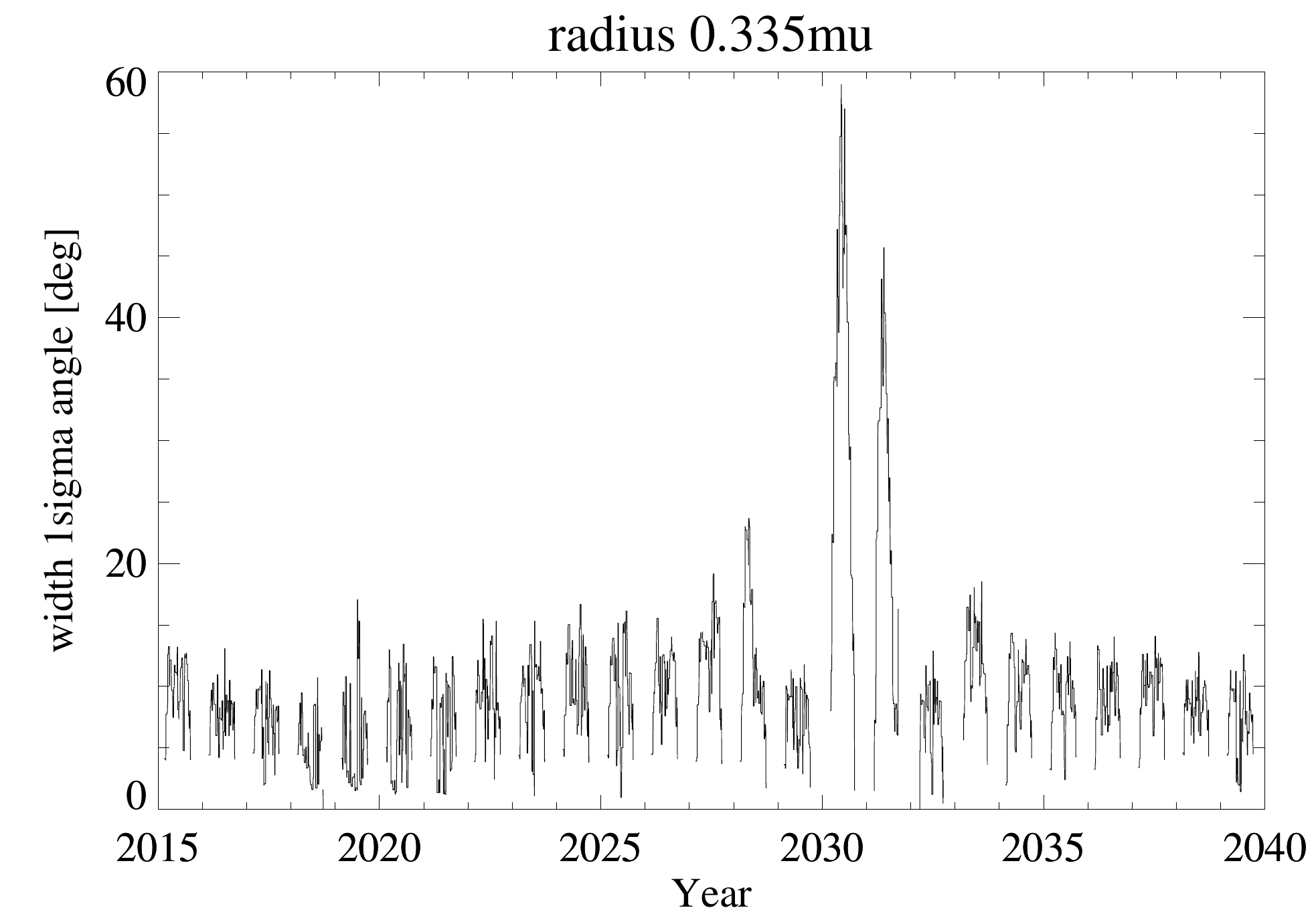} &
\includegraphics[width=0.3\textwidth]{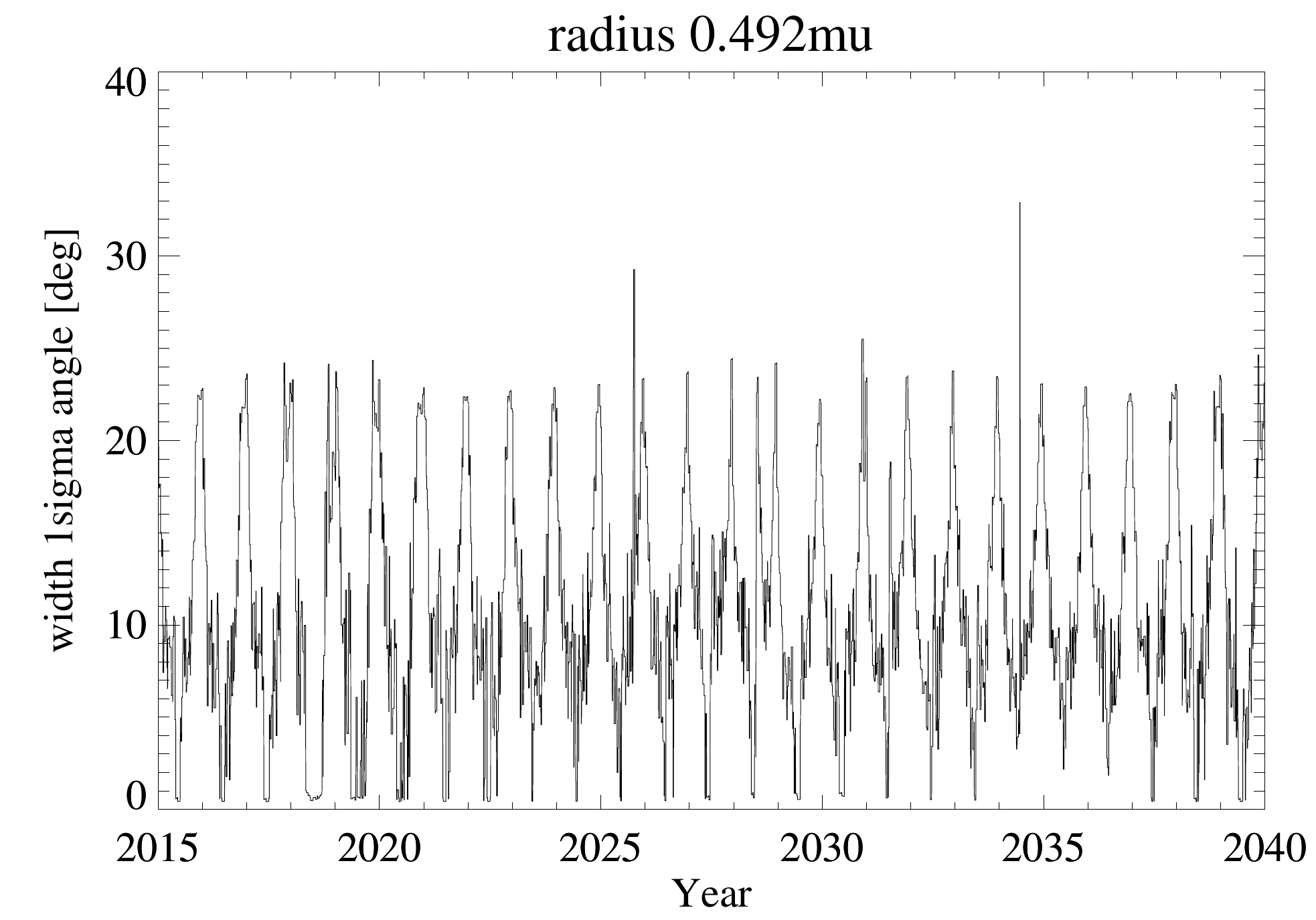} \\
\end{tabular}
\caption{$1\sigma$ stream width at Earth position in the heliocentric reference frame.
}
\label{fig:earthorbit_wid}
\end{figure*}

We compare the heliocentric flow properties of these particles in Figures \ref{fig:earthorbit_dens} to \ref{fig:earthorbit_wid} at the Earth's positions along its orbit. For medium-sized and large particles (0.34\,\mum\ and 0.49\,\mum, respectively), sampling of different regions along the Earth's orbit is the dominant cause for the observed variations, but for small particles ($a_d\lesssim0.1$\,\mum) the ISD flow direction is intrinsically time-dependent due to the strong modulation by the solar wind magnetic field.

For particles with $\beta<1$ the gravitational focussing leads to high-density regions 
downstream of the Sun, where Earth passes on 13 December every year. The point closest to the upstream direction is crossed on 12 June. 

\begin{table}
\begin{center}
\caption{Particle radii $a_d$ and radiation pressure factors $\beta$ used in the simulations. The force dominating the dynamics of a given size can be the radiation pressure (RP), solar gravity, and/or the electromagnetic (EM) force. Particles around $0.2\,\mu$m are not observable in the Earth's orbit because the repulsive force of the radiation pressure leads to a deflection distance $r_h> 1\,$AU from the Sun for $\beta \gtrsim 1.4$. }
\label{tbl:simsizes}
\begin{tabular}{lrll}
\hline \hline
$a_d$ [$\mum$] &  mass [kg] & $\beta$  & dominant force \\
\hline
& & & \vspace{-0.7em}\\
0.049 & $1.00\times10^{-18}$ & 0.55 & EM\\
0.072 & $3.16\times10^{-18}$ & 0.95 & EM\\
0.11 & $1.00\times10^{-17}$ & 1.38 & RP and EM\\
0.16 & $3.16\times10^{-17}$ & 1.59 & (not observable at Earth)\\
0.23 & $1.00\times10^{-16}$ & 1.49 & (not observable at Earth)\\
0.34 & $3.16\times10^{-16}$ & 1.17 & RP and EM\\
0.49 & $1.00\times10^{-15}$ & 0.81 & gravity\\
0.72 & $3.16\times10^{-15}$ & 0.52& gravity\\
1.1  & $1.00\times10^{-14}$ & 0.33 & gravity\\
1.6  & $3.16\times10^{-14}$ & 0.21 & gravity\\
2.3  & $1.00\times10^{-13}$ & 0.14 & gravity\\
3.4  & $3.16\times10^{-13}$ & 0.09 & gravity\\
4.9  & $1.00\times10^{-12}$ & 0.06 & gravity\\
\hline
\end{tabular}
\end{center}
\end{table}

Figure~\ref{fig:earthorbit_dens} shows the spatial density and its fluctuations over the  solar cycle. Besides strong annual variations due to the deflection by radiation pressure, the focussing due to the magnetic field is most pronounced for the smallest particles, and is strongest around the year 2031, in agreement with other studies of ISD in the solar system (further away from the Sun).

This figure also zooms in on the period of moderate electromagnetic focussing between 2024 and 2027, where seasonal variations become obvious. The flux of small particles ($0.07\,\mum$) displays a strong enhancement during times of maximum focussing around 2031, but no annual variation. Conversely, the flux of medium-sized particles, 0.34\,\mum, shows a repeated annual pattern. It becomes zero for several months around December because the Earth is in the $\beta$-cone for these particles, while for a period of several months centred around June the flux is appreciable (on average $\sim 30\%$ of the incoming flux).  
The dust flow lines are compressed near the edge of the $\beta$-cone, leading to an increased density there. Additionally, the flux of large particles is largely constant for most of the year. There is a strong maximum downstream of the Sun around December 13 as a result of the gravitational focussing by the Sun, even though the Earth does not pass centrally through the focussing spot (its orbital plane is inclined by 8\deg\ with respect to the ISD flow). The density in the focussing region shows a typical enhancement by a factor of $\sim$5, and lasts 1-2 months (Fig.~\ref{fig:earthorbit_dens_zoomed}, lower right panel, zoomed particle density at 0.49\,\mum). 

Small particles with $a_d\lesssim 0.1\,\mum$ exhibit an irregular variability on timescales much less than one year (Fig.~\ref{fig:earthorbit_dens}, lower panels). This can be explained by the strong interaction with the magnetic field, which leads to strong concentrations on small spatial and temporal scales. However, the precise location of these density enhancements is very sensitive to $Q/m$, and therefore to the fact that in our simulations we use only a single particle size per bin, rather than a size distribution. With more realistic size distributions some of these density variations are likely to average out.

Figure~\ref{fig:earthorbit_vel} demonstrates the variation in the heliocentric speed of ISD particles at the Earth position. The heliocentric speed of the smallest particles displays strong variations around the speed of the incoming ISD flow ($26\,\mathrm{km\,s^{-1}}$). The mid-range particles are decelerated when they reach 1\,AU because of the dominant radiation pressure. The dynamics of large particles is dominated by solar gravity; at 1AU, particles with sizes of $a_d=0.49\,\mum$  ($a_d\gtrsim 0.72\,\mum)$ experience a net acceleration
to $30\,\mathrm{km\,s^{-1}}$  ($40\,\mathrm{km\,s^{-1}}$, respectively).
In Fig.~\ref{fig:lambet1} we show the inflow direction in terms of ecliptic longitude/latitude. The ISD flow of specific sizes is not highly collimated. Within a volume element of 0.25\,AU on the side, the trajectories of ISD particles vary along the Earth's orbit by up to 30\deg. Due to solar gravity and radiation pressure, the largest directional variation inside a single mesh cell can be seen downstream of the Sun where particles become focussed and reach the Earth's orbit  from a wide range of directions after experiencing a strong gravitational focussing (Fig.~\ref{fig:earthorbit_wid}).

\begin{figure*}
\centering
\begin{tabular}{ccc}
\includegraphics[width=0.3\textwidth]{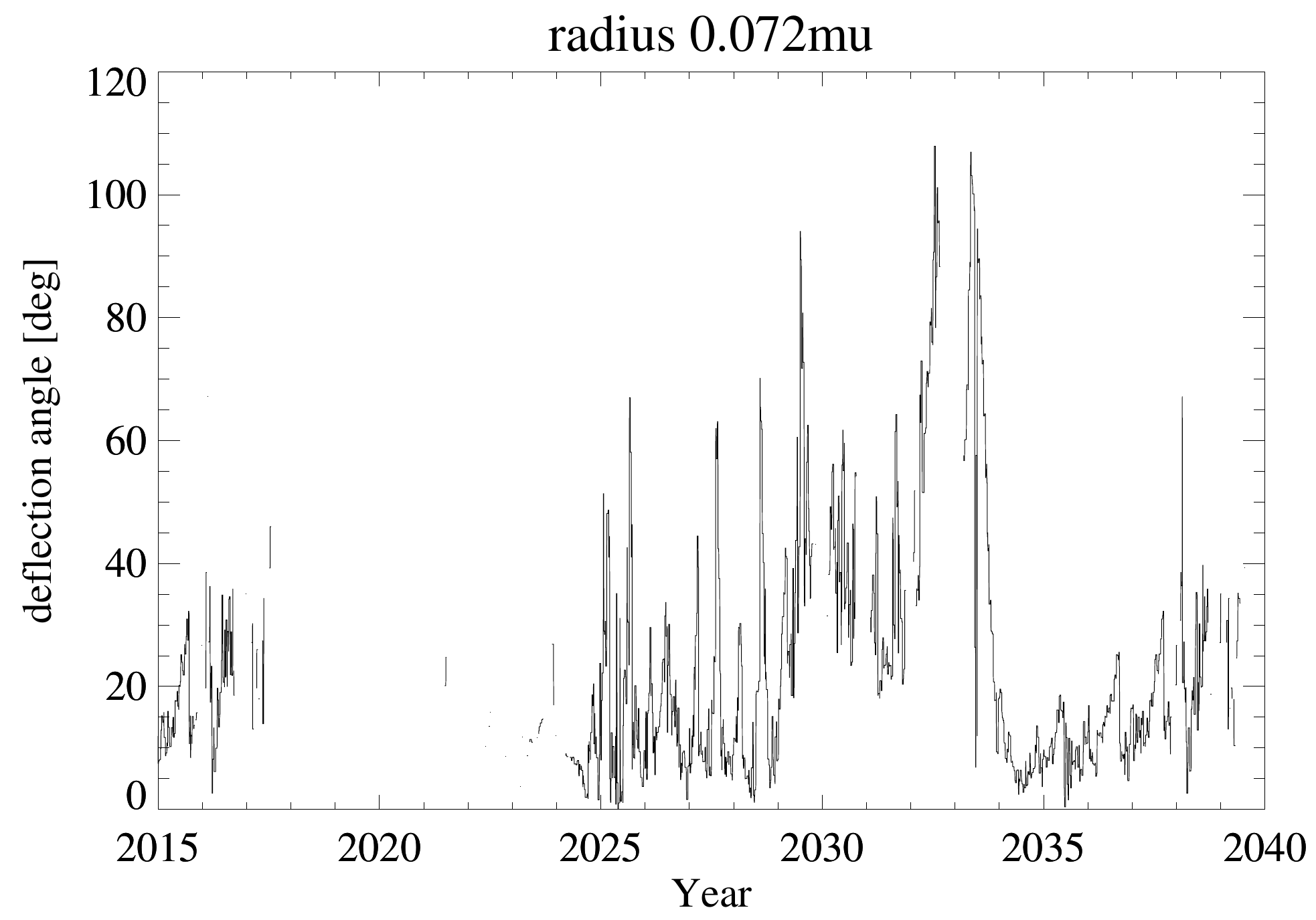} &
\includegraphics[width=0.3\textwidth]{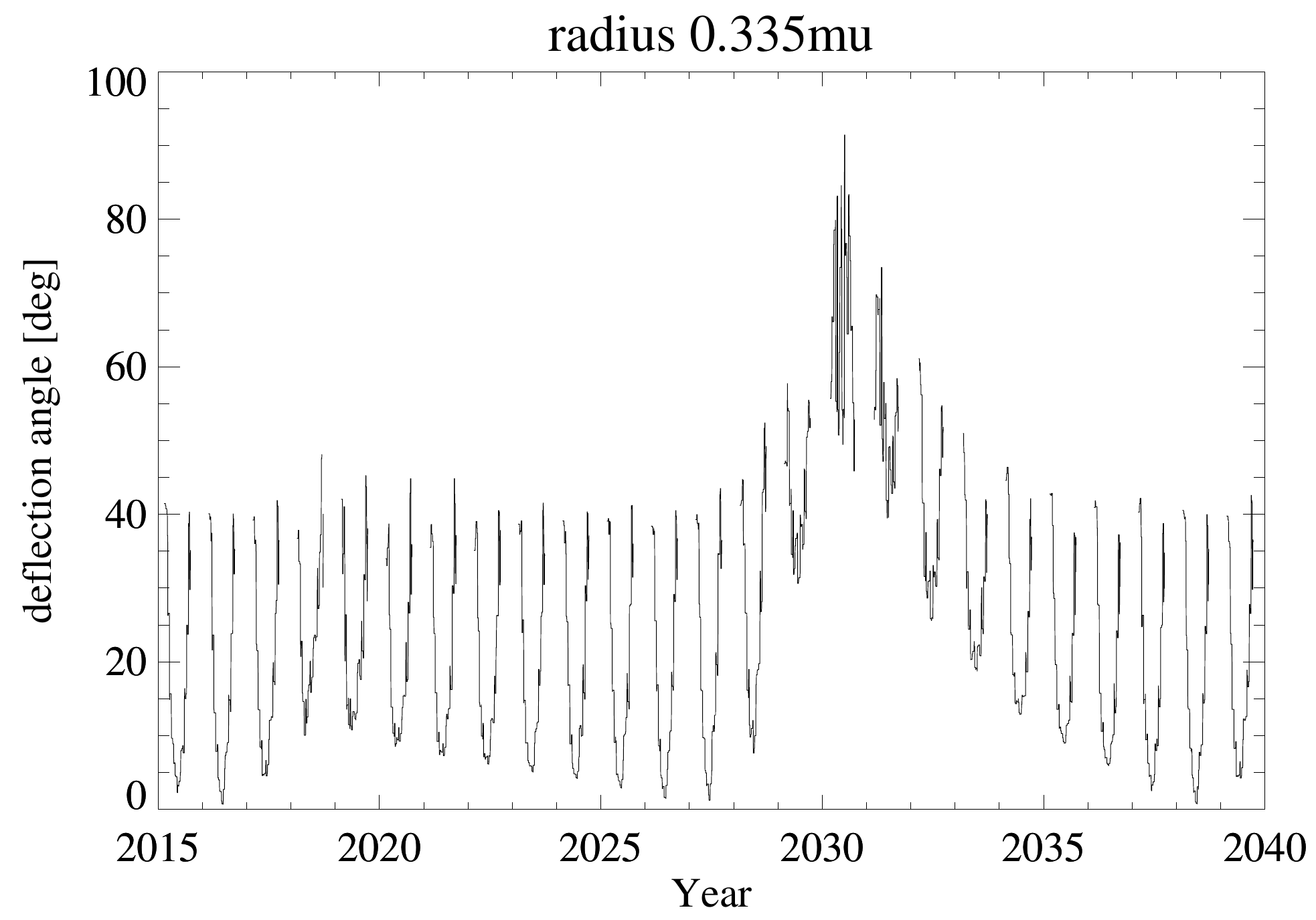} &
\includegraphics[width=0.3\textwidth]{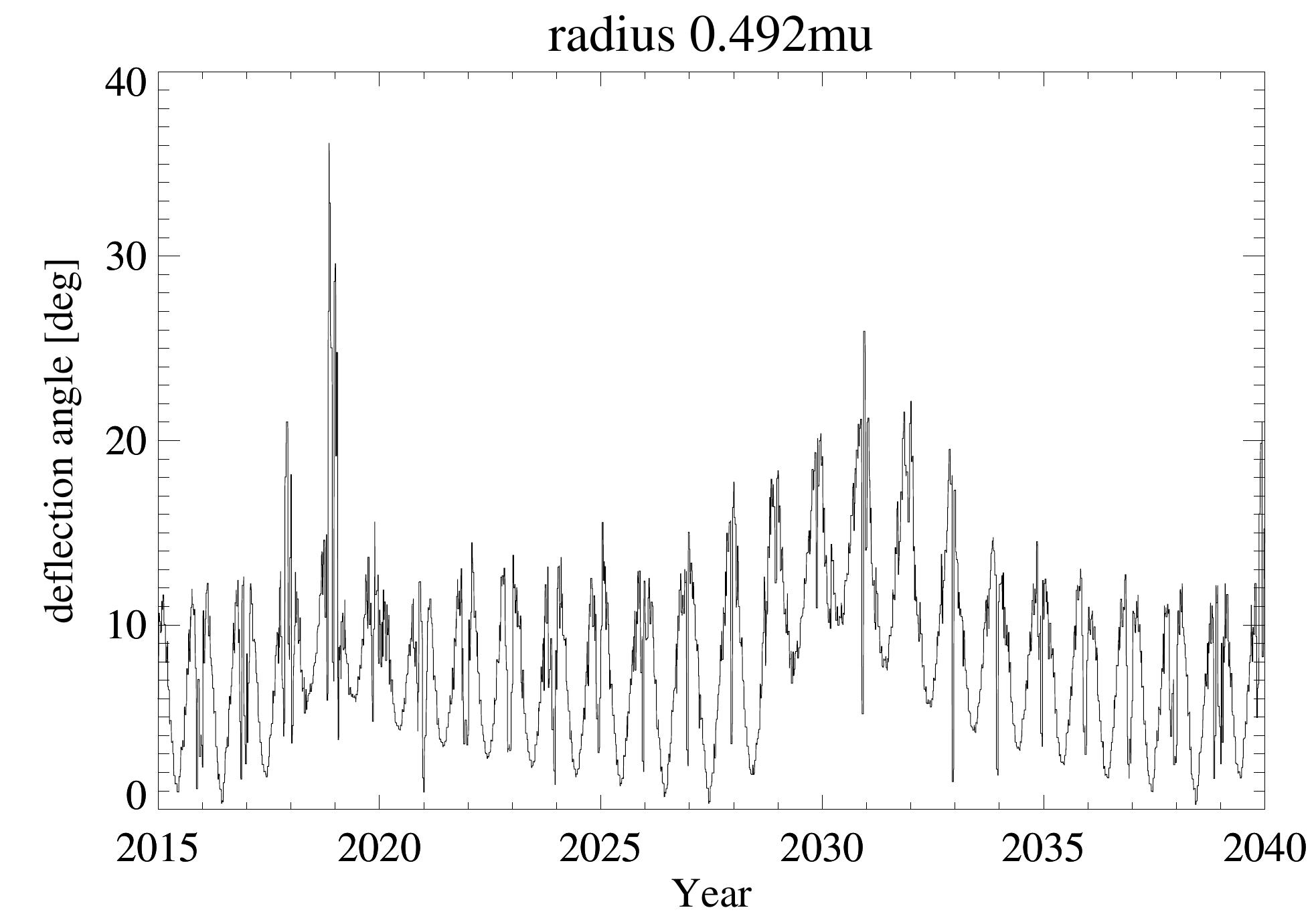} \\
\end{tabular}
\caption{Deflection angle of ISD particles at Earth orbit with respect to the ISD inflow direction into the solar
system.}
\label{fig:earthorbit_ang}
\end{figure*}

Figure~\ref{fig:earthorbit_ang} shows the deflection angle relative to the inflow direction of ISD particles at the Earth's position (in the ecliptic reference frame) over time. The flow of the small particles is mostly irregular, whereas the larger particles (0.34\,\mum\ and 0.49\,\mum) generally follow a yearly modulation. Close to the downstream focussing spot, the deflection is largest because here the effects of gravitational focussing are strongest \citep{sterken2013a}. However, for particles of $a_d = 0.49\,\mum$, the deflection drops steeply inside the focussing spot. This is due to the fact that trajectories from different directions hit the same spot and the deflections cancel out on average. This is reflected in the increase in width of the ISD flow towards the focussing spot (Fig.~\ref{fig:earthorbit_wid}).


\begin{figure*}
\centering
\begin{tabular}{ccc}
\includegraphics[width=0.3\textwidth]{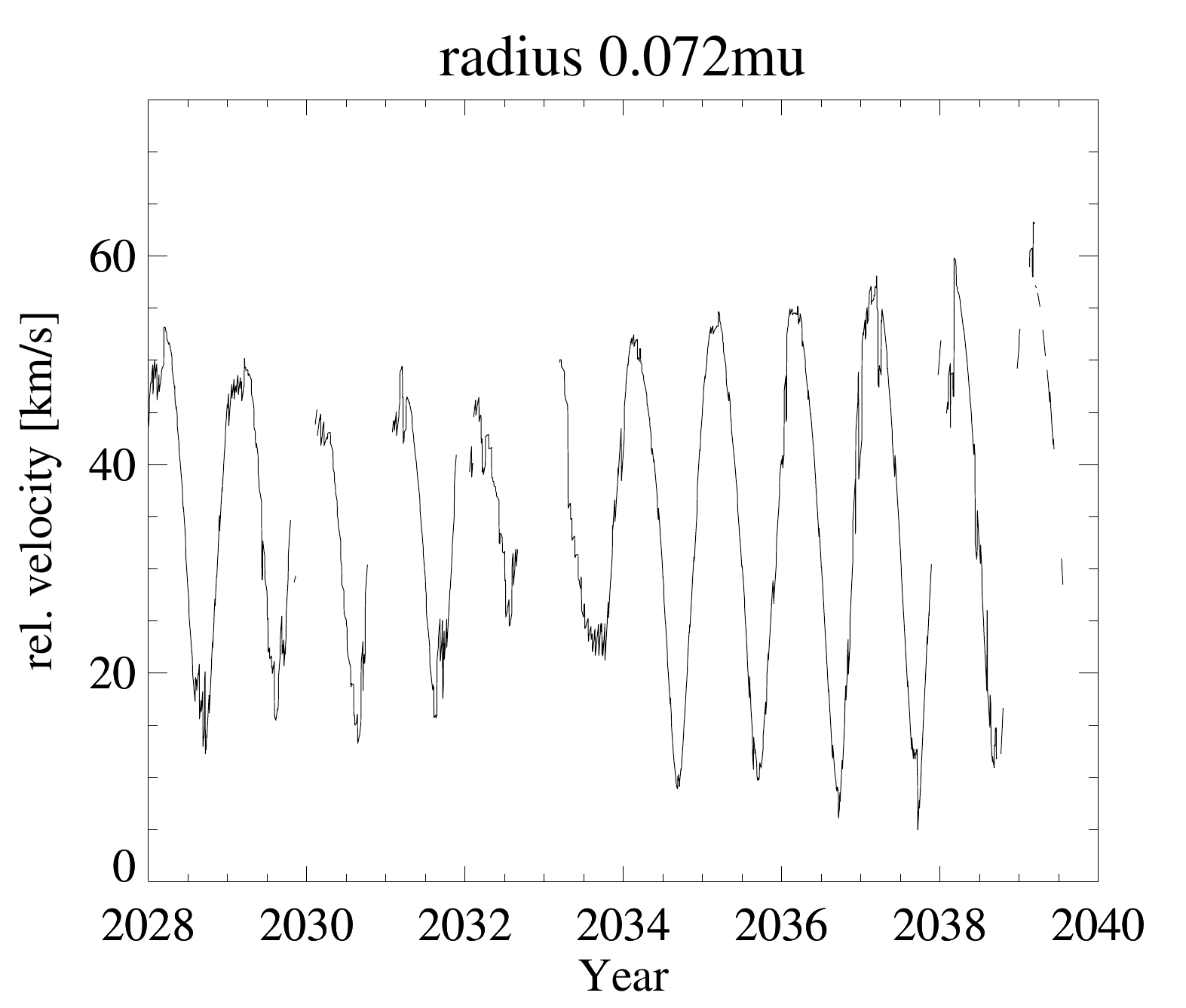} &
\includegraphics[width=0.3\textwidth]{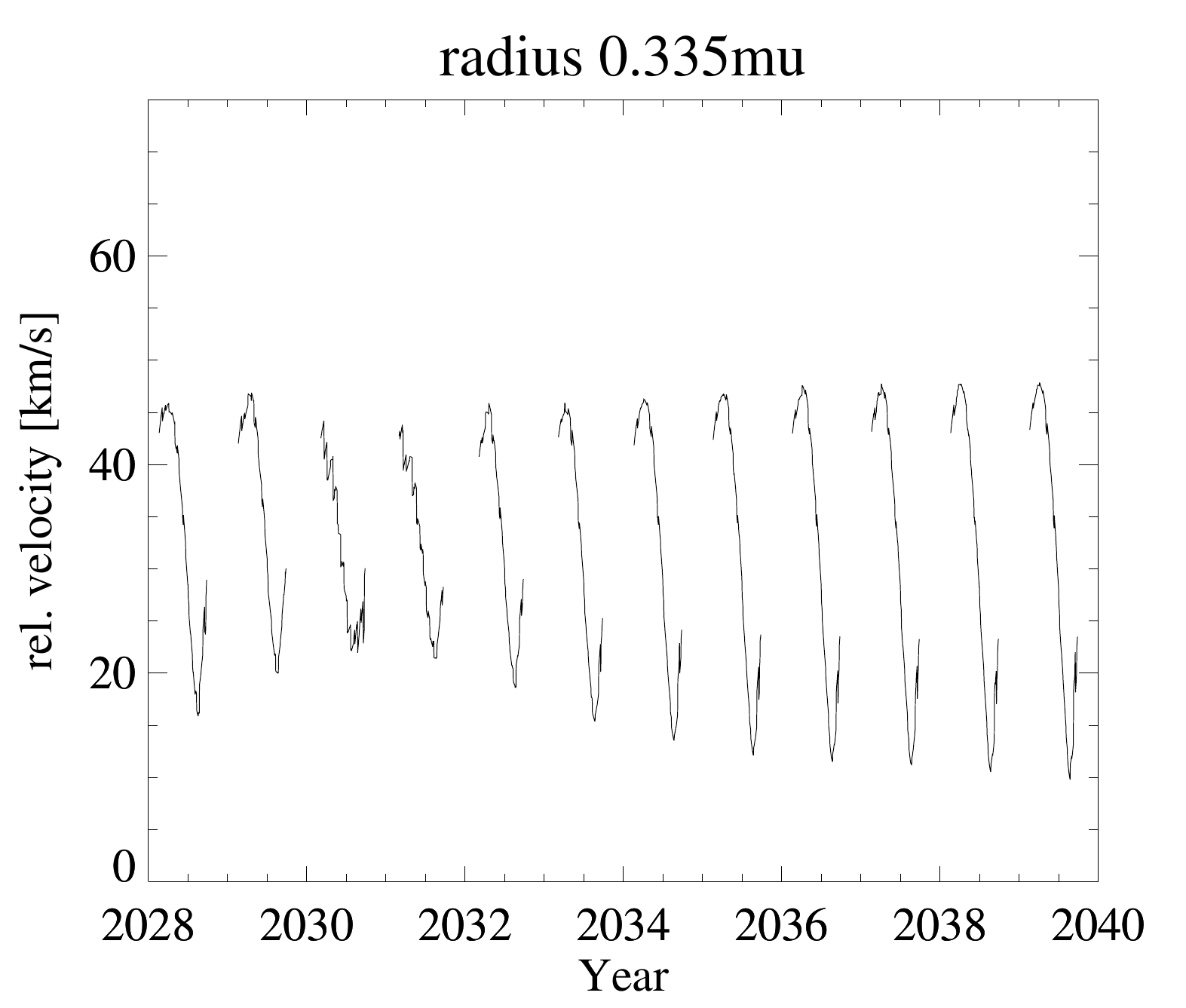} &
\includegraphics[width=0.3\textwidth]{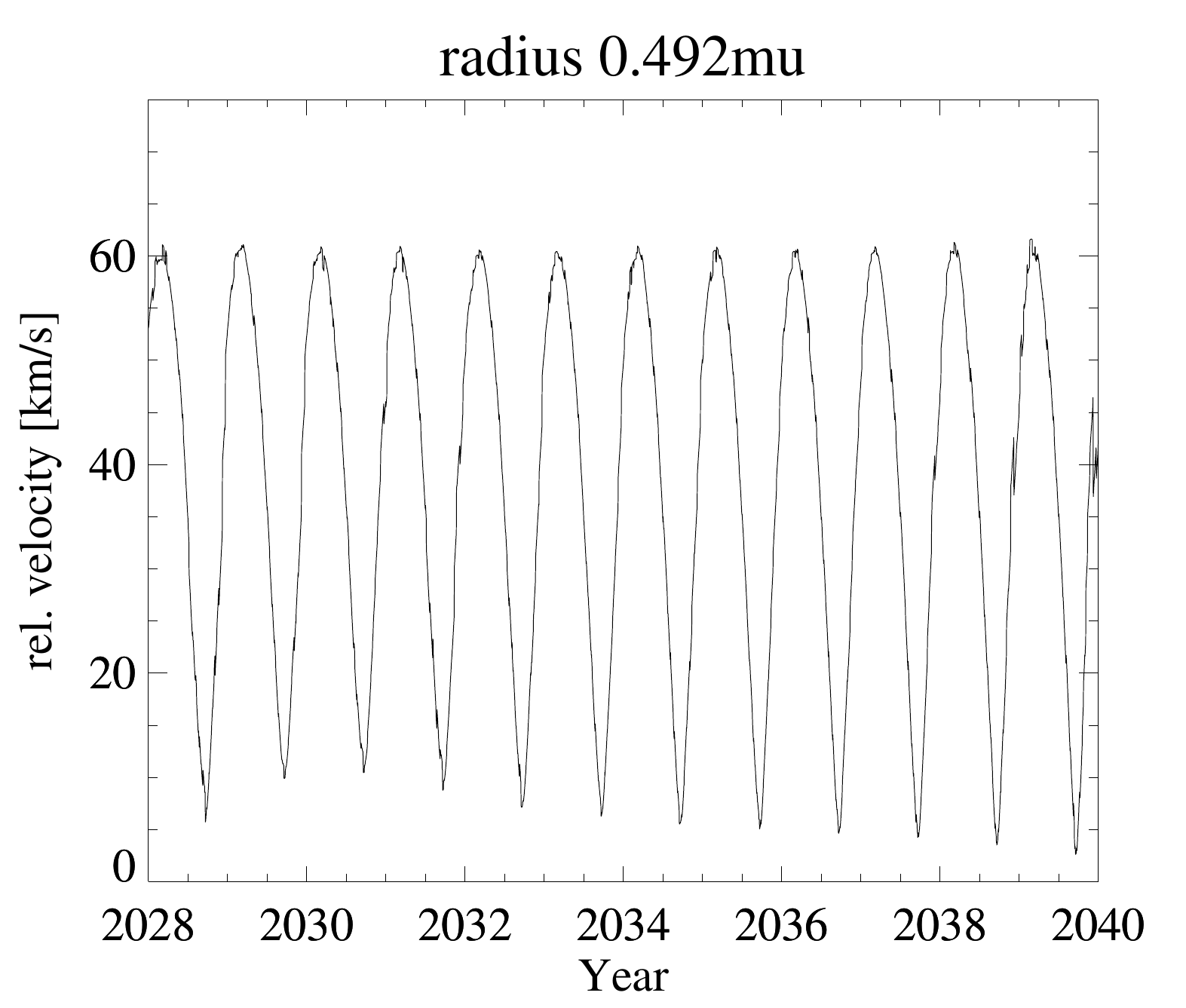} \\
\end{tabular}
\caption{Average velocity of particles in the Earth's reference frame for different selections of particle sizes (from left to right: 0.07\,\mum, 0.34\,\mum, 0.49\,\mum).
}
\label{fig:earthorbit_velearth}
\end{figure*}



\section{The ISD flow as seen by an observer at Earth}
\label{sec:discussion}
We discuss the simulated flow patterns of the ISD near the Earth or near a spacecraft in an orbit around the Sun at 1 AU.
The flow characteristics are modulated by the velocity of the Earth or the spacecraft around the Sun ($\sim 30\, \mathrm{km\,s^{-1}}$) in addition to the effects discussed earlier in this paper. In particular, we concentrate on the impact speed distribution, the inflow direction, the mass density distributions, and the total mass flux on to Earth. We ignore the effects of the Earth's gravity and of the Earth's magnetic field.

\subsection{Speed distribution and relative flow directions}
The impact velocity distribution is required to predict the particle flux and to determine the speed range in which the particles can be observed at the Earth/spacecraft.
Figure~\ref{fig:earthorbit_velearth} shows the average velocity of ISD particles with respect to the Earth for individual particle sizes. The impact velocity is modulated by two different effects: the exact one-year cycle due to the Earth's orbit (including Earth's velocity); and the time-dependent effects of the IMF. 
The speed in the ecliptic reference frame (Fig.~\ref{fig:earthorbit_vel}) only varies strongly for particles of size $a_d <0.1\,\mu$m. Therefore, the amplitude of the Earth's relative velocity dominates the annual variations in the impact speed of the larger particles. When the velocity vectors of ISD particles and the Earth are parallel, the value of the relative velocity vector becomes small (a few $\kms$) in contrast to the situation half a year later when both velocity vectors are anti-parallel and the relative speed becomes 50 to $60\,\kms$. For particles of $0.49$\,$\mu$m in size, the highest impact speed is approximately $60\,\kms$ in December each year.



Figure~\ref{fig:lambet1_earth} shows the inflow direction, in the Earth's reference frame, taking into account the  velocity owing to the Earth's orbital motion. This direction can vary over a full circle (360\deg) (cf. Fig.~\ref{fig:lambet1}). This can be used to identify ISD particles and distinguish them from other dust populations. 


\subsection{Mass flux on to Earth and mass distribution}
The size distribution of ISD particles is also useful for the characterisation of ISD at the Earth and the resulting mass flux into the Earth's atmosphere. The mass distribution at the Earth varies widely during the course of a year. Figure~\ref{fig:orbgeo} shows the positions of the Earth that are used to demonstrate the seasonal changes in the mass distribution. The seasonal mass distributions of the ISD flow at Earth orbit are shown in Fig.~\ref{figMaxDist1}. Variations of the mass distribution at Earth are shown for a period of 1 year, in a defocussing configuration (2018-2019) and in a focussing configuration of the IMF (2029-2030). As the mean IMF is largely constant over a period of this length, the variations reflect changes in the spatial distribution of the ISD particles due to radiation pressure and gravity, as well as the relative velocity observed in the Earth's frame of reference. During the upstream portion of the Earth's motion, small particles do not reach the Earth ($\beta$-cone effect), while during the downstream portion large particles are focussed and their flux is enhanced. In addition to this modulation there is the focussing/defocussing effect for small particles ($<0.3\,\mum$) due to the Lorentz force.

Figure~\ref{figMaxDist2} shows the annual variations over the 22-year solar cycle at four different positions along the Earth's orbit. This illustrates the varying effect of the changing IMF on the mass distributions. The overall patterns of the dust flux are similar to those observable at the orbits of Jupiter and Saturn and in the asteroid belt \citep{sterken2013a}, except for the $\beta$-cone that affects more particles at Earth orbit than further away from the Sun. In addition, due to the orbit of the Earth, the frequency and the amplitude of the modulations are different.

The cumulative mass flux (summed over all masses) on to the Earth reflects this orbital variation (Fig.~\ref{fig:massflow}). The contributions from different particle sizes change during the course of the year. The strong peak at the end of the year is caused by the large particles that are focussed downstream while at other times of the year smaller particles dominate the mass flux.

\subsection{Observing conditions for spacecraft at Earth's orbit}
How can we verify the simulation results of this new model for ISD? Recent in-situ measurements of ISD at 1\,AU were performed primarily by Cassini \citep{altobelli2003}, but observation time was limited to a few weeks. New mission concepts are in development, which allow dust observations for many years. The Destiny$^+$ mission is currently in development with a foreseen launch date of 2022 \citep{arai2018}. It will carry a modern dust telescope to investigate ISD, interplanetary dust, and dust of the active asteroid Phaethon \citep{kobayashi2018b}. The Destiny$^+$ spacecraft will have an Earth-like interplanetary trajectory \citep{sarli2018}. Predictions of the dust flow during the Destiny$^+$ mission are shown in \citet{krueger2018a}.

Here we discuss two distinct populations of particles that can be observed at Earth orbit (Fig.~\ref{fig:numberflux}): the larger particles ($a_d\gtrsim 0.3\,\mum$, corresponding to a particle mass of $m\gtrsim 3\times10^{-16}$\,kg), whose dynamics are dominated by gravity and radiation pressure; and the small particles ($a_d\lesssim 0.2\,\mum$), whose dynamics are dominated by the Lorentz force of the IMF.
While there is a temporal variation caused by the solar
cycle for the smaller particles, the density of the larger particles with  $\beta < 1$ never drops to zero (Fig.~\ref{fig:earthorbit_dens}). However, the flux of dust particles on to a spacecraft at Earth orbit can drop to small values when their velocity vectors become similar (Fig.~\ref{fig:numberflux}). 
For most of the orbit the flux is relatively constant, with a sharp increase when the Earth crosses the gravitational focussing cone downstream of the Sun. At 1 AU the mean interplanetary dust flux of $a_d\lesssim 0.3\,\mum$
particles is ~5 day$^{-1}$m$^{-2}$ \citep{gruen:85a}. The mean ISD flux is therefore 10\% to 80\% of the interplanetary flux, depending on the focussing conditions of the IMF. 

Conversely, the small particles that are dominated by the Lorentz-force can only be observed during the strong focussing configuration of the solar magnetic field in the years after solar minimum. During other time periods, the ISD particles are strongly depleted by the defocussing magnetic field, and their flux drops to zero. 
They were strongly depleted in the Ulysses data for a number of reasons: The interval of maximum focussing was just outside the 17 years of the Ulysses mission, and the intervals closest to the maximum focussing were strongly contaminated by stream particles ejected from the Jovian system, severely hindering the identification of small particles from a potential ISD population \citep{strub2015a}.

\begin{figure*}
\centering
\begin{tabular}{cc}
\includegraphics[width=0.4\textwidth]{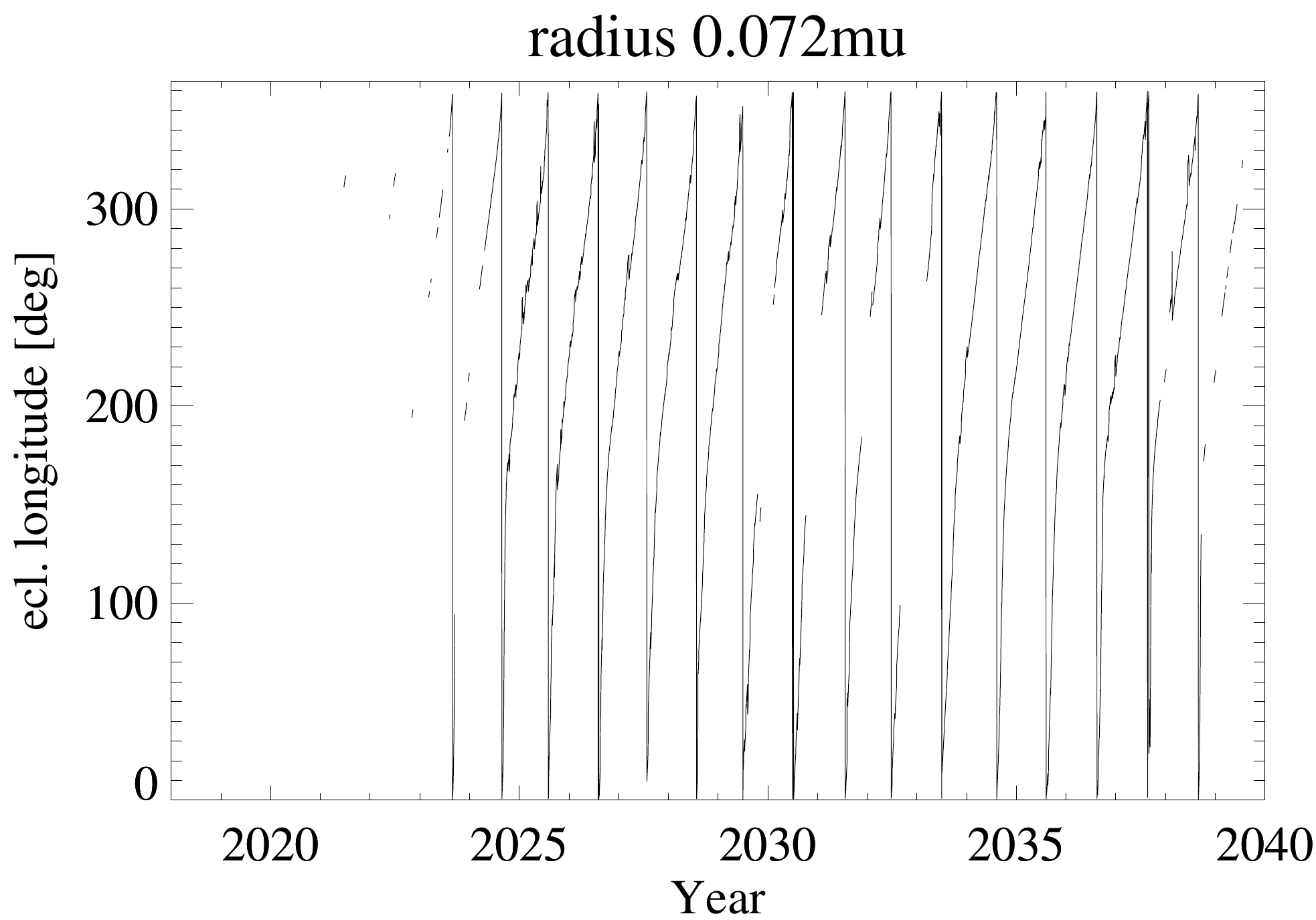} &
\includegraphics[width=0.4\textwidth]{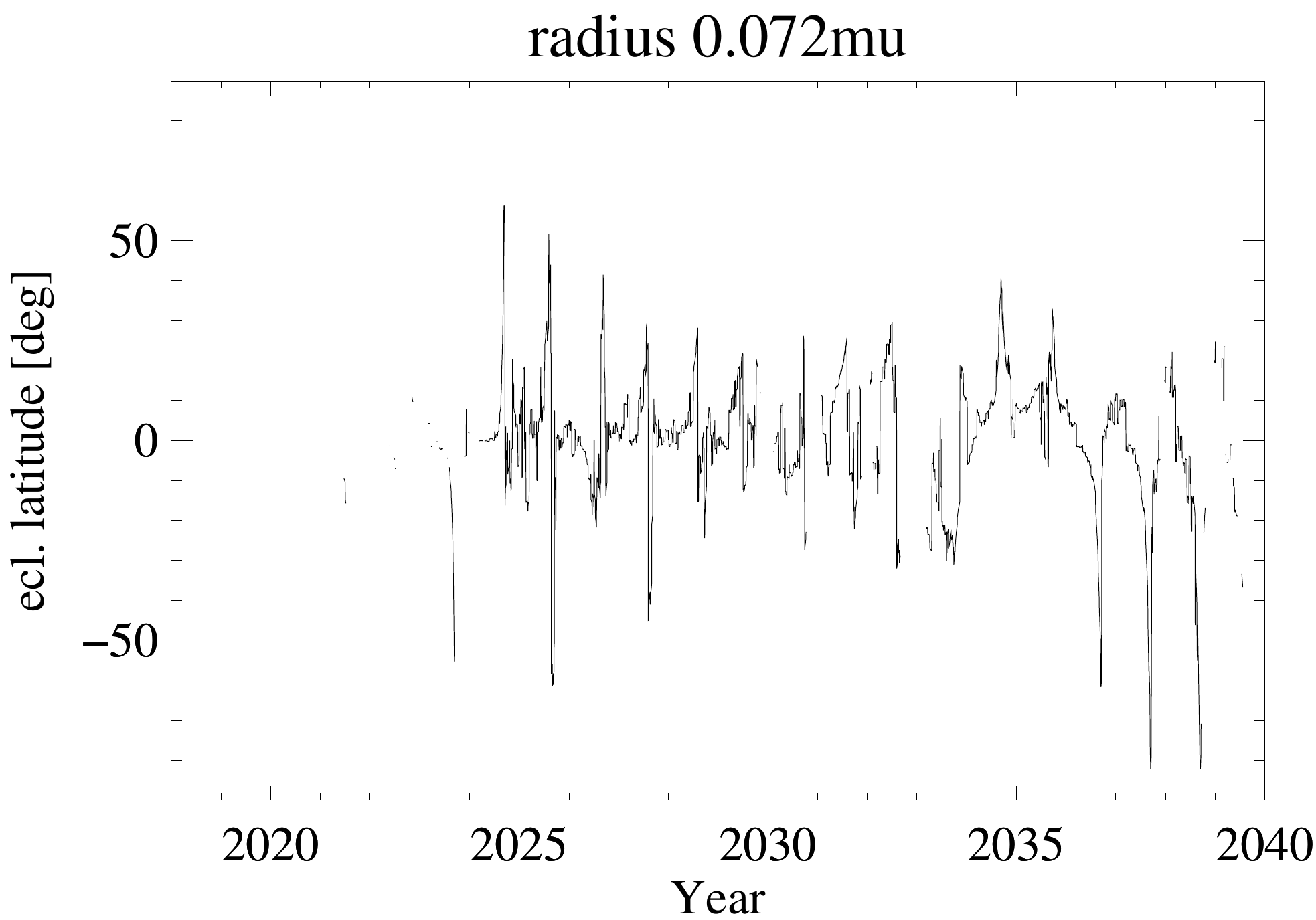}\\
\includegraphics[width=0.4\textwidth]{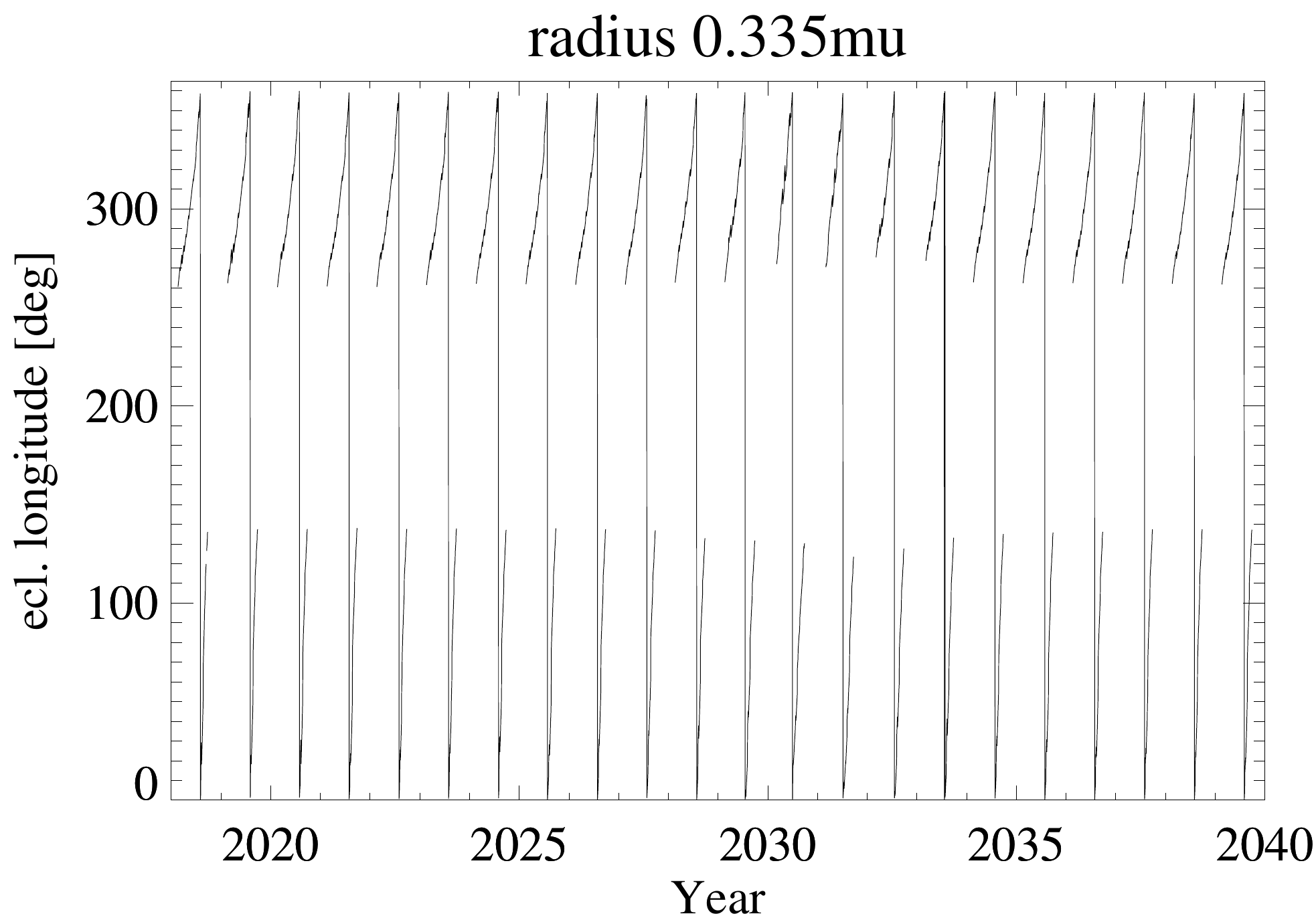} &
\includegraphics[width=0.4\textwidth]{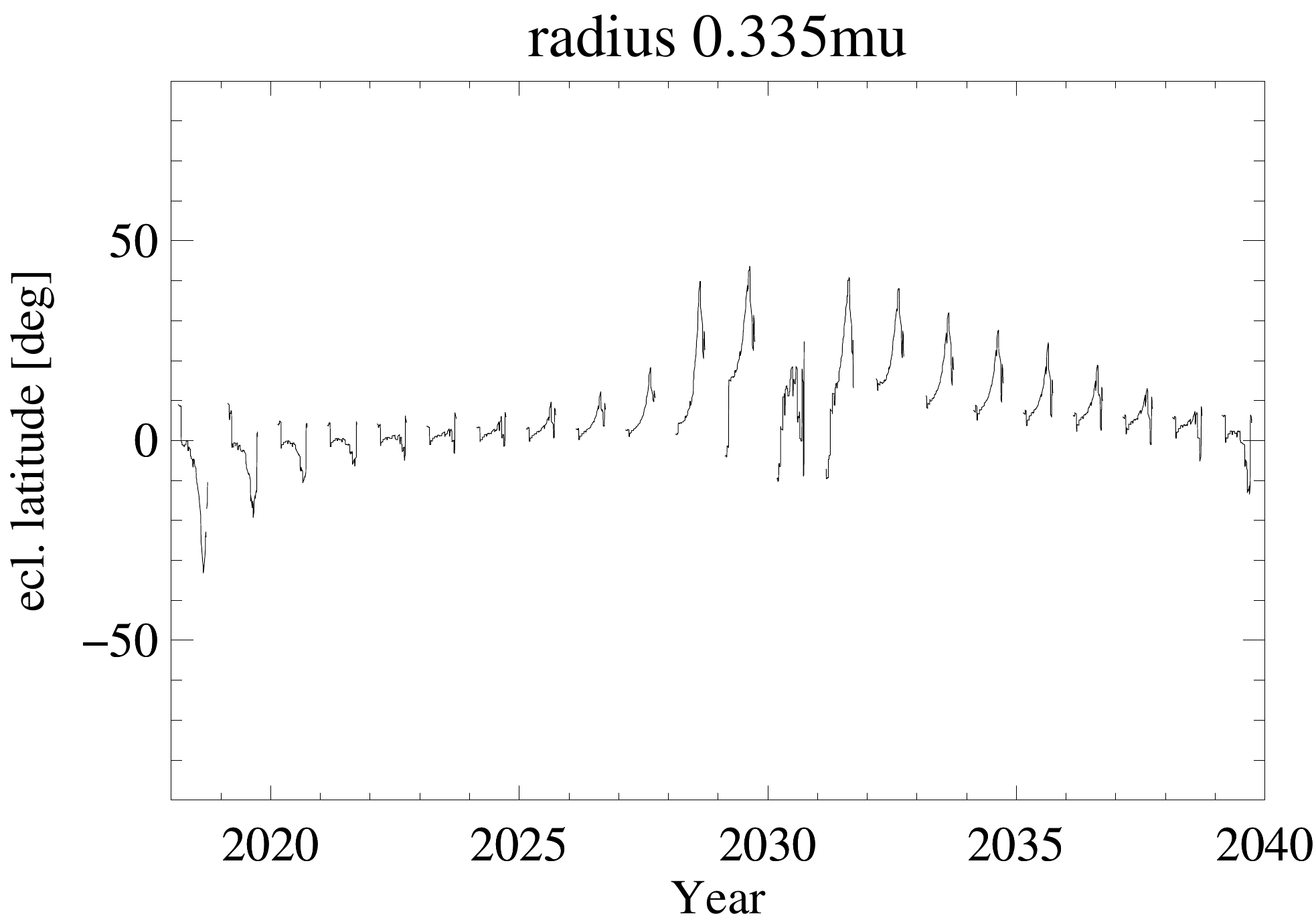}\\
\includegraphics[width=0.4\textwidth]{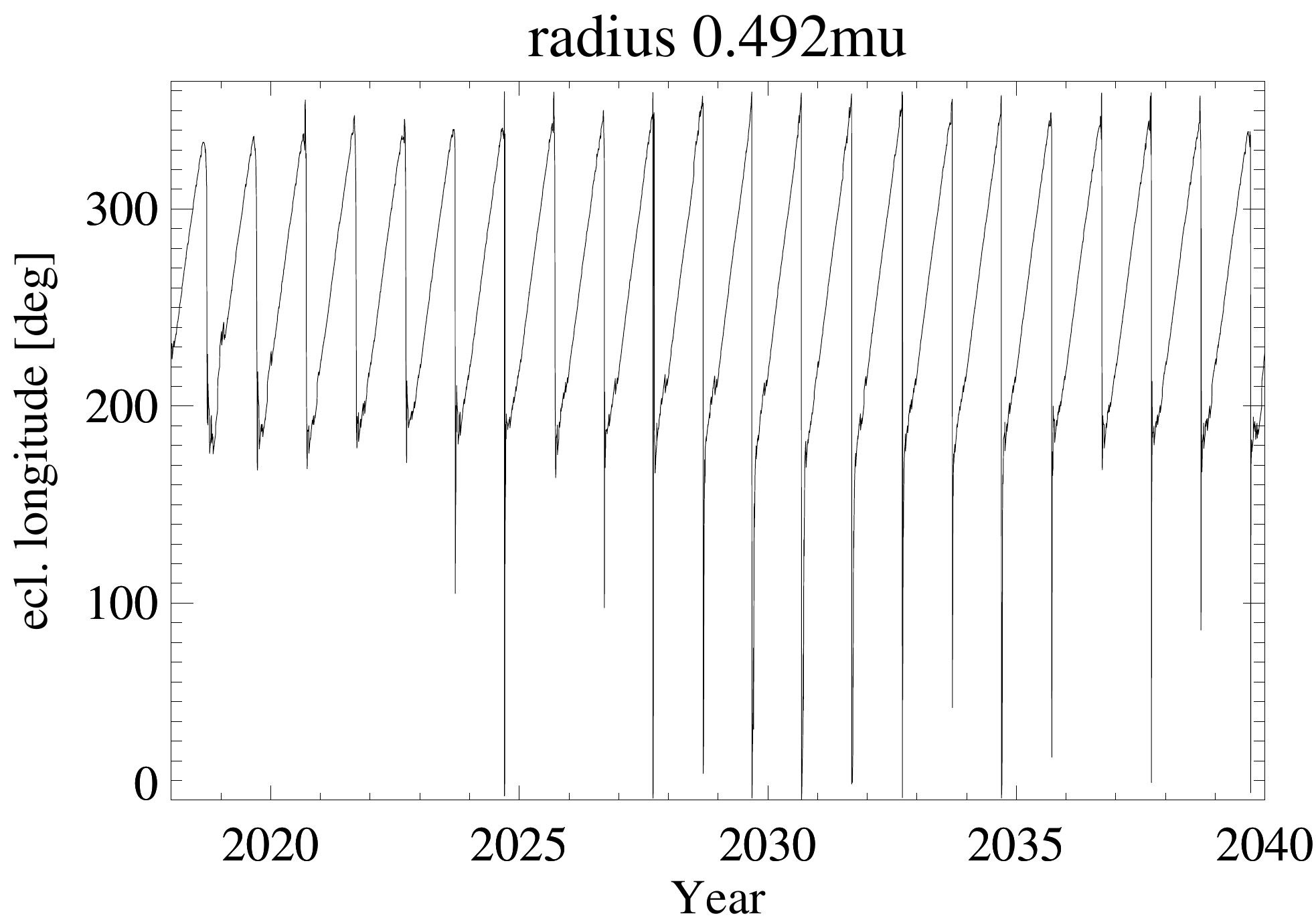} &
\includegraphics[width=0.4\textwidth]{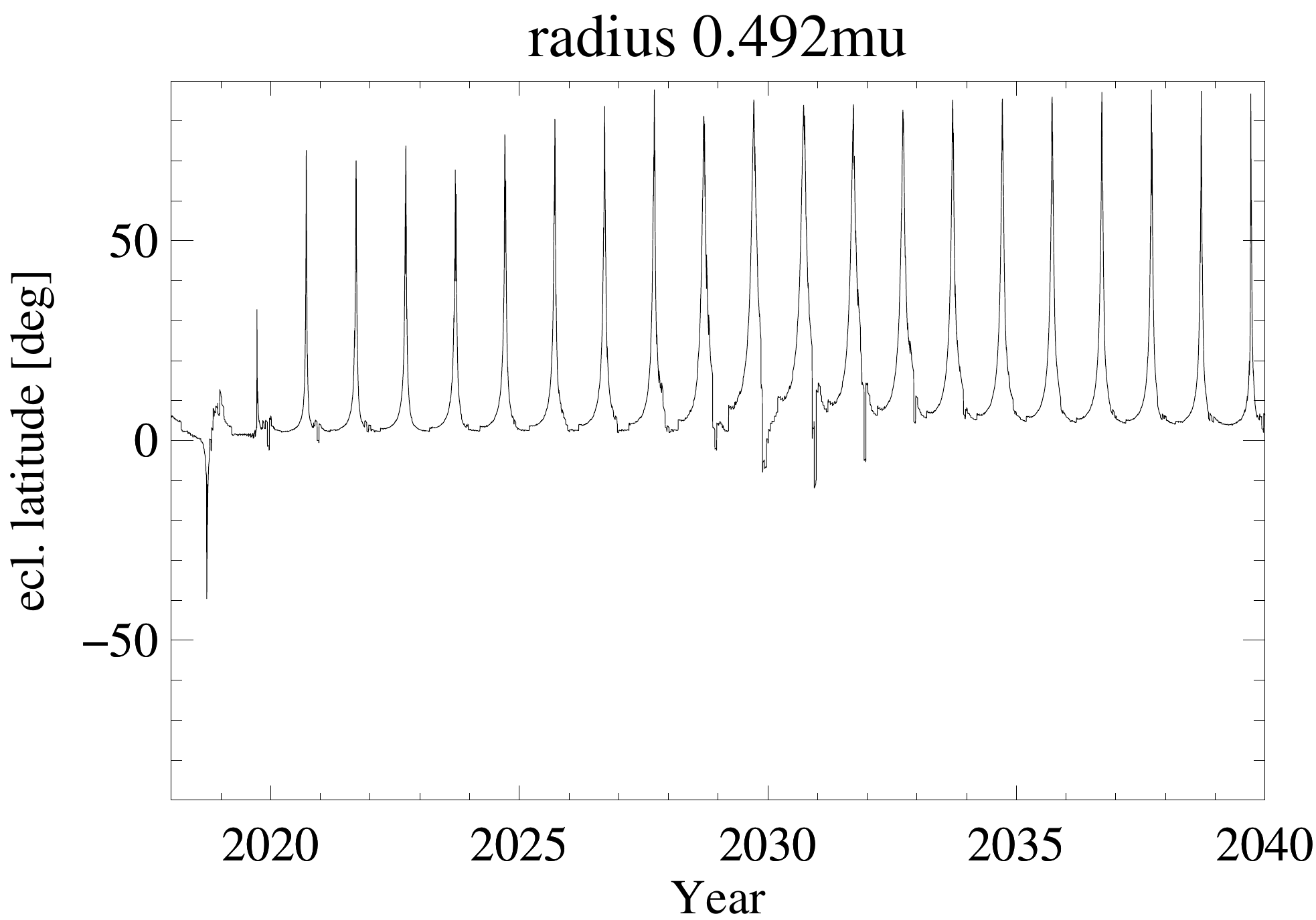}\\
\includegraphics[width=0.4\textwidth]{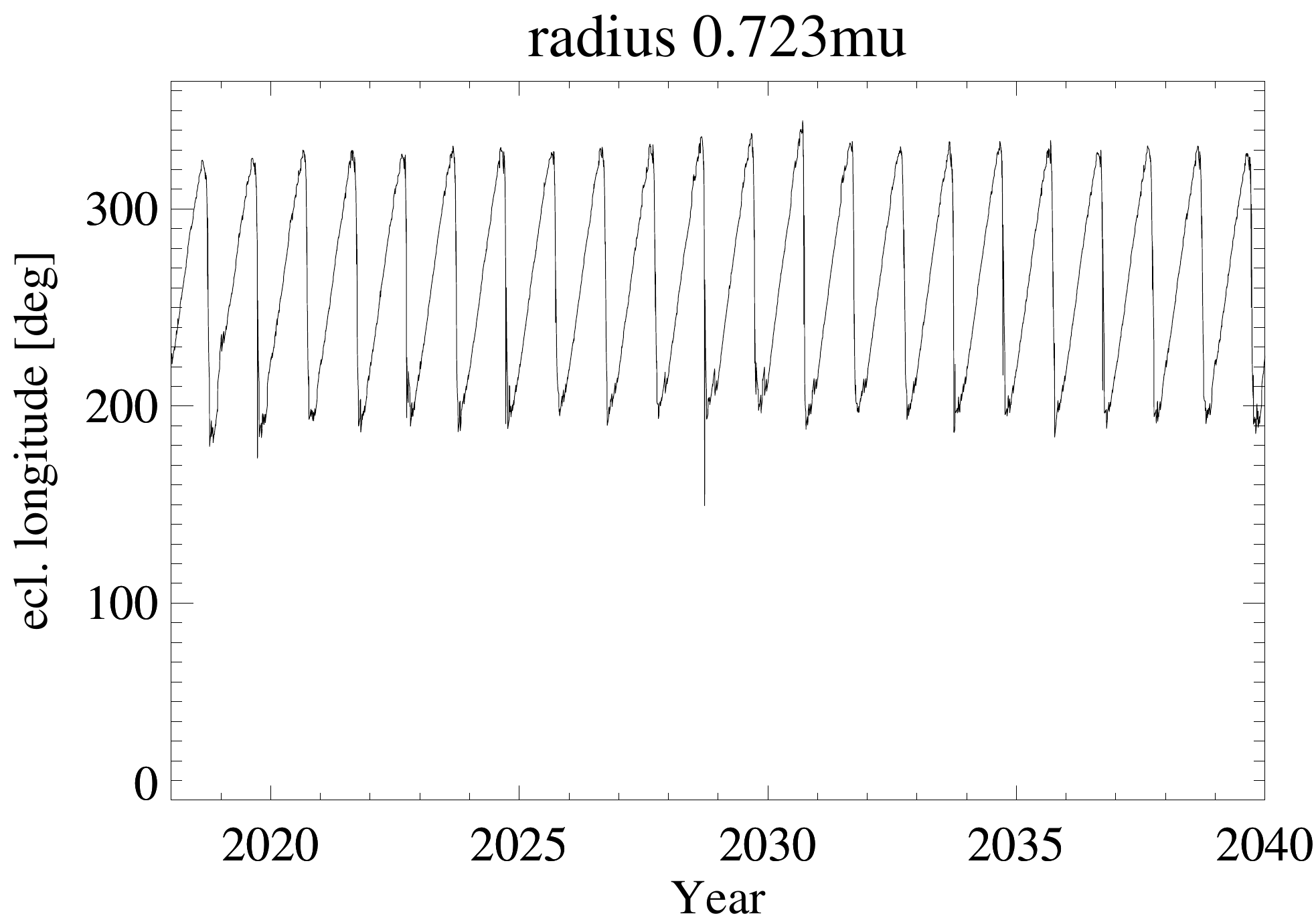} &
\includegraphics[width=0.4\textwidth]{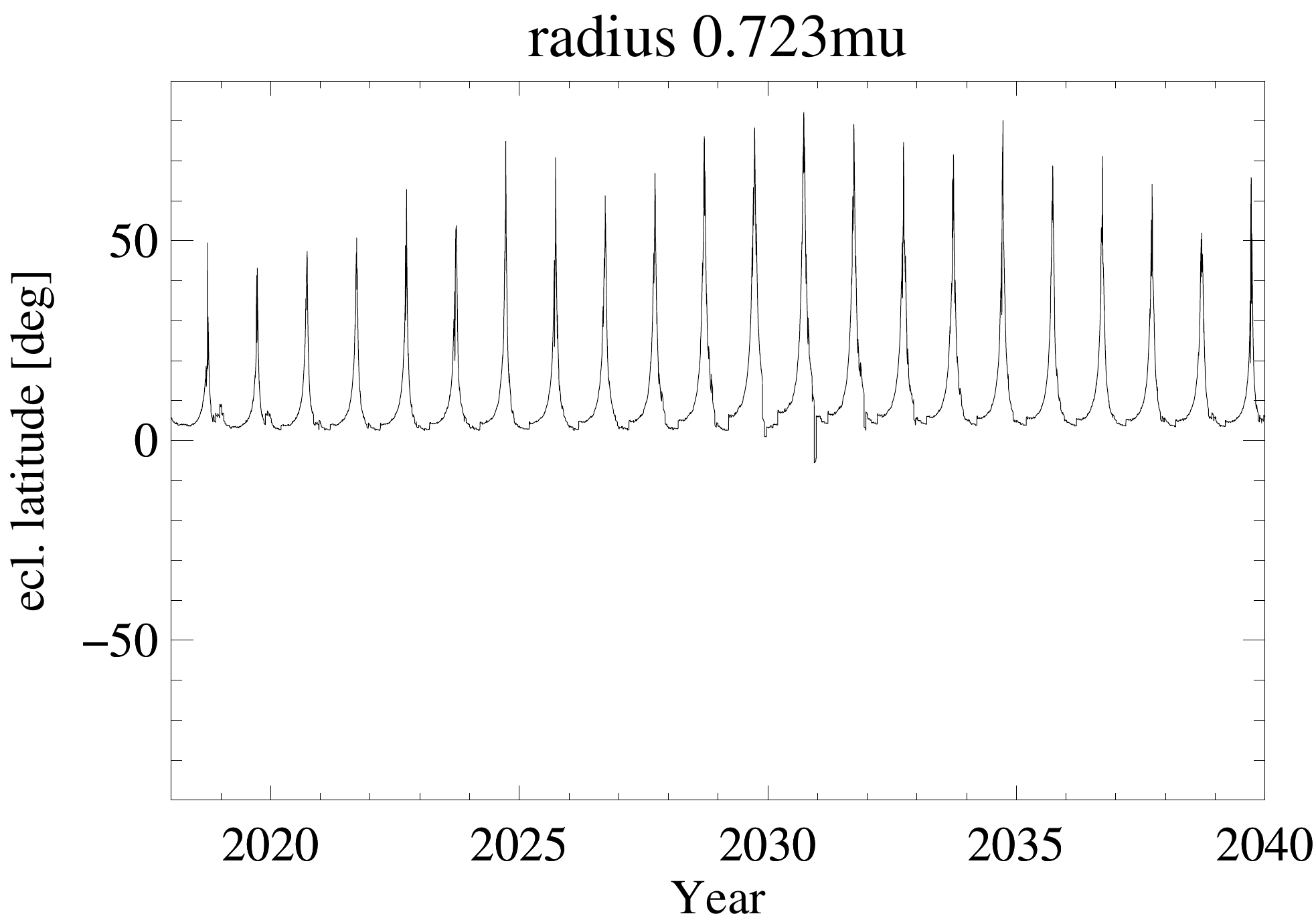} \\
\end{tabular}
\caption{Apparent apex direction in ecliptic coordinates $(\lambda, \beta)$ of ISD particles in the Earth reference frame (taking into account the orbital velocity of the Earth) for particle sizes 0.07\,\mum, 0.34\,\mum, 0.49\,\mum, and 0.72\,\mum.}
\label{fig:lambet1_earth}
\end{figure*}

\begin{figure}[htb] 
\includegraphics[width=\columnwidth]{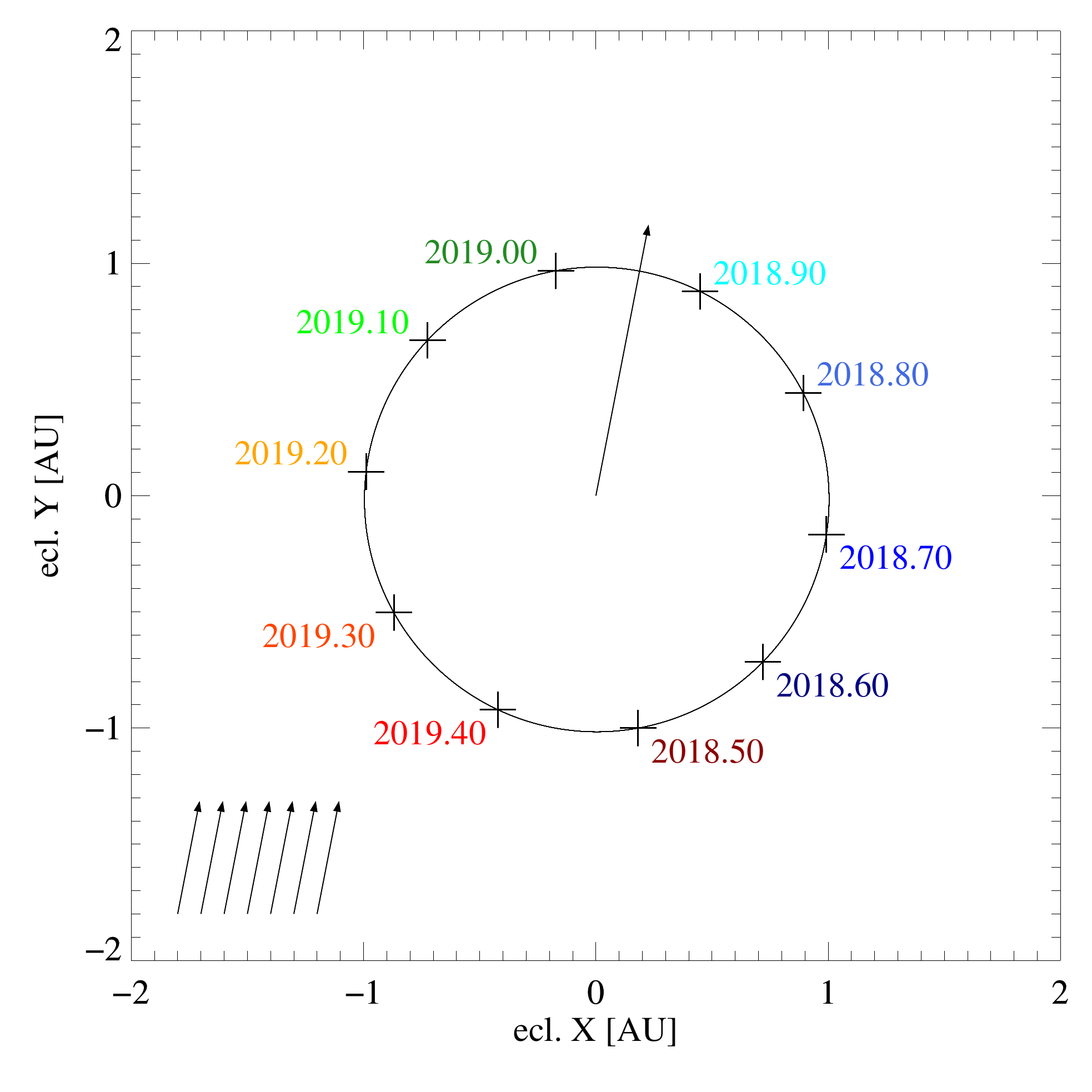}
\caption{
Geometry of the Earth's orbit and points used for extracting the simulated mass distributions (see Fig.~\ref{figMaxDist1}). The arrows indicate the undisturbed flow direction of ISD particles for reference, i.e. the flow direction of the particles outside the solar system. The colours of the dates along the orbit match the colours used in Fig.~\ref{figMaxDist1}. } 
\label{fig:orbgeo}
\end{figure}

\begin{figure*}
\centering
\begin{tabular}{ccc}
\includegraphics[width=0.42\textwidth]{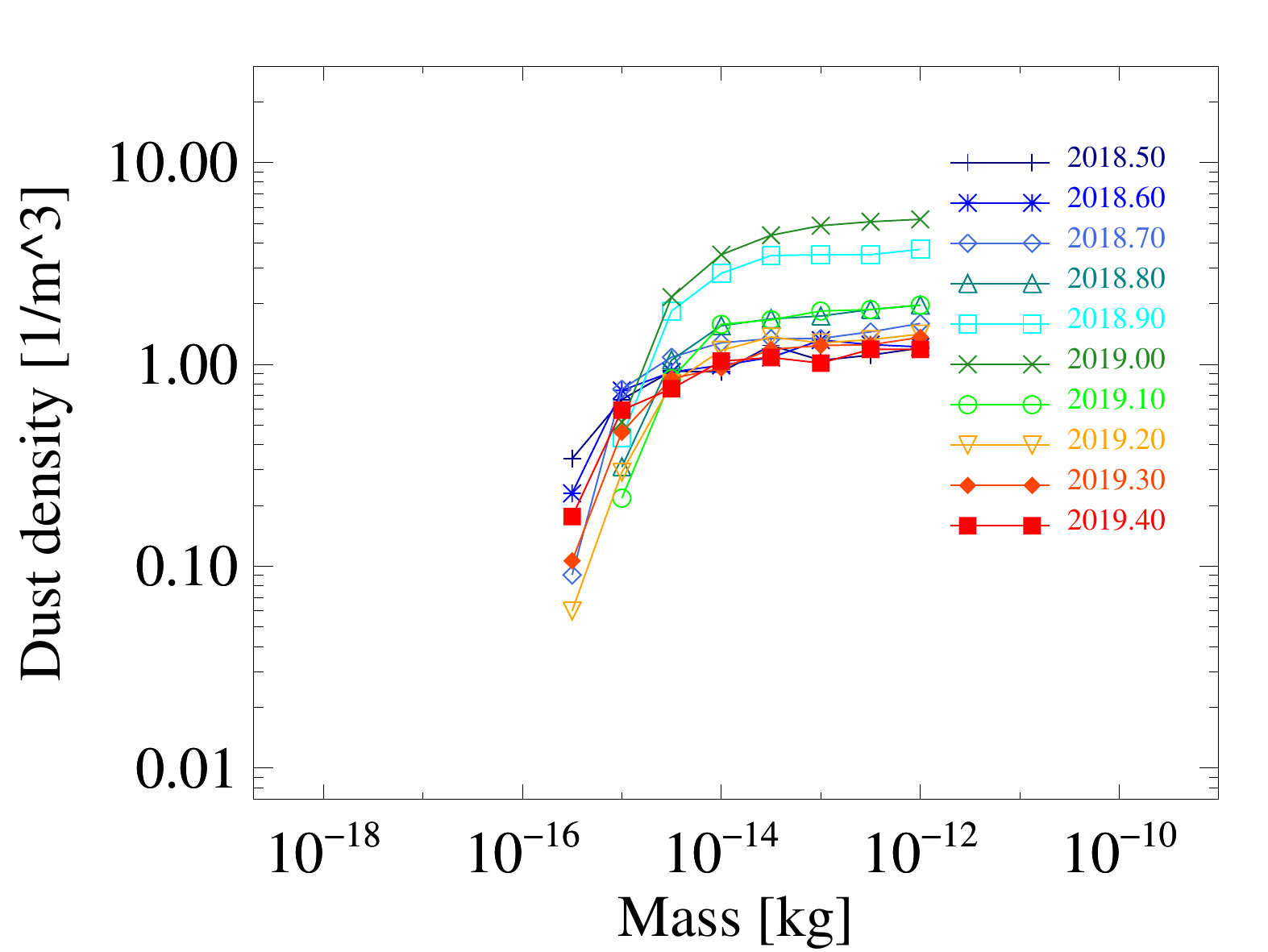} &
\includegraphics[width=0.42\textwidth]{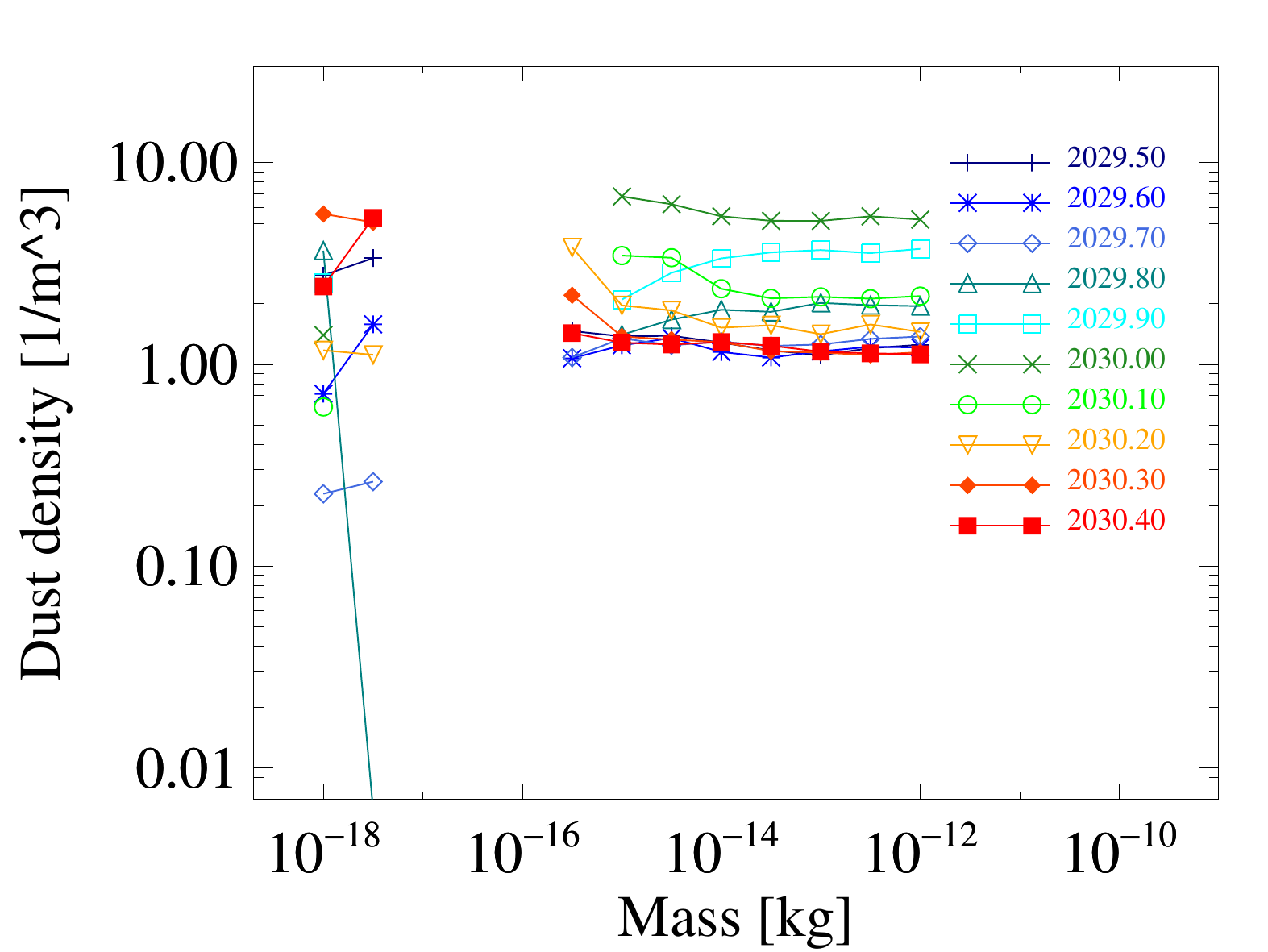} \\
\end{tabular}
\caption{Mass distribution in the ecliptic frame of reference over the course of one year, for a defocussing configuration of the ISM ({\em left panel}, 2018-2019) and for a focussing configuration of the ISM ({\em right panel}, 2029-2030). Since the 3-dimensional flow configuration is virtually constant over these periods of 1 year, the observed changes reflect the spatial variations of the mass distribution along the Earth orbit, in a similar way to the mass distributions for Saturn, Jupiter and asteroid distances in \citet{sterken2013a}.}
\label{figMaxDist1}
\end{figure*}

\begin{figure*}
\centering
\begin{tabular}{ccc}
\includegraphics[width=0.42\textwidth]{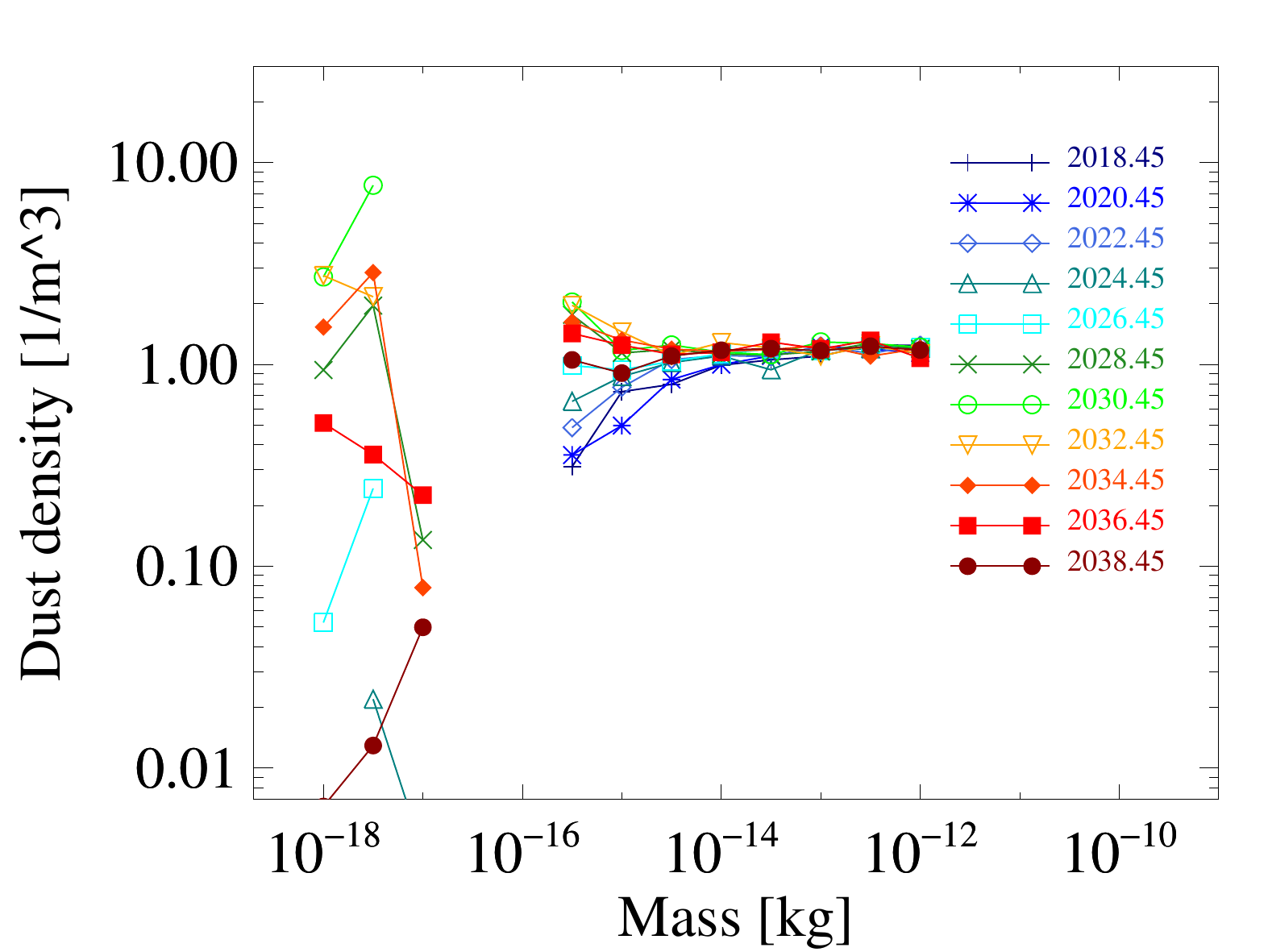} &
\includegraphics[width=0.42\textwidth]{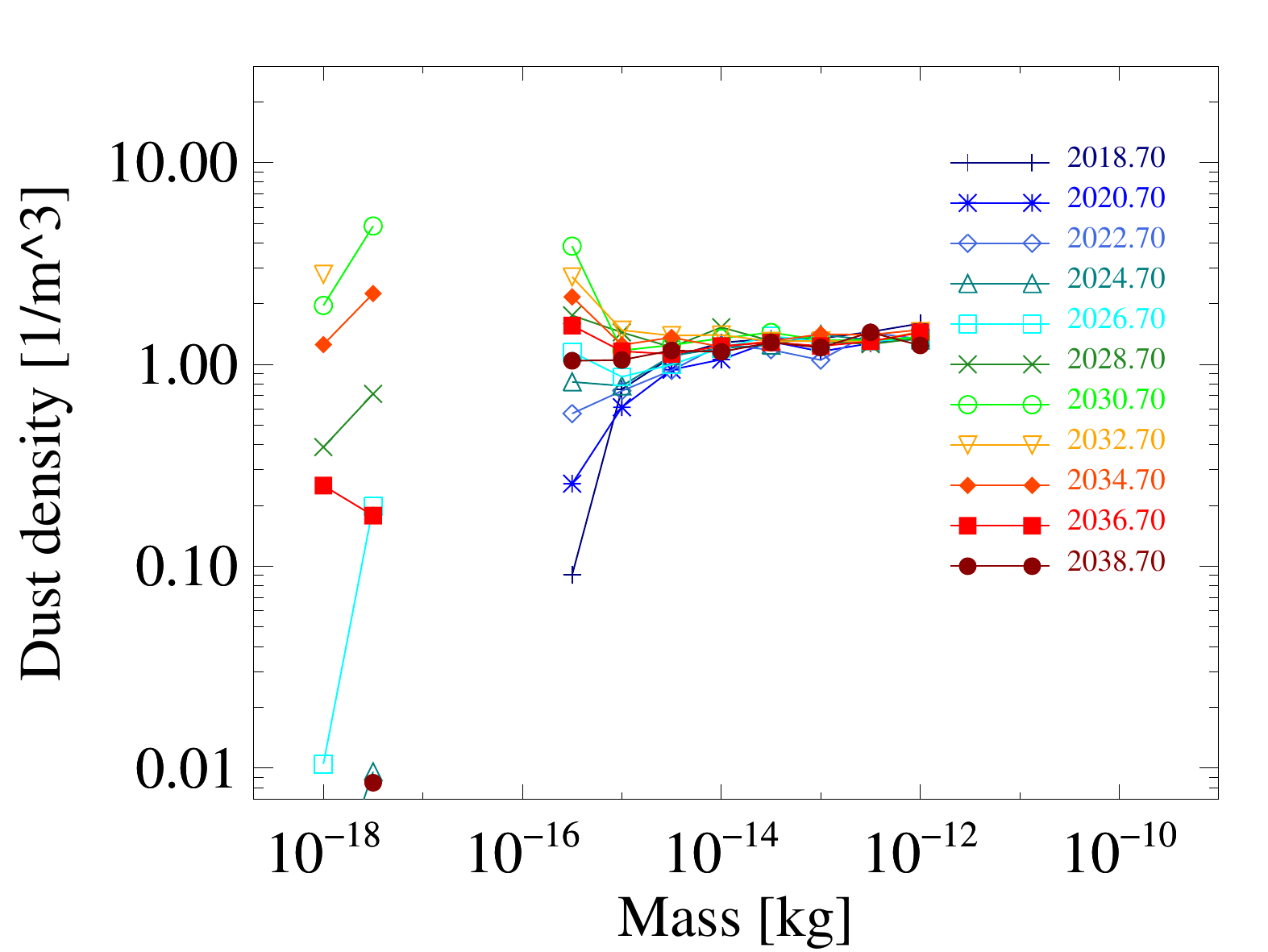} \\
\includegraphics[width=0.42\textwidth]{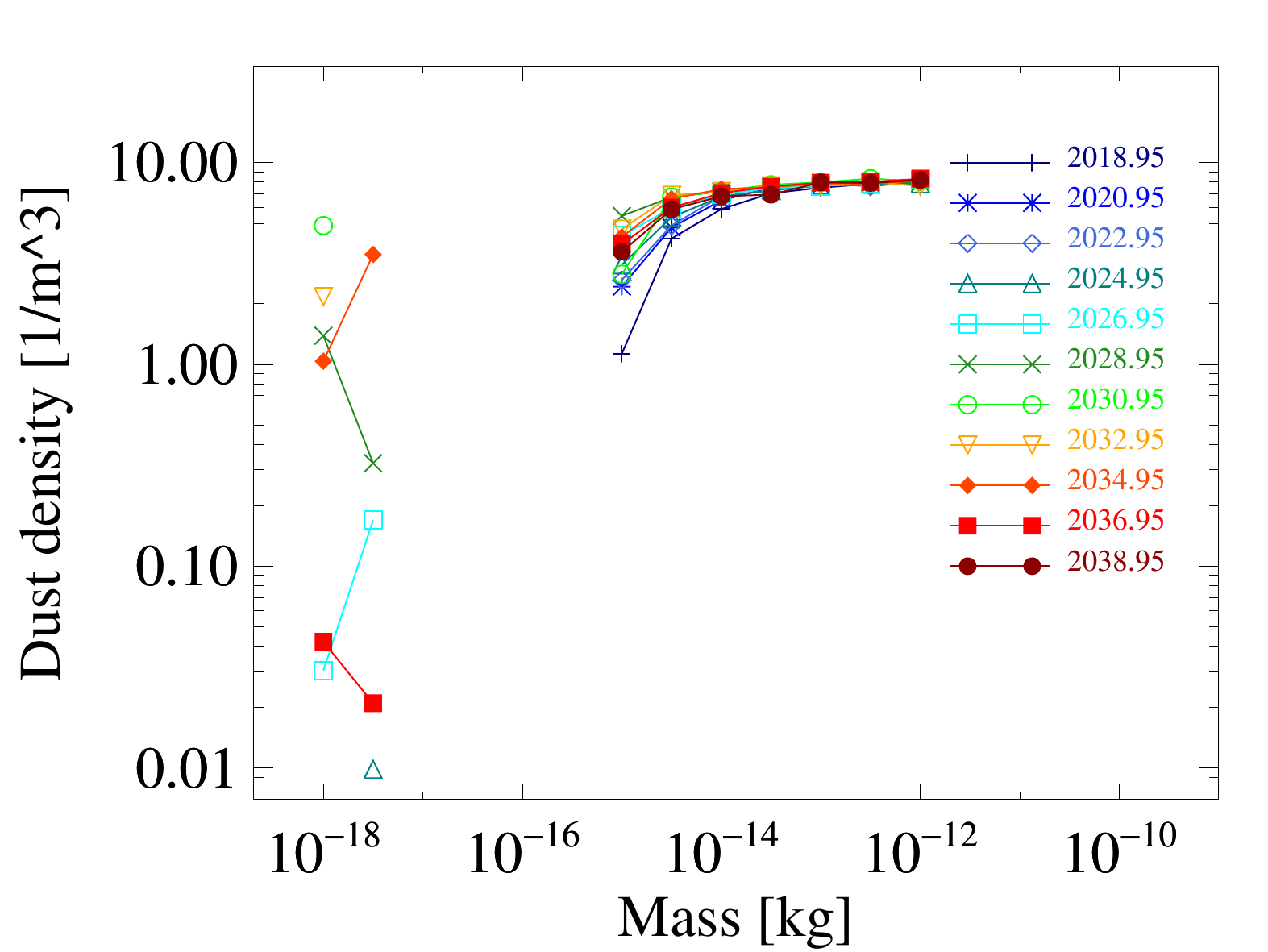} &
\includegraphics[width=0.42\textwidth]{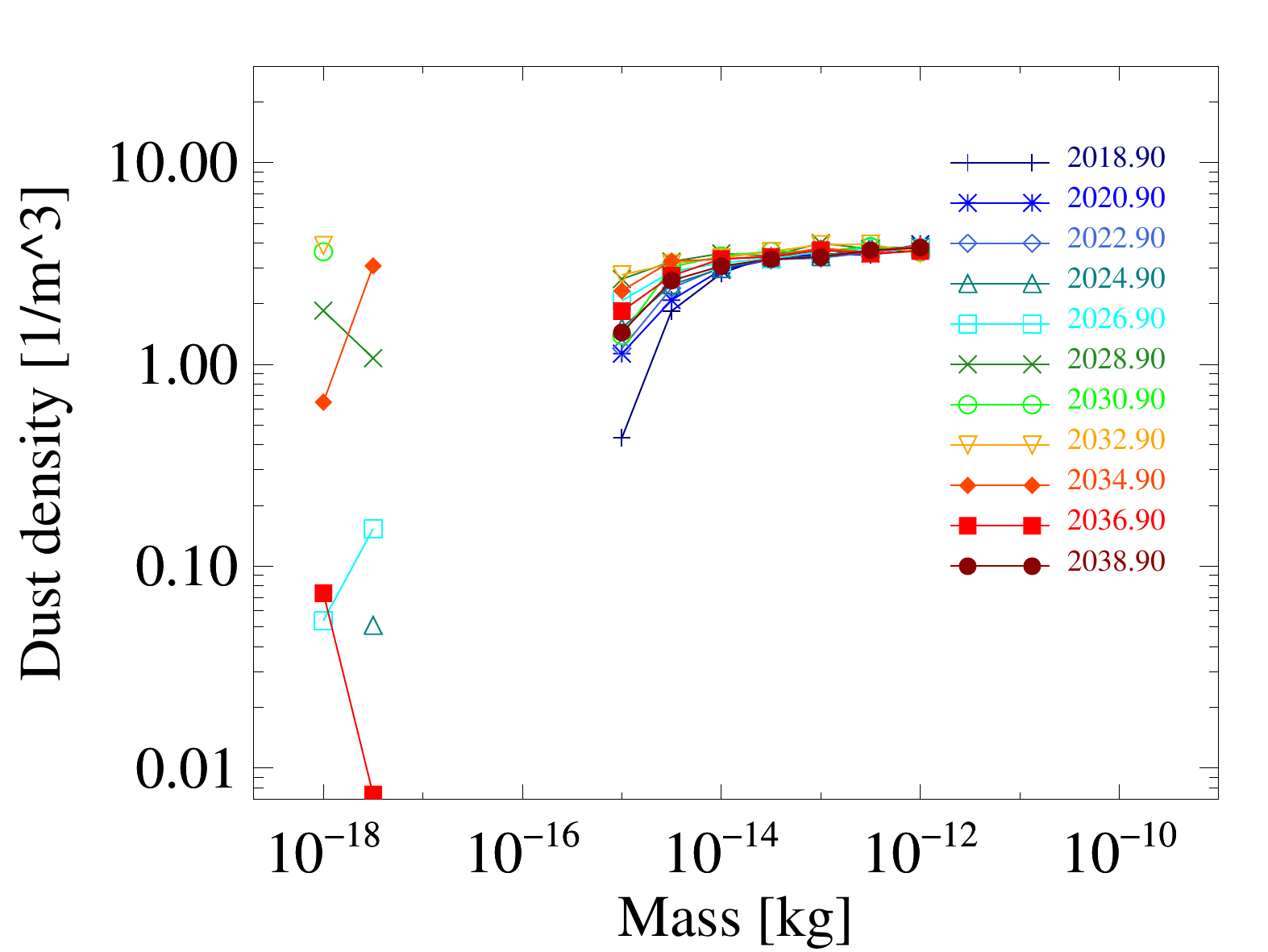} \\
\end{tabular}
\caption{Mass distribution over the whole 22-year solar cycle at four given points along the Earth's orbit (compare to \citet{sterken2013a}). {\em Upper left panel:} upstream of the Sun; {\em Upper right panel:} at an angle of 90 degrees between upstream- and downstream position; {\em Lower left panel:} downstream of the Sun, inside the spot of maximum focussing; {\em Lower right panel:} 18\deg\ away from the downstream downstream direction, outside the strongly localised spot of maximum focussing. The variations seen in each panel reflect the changes due to the focussing/defocussing effects of the IMF throughout the solar cycle.  }
\label{figMaxDist2}
\end{figure*}

\begin{figure*}
\centering
\begin{tabular}{cc}
\includegraphics[width=0.45\textwidth]{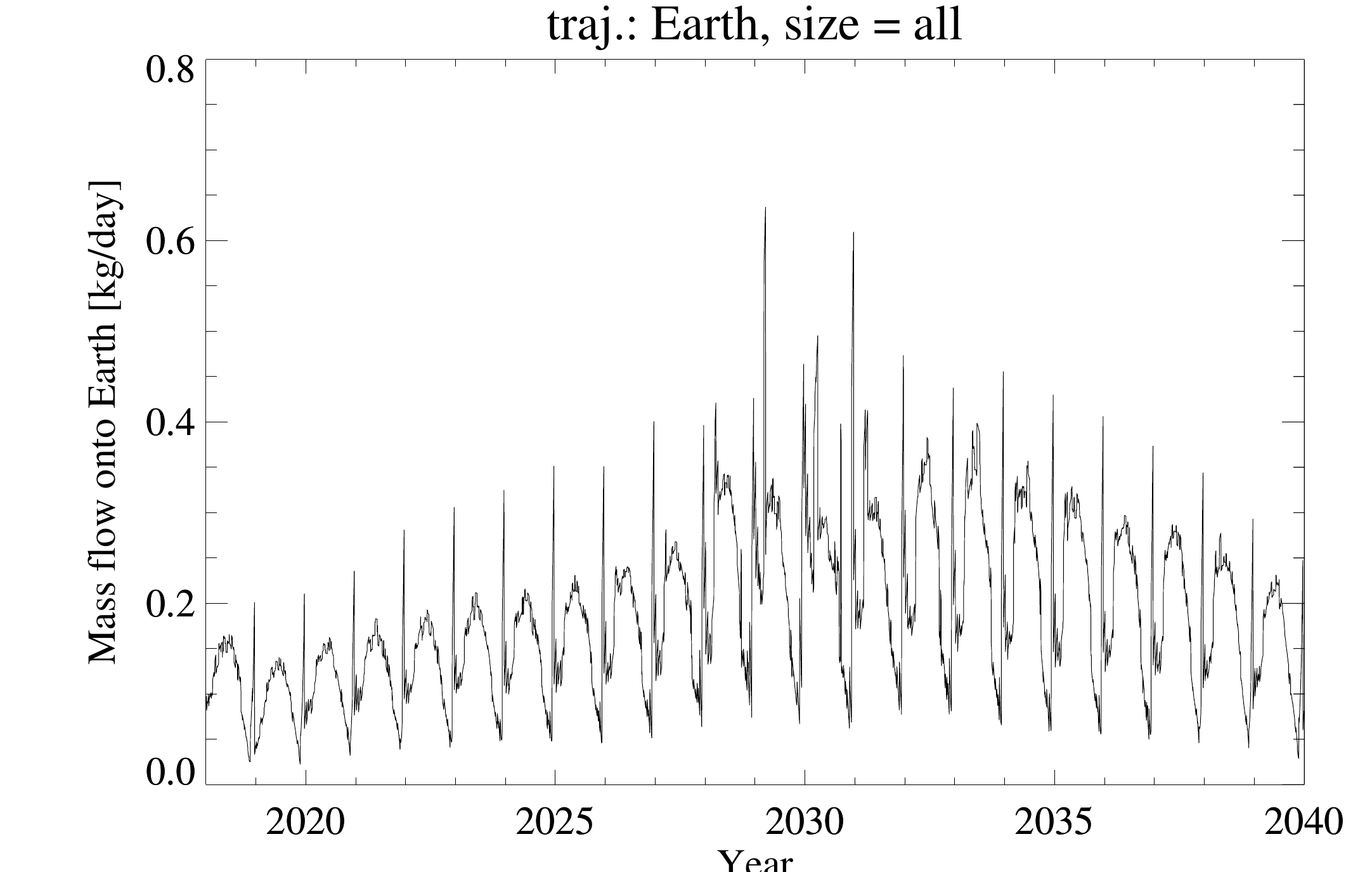} &
\includegraphics[width=0.45\textwidth]{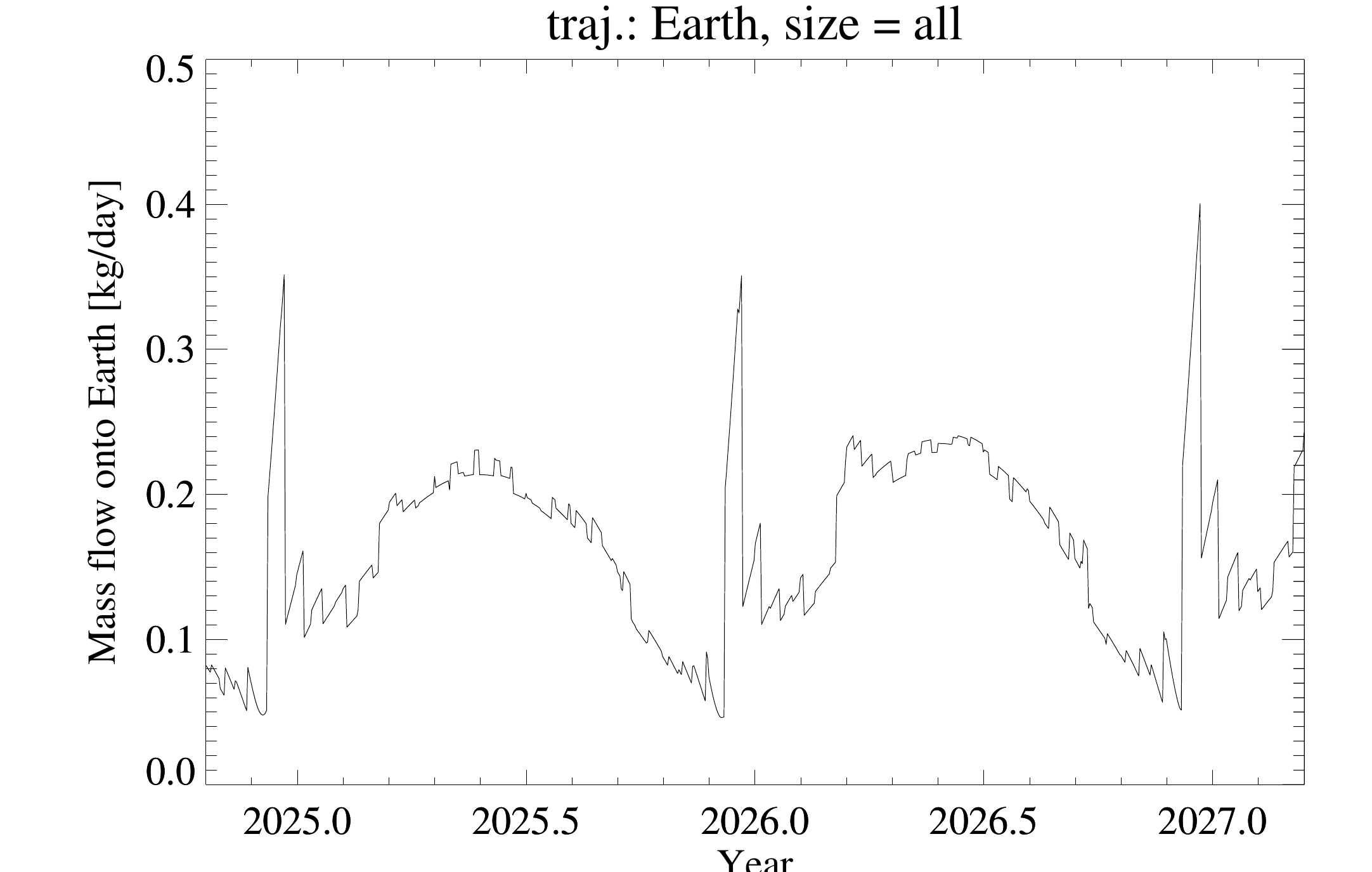} \\

\end{tabular}
\caption{Mass flow on to Earth for all particle sizes. The mass flow takes into account the Earth's velocity relative to the ISD flow and is therefore given in the reference frame of the Earth.}
\label{fig:massflow}
\end{figure*}

\begin{figure*}
\centering
\begin{tabular}{ccc}
\includegraphics[width=0.45\textwidth]{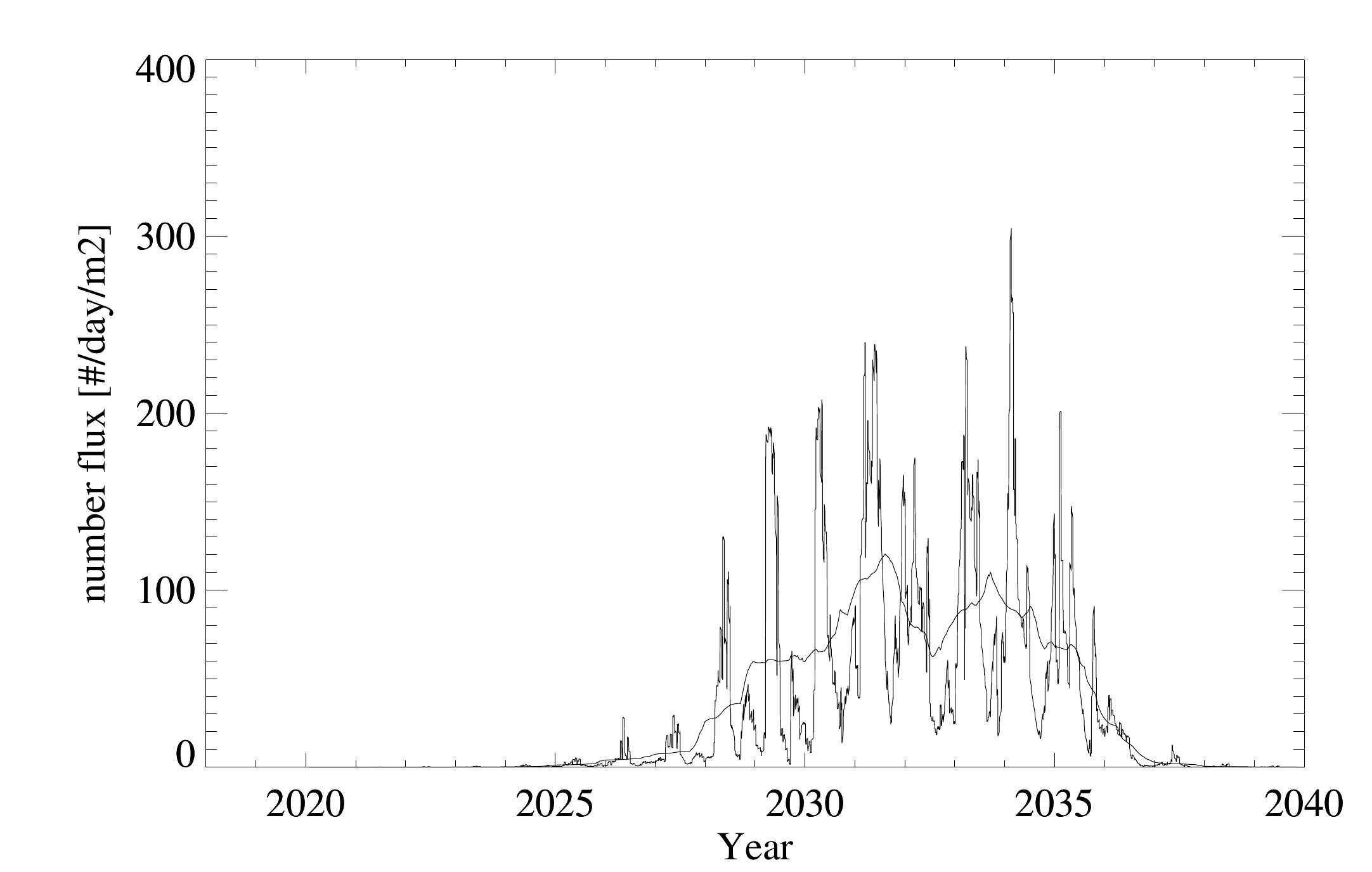} &
\includegraphics[width=0.45\textwidth]{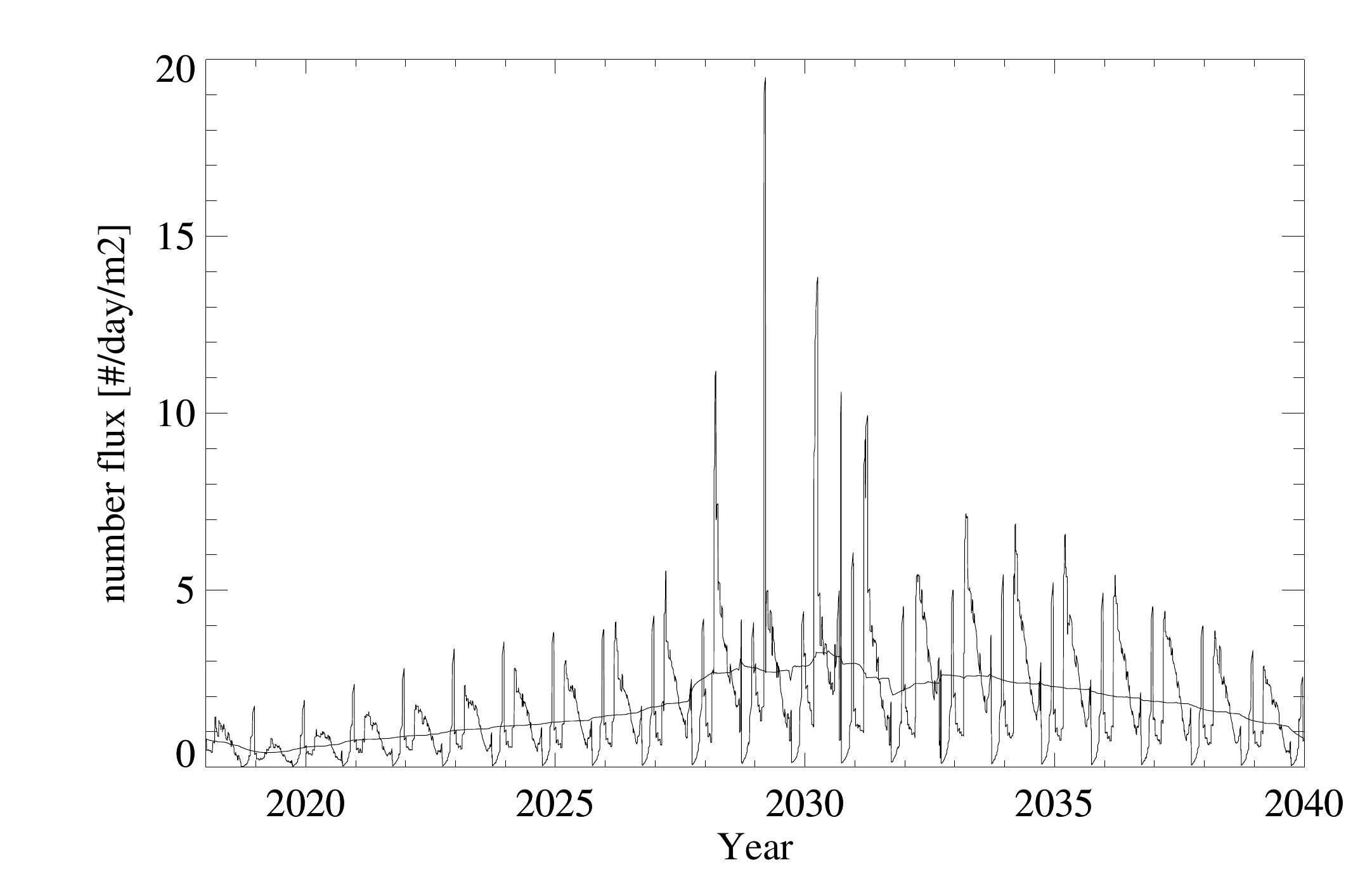} \\
\end{tabular}
\caption{Number flow of the ISD particles over a full solar cycle (2018-2040). {\em Left panel:} Small particles ($a_d<0.2\,\mathrm{\mu m}$). {\em Right panel}:  Large particles, ($a_d>0.3\,\mathrm{\mu m}$). Fluxes are given in $\mathrm{particles\,\,day^{-1}\,\,m^{-2}}$ assuming the detection area is always orthogonal to the ISD flow. The smooth line shows the moving average of the flux over a period of 1 year.
The number flux can reach values close to zero when the Earth and ISD velocity vectors become similar.}
\label{fig:numberflux}
\end{figure*}

\subsection{Limitations of the model}
Our model of the ISD flow is subject to a number of limitations as a result of (1) uncertainty in our understanding of the physical properties of ISD inside the solar system (i.e. composition, porosity, and mass; and therefore different radiation pressure effects), (2) the assumption of a constant electric charge, (3) limited information concerning the actual three-dimensional structure of the IMF on smaller scales (e.g. coronal mass ejections, CMEs), (4) practical considerations of the computation time needed, such as the averaging of the magnetic field averaged over the solar rotation period, and (5) uncertainties in the absolute normalisation of the simulated fluxes based on the Ulysses measurements. 

The uncertainty of physical properties (1) leads to a number of consequences that have been discussed in detail by \citet{sterken2013a}: a change in material properties would lead to a different $\beta$-curve, which in turn modifies the dynamics of particles of a given size. This can be addressed in the future by running further simulations in order to cover a larger area in the parameter space of $\beta$ and mass (i.e. $Q/m$), as has been done in \citet{sterken2012a} and \citet{sterken2013a}
at lower resolution. 
The absolute flux in our ISD model was normalised by the mass calibration of the Ulysses dust detector. Effects of particle structure and particle density on the mass calibration are not considered here.

The approximation (2) of a constant charge of ISD particles is discussed and justified in \citet{horanyi1996b} and in \citet{slavin2012a}, Fig. 2, for distances between one and several tens of astronomical units from the Sun.

In the case of small variations in the magnetic field, the influence of (3) on the simulation result is small, as the coupling scale of ISD particles with sizes $a_d \gtrsim 0.1\,\mum$ is of the order of $\gtrsim$1-10\,AU, and can be neglected for our simulation. Due to the stochastical nature of (3), the effects of smaller-scale structures most likely cancel out on average. Larger deviations from the Parker IMF, however, as can be found inside CMEs, can lead to severe deflections from the simulated flux. Assuming a magnetic field of 30\,nT, as observed inside individual CMEs \citep{wang2005a,obrien2018a}, this is six times higher than the modelled average solar magnetic field at Earth orbit of 5\,nT, and leads to a typical gyro radius of 1\,AU for 0.1\,$\mu$m particles. The simulation results for small particles $\lesssim 0.1\,\mu$m should therefore be treated with caution and are valid for the nominal Parker IMF only. Due to their stochastical nature, the occurrence of CMEs cannot be predicted, and are beyond the scope of this model.

We have verified the validity of (4) by numerically integrating the trajectories of test particles in a rotating magnetic field. The resulting differences compared to an averaged magnetic field are below $0.1\,$AU, which is smaller than the mesh resolution of the density cube, for all but the particles passing within less than 0.25\,AU of the Sun, which constitutes only a negligible fraction of the overall ISD flux.

Concerning the normalisation of the simulated fluxes (5), we note that the temporal variability of the flux and the direction in the Ulysses ISD dataset are not entirely reproduced by the model, and only the overall flux for each particle size bin is taken into account for the normalisation.

The Ulysses ISD data were chosen as the calibration dataset because this dataset contains the most comprehensive homogeneous observation by a single instrument over a period of 16 years, and covers a large portion of the 22 year solar cycle \citep{strub2015a}. With a total of 987 ISD particles used for determining the mass distribution \citep{krueger2015a}, it also has the best statistical errors of all observations. However, \citet{krueger2018b} compared the model predictions to flux measurements from other missions such as Helios, Cassini, and Galileo, and they agree within a factor of typically $\lesssim 2-3$, despite the differences in solar distance covered by these missions. We conclude that this marks the limits of the current understanding of the ISD flow through the heliosphere.

\section{Scenarios for interstellar dust missions}\label{isdmissions}

The investigation of ISD particles requires a statistical significant sample in order to study the dynamics, mass distribution and composition over time and space. 
What is the composition of the ISD particles and how did they form? What are their dynamical parameters, and what can be learned from a test and refinement of the model based on improved measurements? How do ISD particles interact with the heliosphere, and how does their dynamics and overall flux modulation depend on the location in our solar system?

The ISD dynamics and flux described above allow for a detailed study of the ISD as long as observational campaigns are carefully planned. This is a lesson learned from the Cassini mission, where a special observation campaign in 1999 led to the discovery of interstellar particles as close as 1 AU to the Sun \citep{altobelli:04}. Later, only the long integration time during ISD campaigns in Saturn's orbit allowed for an analysis of the composition of 36 ISD particles \citep{altobelli2016a}. In order to plan observational strategies, the modelling and prediction of ISD particle properties and their variations over time and space are essential.
What is the best observational strategy and what are the requirements for a mission close to Earth's orbit?
What is the ideal pointing scenario and what are the related observational times? 
\newline

Our model indicates that observations at 1\,AU are well suited to study the ISD in the solar system including its varying dynamics during different phases of the solar cycle. 
This can be achieved on a high Earth orbit, but a lunar orbit or an orbit about the libration point L2 of the Earth-Sun system would be viable options as well. 

\begin{figure}[pth]
\centering
\includegraphics[width=\columnwidth]{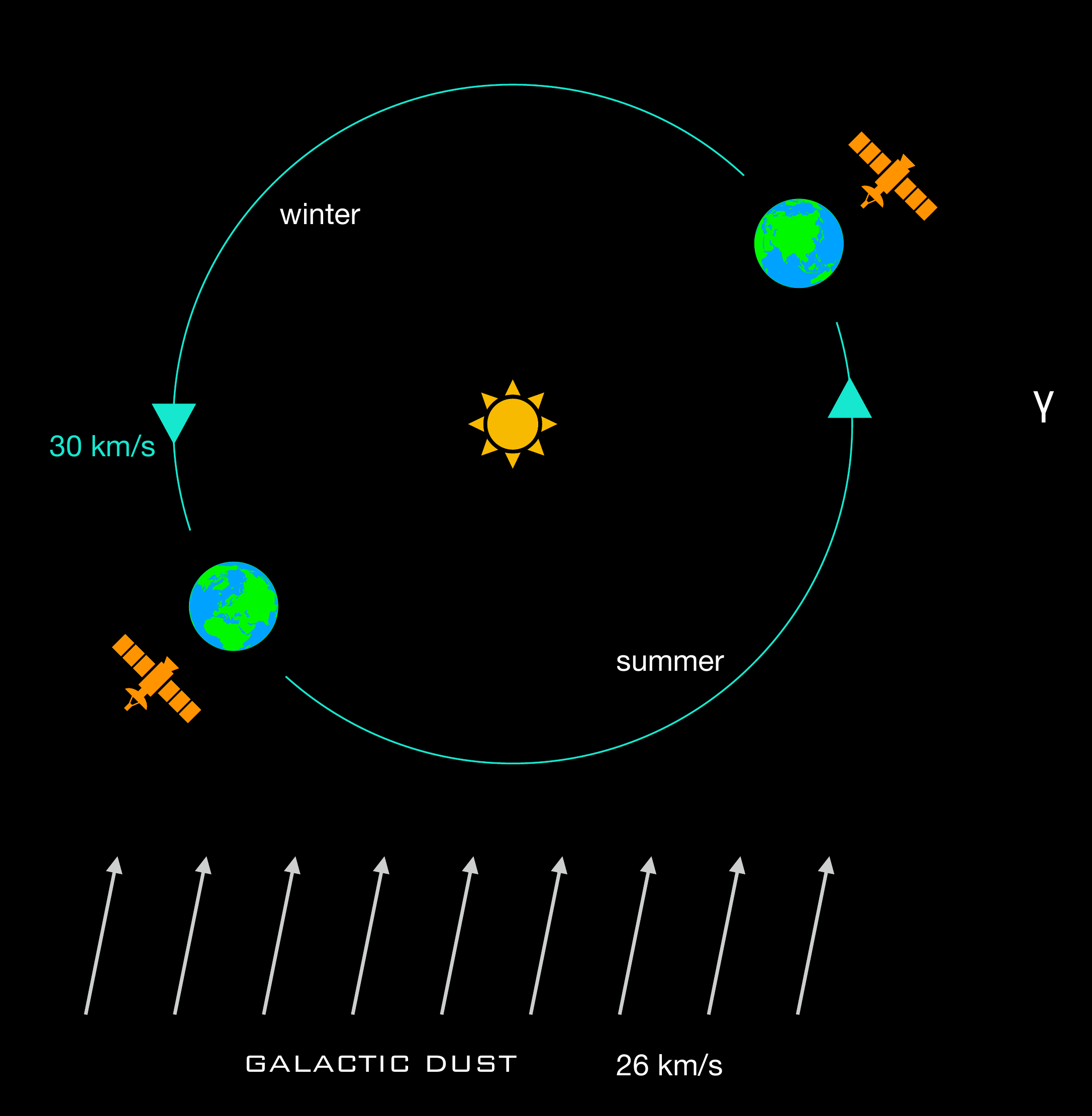} 
\caption{Mission concept for the characterisation of galactic and interplanetary dust. The spacecraft is located in the Lagrange point L2 of the Earth-Sun system. Two positions of the Earth and the satellite are shown (spring and autumn). In spring and summer the relative impact speed and the related dust fluxes of ISD are high.
}
\label{fig:earth_relvel}
\end{figure}

As discussed in Sect.~4, the Earth speed of about 30\,\kms\ leads to a strong modulation of the ISD flow which has a typical velocity of 26\,\kms. This strongly affects both the registered dust flux and the impact velocity along the Earth's orbit. 
These high-amplitude variations in flux and velocity can be used to distinguish the interplanetary from the ISD population \citep{gruen2009}. 
Figure~\ref{fig:earth_relvel} schematically shows the Earth's orbit with the ISD observatory at the libration point L2. Highest fluxes and impact velocities are measured when the Earth moves anti-parallel to the ISD flow (left), and lowest fluxes and impact speeds occur when the spacecraft moves parallel to the ISD direction (right). As the in-situ detectors typically have large but limited opening angles, the simulated flow directions can be used to 
optimise observation campaigns and to align the instrument boresight to the ISD ram direction
(cf. Fig.~\ref{fig:lambet1_earth}).

Generally, two different methods are available for ISD observations:
in-situ measurements, and sample return with a subsequent in-depth laboratory analysis of the samples here on Earth. Both options require different operational scenarios. In the past several in-situ missions serendipitously detected and analysed ISD: Ulysses, Galileo, Cassini and Helios. The latter two provided the first compositional analysis of ISD grains \citep{altobelli2006,altobelli2016a}. Sample return and subsequent analysis can make use of the latest laboratory techniques that are not available for space instrumentation, promising very accurate results.  
The Stardust mission is an example of such a strategy, and successfully brought back ISD samples to the Earth for laboratory analysis \citep{westphal2014a}. However, the laboratory analysis took many years to complete because the process to identify a few submicron-sized particles on a 0.1\,m$^2$ collector is very challenging. We briefly discuss both mission scenarios.

\subsection{In-situ observations}
Reliable in-situ dust instrumentation is available for the analysis of dust fluxes, masses, electrical charges and, in particular, composition \citep{auer2002,auer2001, srama2004,sternovsky2017,krueger2017b}.
The instruments are typically scalable in size to allow for the large sensitive areas that are needed to detect a statistically meaningful number of particles. As can be seen in Fig.~\ref{fig:numberflux}, the flux of interstellar particles of size $a_d>0.3\,\mum$ at 1 AU ranges between approximately $0.01\,\mathrm{m^{-2}\,day^{-1}}$ 
($1.3\times 10^{-7}\,\mathrm{m}^{-2}\,\mathrm{s}^{-1}$) and $20\,\mathrm{m^{-2}\,day^{-1}}$
($2.3\times10^{-4}\,\mathrm{m}^{-2}\,\mathrm{s}^{-1}$),
which is only a fraction of the flux of interplanetary particles. Therefore, accurate trajectory information is necessary in order to reliably distinguish both types of dust particles.

The following considerations assume a sensitive detector area of $0.1\,\mathrm{m}^2$, which is the same as in the dust detectors on board Ulysses and Galileo. The boresight however is assumed to point constantly towards the ISD flow. This is equivalent to a 4$\pi$ field of view with a constant sensitive area.  For a given dust flow, this leads to approximately three times higher dust fluxes than the rotating scanning pattern 
used by Ulysses.

The year-integrated flux of large particles with $a_d\gtrsim 0.3\,\mum$ (i.e. above the $\beta$-cone cutoff size) exhibits marked variations by a factor of approximately seven with the (de-)focussing cycle of the solar magnetic field, and reaches a maximum of 116 particles in 2030, and a minimum of 16 particles in 2041. We note that due to the steep mass distribution these particle counts are dominated by the number in the smallest size bin. 

Small dust particles with $a_d\lesssim 0.11\,\mum$ are primarily observable at Earth's orbit close to the phase of strongest focussing. A number of 2610 particles is expected in 2030 which is maximum focussing.


In-situ observations allow us to link optical particle properties with dynamical properties. 
For a limited range of $1<\beta<1.4$, a direct measurement of the $\beta$-curve would be possible through $\beta$-spectroscopy \citep{altobelli2005a}. For a given particle mass, the size of the $\beta$-cone (the zone devoid of particles) largely depends on $\beta$. Therefore, by measuring the size of this exclusion zone for different masses, the $\beta$-mass-relation, a determining factor for the ISD dynamics, can be established. 
\newline

With new instrumentation, the mass distribution of the ISD can be measured with a better accuracy. At the moment, this is a major uncertainty in our model: The normalisation of the incoming flux is based on the mass distribution measured by Ulysses, which is limited by relatively low number counts per bin. A multi-year mission or an increased sensitive area ($>0.1 \mathrm{m}^2$) is therefore required in order to analyse a sufficient number of ISD particles. A new mission may also provide much higher sensitivities for small particles. During a period of electromagnetic focussing, a dust instrument could possibly measure ISD particles $\lesssim 0.1\mum$ if it is able to distinguish them from sub-micron sized interplanetary dust particles. 
If successful, such extra measurements may strongly support understanding the filtering of ISD - especially in the boundary regions of the heliosphere (heliosheath, etc.).

One significant point along the orbit is the gravitational focussing region downstream of the Sun. There the density and flux of large ISD particles are about one order of magnitude higher than the typical values along the rest of the orbit. This would provide an important opportunity to detect larger ISD particles with $\beta<1$. 
The simulation results allow for detailed investigations of the required ISD ram direction, which deviates from the direct Sun direction.

A nanodust analyser (NDA) has been developed for the detection and compositional analysis of dust particles originating in the inner heliosphere \citep{obrien2014a}. The NDA is the first instrument optimised for the detection and compositional analysis of nanodust particles that are arriving from close to the Sun's direction. The NDA is a linear time-of-flight impact mass spectrometer derived from the successful Cassini CDA instrument. The operation and performance of the NDA instrument, including the efficiency of the solar wind rejection grids, have been verified in the laboratory using a dust accelerator and a solar wind simulator. This instrument is well suited to analysing the down-stream focused ISD flux from the solar direction.

\subsection{Dust sample return}

The second method to analyse ISD particles is dust collection with sample return. 
While Stardust has collected and brought back to Earth some ISD, the process to identify the few submicron sized dust particles on a 0.1 m$^2$ collector was very challenging \citep{westphal2014a}.
An active cosmic dust collector \citep{gruen2012} would tremendously improve future dust collections in interplanetary space by determining the impact position and time together with the velocity vector of the impacting particle. 

A sample return of interstellar matter mission (SARIM) was described by \citet{srama2009a}. In this scenario the spacecraft would be placed at the L2 libration point of the Sun/Earth system, outside the Earth's debris belts and inside the solar-wind charging environment. SARIM is a three-axes stabilised spacecraft and collects interstellar particles between July and October when the relative encounter speeds with ISD particles are lowest (4 to $20\,\mathrm{km\,s^{-1}}$, Fig. \ref{fig:earth_relvel}). 
Active dust collectors with a total sensitive area of 1\,m$^2$ determine the trajectory, speed, mass, and impact location of individual dust impacts. This allows for a discrimination between interstellar and interplanetary dust particles. During a three-year dust collection period, several hundred interstellar and several thousand interplanetary particles can be collected by such a detector. 
At the end of the collection phase, collector modules would be stored and sealed in a sample return capsule. The probe with the capsule would then return to Earth and an in-depth dust analysis with advanced laboratory methods would then be possible.


\section{Conclusions}
The Earth is currently collecting an average of about 100\,kg per year of interstellar material in the form of ISD particles. This flow is small compared to the influx of $\sim30\,000\,\mathrm{t\,\,yr^{-1}}$ of interplanetary dust on to the Earth \citep{love1993}. However, there is significant interest in this exotic material since interstellar particles are the raw material
from which a protoplanetary disk and subsequently planets, asteroids and comets form.

The flow of ISD particles near the Earth is strongly modulated by the Earth's orbit around the Sun and by the solar cycle variation of the interplanetary magnetic field. While the flow pattern generated by gravity and the radiation pressure are constant in time, the solar magnetic field leads to significant variations in flux, direction, velocity, and size distribution with a periodicity of 22 years. This must be considered when planning observations. The identification of ISD and its distinction from interplanetary dust can be accomplished by using modern dust instruments that characterise both the dynamical state as well as the composition of the analysed particles.

\begin{acknowledgements}
      This work was funded under ESA contract 4000106316/12/NL/MV – IMEX. Wilcox Solar Observatory data used in this study was obtained via the web site http://wso.stanford.edu at 2017:10:04\_05:08:21 PDT courtesy of J.T. Hoeksema. 
      
We would like to thank our referee,  Mihály Horányi, for his valuable comments helping to improve the presentation of our results.
\end{acknowledgements}

\bibliographystyle{aa}

\bibliography{literatur,lit_imextn4veerle}  

\begin{thebibliography}{57}
\expandafter\ifx\csname natexlab\endcsname\relax\def\natexlab#1{#1}\fi

\bibitem[{Altobelli(2004)}]{altobelli:04}
Altobelli, N. 2004, PhD thesis, Ruprecht-Karls-Universit{\"a}t Heidelberg

\bibitem[{{Altobelli} {et~al.}(2006){Altobelli}, {Gr{\"u}n}, \&
  {Landgraf}}]{altobelli2006}
{Altobelli}, N., {Gr{\"u}n}, E., \& {Landgraf}, M. 2006, \aap, 448, 243

\bibitem[{{Altobelli} {et~al.}(2005){Altobelli}, {Kempf}, {Kr{\"u}ger},
  {Landgraf}, {Roy}, \& {Gr{\"u}n}}]{altobelli2005a}
{Altobelli}, N., {Kempf}, S., {Kr{\"u}ger}, H., {et~al.} 2005, \jgr, 110, 7102

\bibitem[{{Altobelli} {et~al.}(2003){Altobelli}, {Kempf}, {Landgraf}, {Srama},
  {Dikarev}, {Kr\"uger}, {Moragas-Klostermeyer}, \& {Gr{\"u}n}}]{altobelli2003}
{Altobelli}, N., {Kempf}, S., {Landgraf}, M., {et~al.} 2003, \jgr, 108, A10, 7

\bibitem[{{Altobelli} {et~al.}(2016){Altobelli}, {Postberg}, {Fiege},
  {Trieloff}, {Kimura}, {Sterken}, {Hsu}, {Hillier}, {Khawaja},
  {Moragas-Klostermeyer}, {Blum}, {Burton}, {Srama}, {Kempf}, \&
  {Gruen}}]{altobelli2016a}
{Altobelli}, N., {Postberg}, F., {Fiege}, K., {et~al.} 2016, Science, 352, 312

\bibitem[{{Arai} {et~al.}(2018){Arai}, {Kobayashi}, {Ishibashi}, {Yoshida},
  {Kimura}, {Wada}, {Senshu}, {Yamada}, {Okudaira}, {Okamoto}, {Kameda},
  {Srama}, {Kr\"uger}, {Ishiguro}, {Yabuta}, {Nakamura}, {Watanabe}, {Ito},
  {Ohtsuka}, {Tachibana}, {Mikouchi}, {Komatsu}, {Nakamura-Messenger},
  {Sasaki}, {Hiroi}, {Abe}, {Urakawa}, {Hirata}, {Demura}, {Komatsu},
  {Noguchi}, {Sekiguchi}, {Inamori}, {Yano}, {Yoshikawa}, {Ohtsubo}, {Okada},
  {Iwata}, {Nishiyama}, {Toyota}, {Kawakatsu}., \& {Takashima}}]{arai2018}
{Arai}, T., {Kobayashi}, M., {Ishibashi}, K., {et~al.} 2018, in Lunar and
  Planetary Institute Science Conference Abstracts, Vol.~49, Lunar and
  Planetary Institute Science Conference Abstracts, 2570

\bibitem[{{Auer}(2001)}]{auer2001}
{Auer}, S. 2001, in Interplanetary Dust, ed. E.~{Gr{\"u}n}, B.~A.~S.
  {Gustafson}, S.~F. {Dermott}, \& H.~{Fechtig} (Springer Verlag, Berlin
  Heidelberg New York), 385--444

\bibitem[{{Auer} {et~al.}(2002){Auer}, {Gr{\"u}n}, {Srama}, {Kempf}, \&
  {Auer}}]{auer2002}
{Auer}, S., {Gr{\"u}n}, E., {Srama}, R., {Kempf}, S., \& {Auer}, R. 2002,
  \planss, 50, 773

\bibitem[{{Burns} {et~al.}(1979){Burns}, {Lamy}, \& {Soter}}]{burns1979a}
{Burns}, J.~A., {Lamy}, P.~L., \& {Soter}, S. 1979, Icarus, 40, 1

\bibitem[{Cravens(1997)}]{cravens1997}
Cravens, T.~E. 1997, Physics of Solar System Plasmas, Cambridge Atmospheric and
  Space Science Series (Cambridge University Press)

\bibitem[{{Draine} \& {Lee}(1984)}]{draine1984}
{Draine}, B.~T. \& {Lee}, H.~M. 1984, \apj, 285, 89

\bibitem[{{Frisch} {et~al.}(1999{\natexlab{a}}){Frisch}, {Dorschner},
  {Gei{\ss}}, {Greenberg}, {Gr{\"u}n}, {Landgraf}, {Hoppe}, {Jones},
  {Kr{\"a}tschmer}, {Linde}, {Morfill}, {Reach}, {Slavin}, {Svestka}, {Witt},
  \& {Zank}}]{frisch1999a}
{Frisch}, P.~C., {Dorschner}, J., {Gei{\ss}}, J., {et~al.} 1999{\natexlab{a}},
  \apj, 525, 492

\bibitem[{{Frisch} {et~al.}(1999{\natexlab{b}}){Frisch}, {Dorschner}, {Geiss},
  {Greenberg}, {Gr{\"u}n}, {Landgraf}, {Hoppe}, {Jones}, {Kr{\"a}tschmer},
  {Linde}, {Morfill}, {Reach}, {Slavin}, {Svestka}, {Witt}, \&
  {Zank}}]{frisch:99a}
{Frisch}, P.~C., {Dorschner}, J.~M., {Geiss}, J., {et~al.} 1999{\natexlab{b}},
  Astrophys. J., 525, 492

\bibitem[{{Gr{\"u}n} {et~al.}(1994){Gr{\"u}n}, {Gustafson}, {Mann}, {Baguhl},
  {Morfill}, {Staubach}, {Taylor}, \& {Zook}}]{gruen1994a}
{Gr{\"u}n}, E., {Gustafson}, B.~E., {Mann}, I., {et~al.} 1994, \aap, 286, 915

\bibitem[{{Gr{\"u}n} {et~al.}(2009){Gr{\"u}n}, {Srama}, {Altobelli}, {Altwegg},
  {Carpenter}, {Colangeli}, {Glassmeier}, {Helfert}, {Henkel}, {Horanyi},
  {J{\"a}ckel}, {Kempf}, {Landgraf}, {McBride}, {Moragas-Klostermeyer},
  {Palumbo}, {Scholten}, {Srowig}, {Sternovsky}, \& {Vo}}]{gruen2009}
{Gr{\"u}n}, E., {Srama}, R., {Altobelli}, N., {et~al.} 2009, Experimental
  Astronomy, 23, 981

\bibitem[{{Gr{\"u}n} {et~al.}(2012){Gr{\"u}n}, {Sternovsky}, {Horanyi},
  {Hoxie}, {Robertson}, {Xi}, {Auer}, {Landgraf}, {Postberg}, {Price}, {Srama},
  {Starkey}, {Hillier}, {Franchi}, {Tsou}, {Westphal}, \&
  {Gainsforth}}]{gruen2012}
{Gr{\"u}n}, E., {Sternovsky}, Z., {Horanyi}, M., {et~al.} 2012, \planss, 60,
  261

\bibitem[{Gr{\"u}n {et~al.}(1985)Gr{\"u}n, {Zook}, {Fechtig}, \&
  {Giese}}]{gruen:85a}
Gr{\"u}n, E., {Zook}, H., {Fechtig}, H., \& {Giese}, R. 1985, Icarus (ISSN
  0019-1035), vol.62, May 1985, p.244-272., 62, 244

\bibitem[{{Gr{\"u}n} {et~al.}(1993){Gr{\"u}n}, {Zook}, {Baguhl}, {Balogh},
  {Bame}, {Fechtig}, {Forsyth}, {Hanner}, {Hor\'anyi}, {Kissel}, {Lindblad},
  {Linkert}, {Linkert}, {Mann}, {McDonnell}, {Morfill}, {Phillips},
  {Polanskey}, {Schwehm}, {Siddique}, {Staubach}, {Svestka}, \&
  {Taylor}}]{gruen1993a}
{Gr{\"u}n}, E., {Zook}, H.~A., {Baguhl}, M., {et~al.} 1993, Nature, 362, 428

\bibitem[{{Gustafson}(1994)}]{gustafson1994a}
{Gustafson}, B.~A.~S. 1994, Annual Review of Earth and Planetary Sciences, 22,
  553

\bibitem[{Gustafson(1994)}]{gustafson:94a}
Gustafson, B.~S. 1994, Ann. Rev. Earth Planet. Sci., 22, 553

\bibitem[{{Hoeksema}(2018)}]{WSO:18}
{Hoeksema}, J. 2018, Wilcox Solar Observatory, http://wso.stanford.edu

\bibitem[{{Hor{\'a}nyi}(1996)}]{horanyi1996b}
{Hor{\'a}nyi}, M. 1996, \araa, 34, 383

\bibitem[{Kempf {et~al.}(2004)Kempf, Srama, Altobelli, Auer, Tschernjawski,
  Bradley, Burton, Helfert, Johnson, Kr{\"u}ger, Moragas-Klostermeyer, \&
  Gr{\"un}}]{kempf:04a}
Kempf, S., Srama, R., Altobelli, N., {et~al.} 2004, Icarus, 171, 317

\bibitem[{Kimura \& Mann(1998)}]{kimura1998a}
Kimura, H. \& Mann, I. 1998, ApJ, 499, 454

\bibitem[{{Kobayashi} {et~al.}(2018){Kobayashi}, {Srama}, {Kr\"uger}, {Arai},
  \& {Kimura}}]{kobayashi2018b}
{Kobayashi}, M., {Srama}, R., {Kr\"uger}, H., {Arai}, T., \& {Kimura}, H. 2018,
  in Lunar and Planetary Institute Science Conference Abstracts, Vol.~49, Lunar
  and Planetary Institute Science Conference Abstracts, 2050

\bibitem[{{Kr{\"u}ger} {et~al.}(2018{\natexlab{a}}){Kr{\"u}ger}, {Altobelli},
  {Strub}, {Sterken}, {Srama}, \& {Gr\"un}}]{krueger2018b}
{Kr{\"u}ger}, H., {Altobelli}, N., {Strub}, P., {et~al.} 2018{\natexlab{a}},
  \aap, submitted

\bibitem[{{Kr{\"u}ger} {et~al.}(2017){Kr{\"u}ger}, {Kobayashi}, {Arai},
  {Srama}, {Sarli}, {Kimura}, {Moragas-Klostermeyer}, {Soja}, {Altobelli}, \&
  {Gr{\"u}n}}]{krueger2017b}
{Kr{\"u}ger}, H., {Kobayashi}, M., {Arai}, T., {et~al.} 2017, European
  Planetary Science Congress, 11, EPSC2017

\bibitem[{{Kr{\"u}ger} {et~al.}(2015){Kr{\"u}ger}, {Strub}, {Gr{\"u}n}, \&
  {Sterken}}]{krueger2015a}
{Kr{\"u}ger}, H., {Strub}, P., {Gr{\"u}n}, E., \& {Sterken}, V.~J. 2015, \apj,
  812, 139

\bibitem[{{Kr{\"u}ger} {et~al.}(2018{\natexlab{b}}){Kr{\"u}ger}, {Strub},
  {Srama}, {Kobayashi}, {Arai}, {Kimura}, {Moragas-Klostermeyer}, {Altobelli},
  {Sterken}, {Agarwal}, \& {Gr\"un}}]{krueger2018a}
{Kr{\"u}ger}, H., {Strub}, P., {Srama}, R., {et~al.} 2018{\natexlab{b}},
  \planss, submitted

\bibitem[{{Lallement} \& {Bertaux}(2014)}]{lallement2014a}
{Lallement}, R. \& {Bertaux}, J.~L. 2014, A{\&}A, 565, A41

\bibitem[{{Landgraf}(1998)}]{landgraf1998a}
{Landgraf}, M. 1998, PhD thesis, Ruprecht-Karls-Universit{\"a}t Heidelberg

\bibitem[{{Landgraf}(2000{\natexlab{a}})}]{landgraf:00a}
{Landgraf}, M. 2000{\natexlab{a}}, J. Geophys. Res., 105, 10303

\bibitem[{{Landgraf}(2000{\natexlab{b}})}]{landgraf2000c}
{Landgraf}, M. 2000{\natexlab{b}}, \jgr, 105, no. A5, 10,303

\bibitem[{{Landgraf} {et~al.}(1999){Landgraf}, {Augustsson}, {Gr{\"u}n}, \&
  {Gustafson}}]{landgraf:99a}
{Landgraf}, M., {Augustsson}, K., {Gr{\"u}n}, E., \& {Gustafson}, B.~A.~S.
  1999, Science, 286, 2319

\bibitem[{{Linde} \& {Gombosi}(2000)}]{linde2000a}
{Linde}, T.~J. \& {Gombosi}, T.~I. 2000, \jgr, 105, 10411

\bibitem[{{Love} \& {Brownlee}(1993)}]{love1993}
{Love}, S.~G. \& {Brownlee}, D.~E. 1993, Science, 262, 550

\bibitem[{{Mathis} {et~al.}(1977){Mathis}, {Rumpl}, \&
  {Nordsieck}}]{mathis1977}
{Mathis}, J.~S., {Rumpl}, W., \& {Nordsieck}, K.~H. 1977, \apj, 217, 425

\bibitem[{{Morfill} \& {Gr{\"u}n}(1979)}]{morfill1979a}
{Morfill}, G.~E. \& {Gr{\"u}n}, E. 1979, \planss, 27, 1269

\bibitem[{{Mukai}(1981)}]{mukai1981a}
{Mukai}, T. 1981, \aap, 99, 1

\bibitem[{{O'Brien} {et~al.}(2014){O'Brien}, {Auer}, {Gemer}, {Gr{\"u}n},
  {Horanyi}, {Juhasz}, {Kempf}, {Malaspina}, {Mocker}, {Moebius}, {Srama}, \&
  {Sternovsky}}]{obrien2014a}
{O'Brien}, L., {Auer}, S., {Gemer}, A., {et~al.} 2014, Review of Scientific
  Instruments, 85, 035113

\bibitem[{{O'Brien} {et~al.}(2018){O'Brien}, {Juh{\'a}sz}, {Sternovsky}, \&
  {Hor{\'a}nyi}}]{obrien2018a}
{O'Brien}, L., {Juh{\'a}sz}, A., {Sternovsky}, Z., \& {Hor{\'a}nyi}, M. 2018,
  \planss, 156, 7

\bibitem[{{Parker}(1958)}]{parker1958}
{Parker}, E.~N. 1958, \apj, 128, 664

\bibitem[{{Sarli} {et~al.}(2018){Sarli}, {Horikawa}, {Yam}, {Kawakatsu}, \&
  {Yamamoto}}]{sarli2018}
{Sarli}, B.~V., {Horikawa}, M., {Yam}, C.~H., {Kawakatsu}, Y., \& {Yamamoto},
  T. 2018, Journal of the Astronautical Sciences, 65, 82

\bibitem[{{Slavin} {et~al.}(2012){Slavin}, {Frisch}, {M{\"u}ller},
  {Heerikhuisen}, {Pogorelov}, {Reach}, \& {Zank}}]{slavin2012a}
{Slavin}, J.~D., {Frisch}, P.~C., {M{\"u}ller}, H.-R., {et~al.} 2012, \apj,
  760, 46

\bibitem[{{Srama} {et~al.}(2004){Srama}, {Ahrens}, {Auer}, {Bradley},
  {Dikarev}, {Economou}, {Fechtig}, {G\"orlich}, {Grande}, {Graps}, {Gr\"un},
  {Havnes}, {Helfert}, {Hor\'anyi}, {Igenbergs}, {Je{\ss}berger}, {Johnson},
  {Kempf}, {Krivov}, {Kr\"uger}, {Moragas-Klostermeyer}, {Lamy}, {Landgraf},
  {Linkert}, {Linkert}, {Lura}, {Mocker-Ahlreep}, {McDonnell}, {M\"ohlmann},
  {Morfill}, {M\"uller}, {Roy}, {Sch\"afer}, {Schlotzhauer}, {Schwehm},
  {Spahn}, {St\"ubig}, {Svestka}, {Tschernjawski}, {Tuzzolino}, {W\"asch}, \&
  {Zook}}]{srama2004}
{Srama}, R., {Ahrens}, T. J.~{Altobelli}, N., {Auer}, S., {et~al.} 2004, \ssr,
  114, 465

\bibitem[{{Srama} {et~al.}(2009){Srama}, {Stephan}, {Gr{\"u}n}, {Pailer},
  {Kearsley}, {Graps}, {Laufer}, {Ehrenfreund}, {Altobelli}, {Altwegg}, {Auer},
  {Baggaley}, {Burchell}, {Carpenter}, {Colangeli}, {Esposito}, {Green},
  {Henkel}, {Horanyi}, {J{\"a}ckel}, {Kempf}, {McBride},
  {Moragas-Klostermeyer}, {Kr{\"u}ger}, {Palumbo}, {Srowig}, {Trieloff},
  {Tsou}, {Sternovsky}, {Zeile}, \& {R{\"o}ser}}]{srama2009a}
{Srama}, R., {Stephan}, T., {Gr{\"u}n}, E., {et~al.} 2009, Experimental
  Astronomy, 23, 303

\bibitem[{{Sterken} {et~al.}(2013){Sterken}, {Altobelli}, {Kempf},
  {Kr{\"u}ger}, {Srama}, {Strub}, \& {Gr{\"u}n}}]{sterken2013a}
{Sterken}, V.~J., {Altobelli}, N., {Kempf}, S., {et~al.} 2013, \aap, 552, A130

\bibitem[{{Sterken} {et~al.}(2012){Sterken}, {Altobelli}, {Kempf}, {Schwehm},
  {Srama}, \& {Gr{\"u}n}}]{sterken2012a}
{Sterken}, V.~J., {Altobelli}, N., {Kempf}, S., {et~al.} 2012, A{\&}A, 538,
  A102

\bibitem[{{Sterken} {et~al.}(2015){Sterken}, {Strub}, {Kr{\"u}ger}, {von
  Steiger}, \& {Frisch}}]{sterken2015a}
{Sterken}, V.~J., {Strub}, P., {Kr{\"u}ger}, H., {von Steiger}, R., \&
  {Frisch}, P. 2015, \apj, 812, 141

\bibitem[{{Sternovsky} {et~al.}(2007){Sternovsky}, {Amyx}, \&
  {Bano}}]{sternovsky2017}
{Sternovsky}, Z., {Amyx}, K., \& {Bano}, G. e.~a. 2007, Rev. Sci. Instrum., 78

\bibitem[{{Strub} {et~al.}(2015){Strub}, {Kr{\"u}ger}, \&
  {Sterken}}]{strub2015a}
{Strub}, P., {Kr{\"u}ger}, H., \& {Sterken}, V.~J. 2015, \apj, 812, 140

\bibitem[{{Wang} {et~al.}(2005){Wang}, {Du}, \& {Richardson}}]{wang2005a}
{Wang}, C., {Du}, D., \& {Richardson}, J.~D. 2005, Journal of Geophysical
  Research (Space Physics), 110, A10107

\bibitem[{{Weingartner} \& {Draine}(2001)}]{weingartner2001}
{Weingartner}, J.~C. \& {Draine}, B.~T. 2001, \apj, 548, 296

\bibitem[{{Westphal} {et~al.}(2014){Westphal}, {Stroud}, {Bechtel}, {Brenker},
  {Butterworth}, {Flynn}, {Frank}, {Gainsforth}, {Hillier}, {Postberg},
  {Simionovici}, {Sterken}, {Allen}, {Anderson}, {Ansari}, {Bajt}, {Bastien},
  {Bassim}, {Bridges}, {Brownlee}, {Burchell}, {Burghammer}, {Changela},
  {Cloetens}, {Davis}, {Doll}, {Floss}, {Gruen}, {Heck}, {Hoppe}, {Hudson},
  {Huth}, {Kearsley}, {King}, {Lai}, {Leitner}, {Lemelle}, {Leonard}, {Leroux},
  {Lettieri}, {Marchant}, {Nittler}, {Ogliore}, {Ong}, {Price}, {Sandford},
  {Sans Tresseras}, {Schmitz}, {Schoonjans}, {Schreiber}, {Silversmit}, {Sole},
  {Srama}, {Stadermann}, {Stephan}, {Stodolna}, {Sutton}, {Trieloff}, {Tsou},
  {Tyliszczak}, {Vekemans}, {Vincze}, {Von Korff}, {Wordsworth}, {Zevin},
  {Zolensky}, \& {Stardust{@}home dusters}}]{westphal2014a}
{Westphal}, A.~J., {Stroud}, R.~M., {Bechtel}, H.~A., {et~al.} 2014, in Lunar
  and Planetary Inst. Technical Report, Vol.~45, Lunar and Planetary Institute
  Science Conference Abstracts, 2269

\bibitem[{{Witte} {et~al.}(1993){Witte}, {Rosenbauer}, {Banaszkiewicz}, \&
  {Fahr}}]{witte:93a}
{Witte}, M., {Rosenbauer}, H., {Banaszkiewicz}, M., \& {Fahr}, H. 1993,
  Advances in Space Research, 13, 121

\bibitem[{{Wood} {et~al.}(2015){Wood}, {M{\"u}ller}, \& {Witte}}]{wood2015a}
{Wood}, B.~E., {M{\"u}ller}, H.-R., \& {Witte}, M. 2015, ApJ, 801, 62

\bibitem[{{Zubko} {et~al.}(2004){Zubko}, {Dwek}, \& {Arendt}}]{zubko2004}
{Zubko}, V., {Dwek}, E., \& {Arendt}, R.~G. 2004, \apjs, 152, 211

\end{thebibliography}

\clearpage
\begin{appendix}

\section{The ISD flow through the inner planetary system} \label{sec:general_flow}

\begin{figure*}
\noindent
\centering
\begin{tabular}{ccc}
\includegraphics[width=0.3\textwidth]{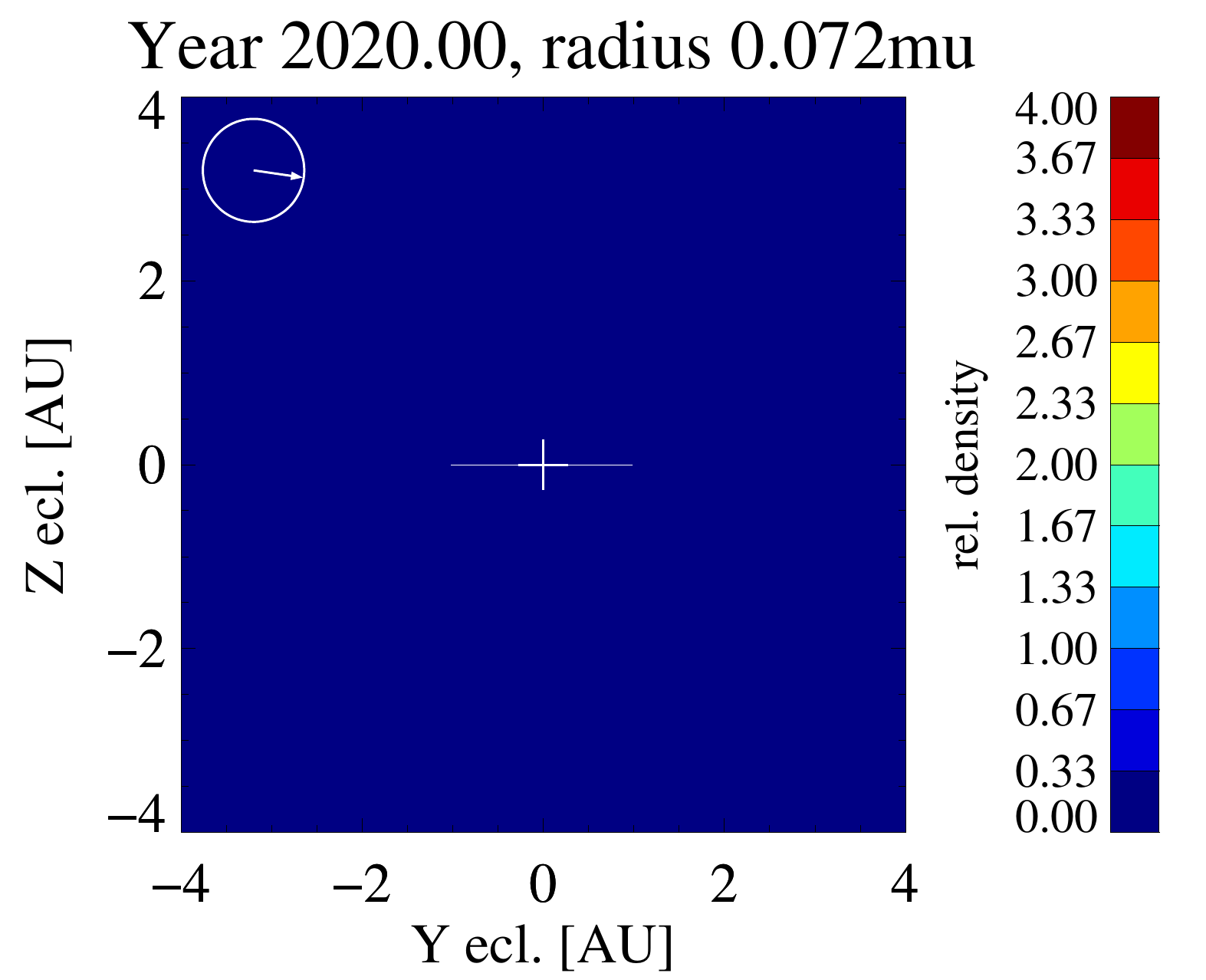} &
\includegraphics[width=0.3\textwidth]{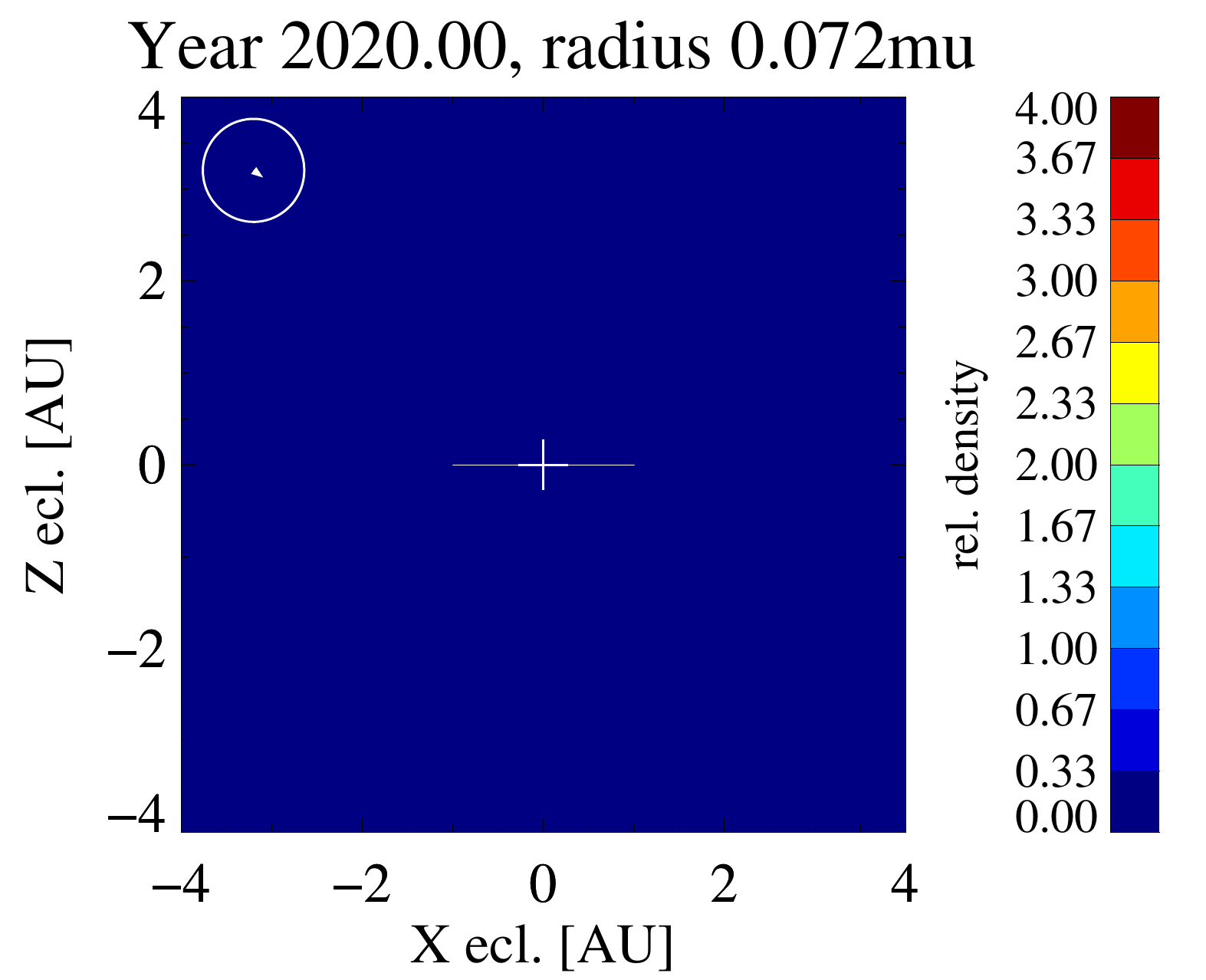} &
\includegraphics[width=0.3\textwidth]{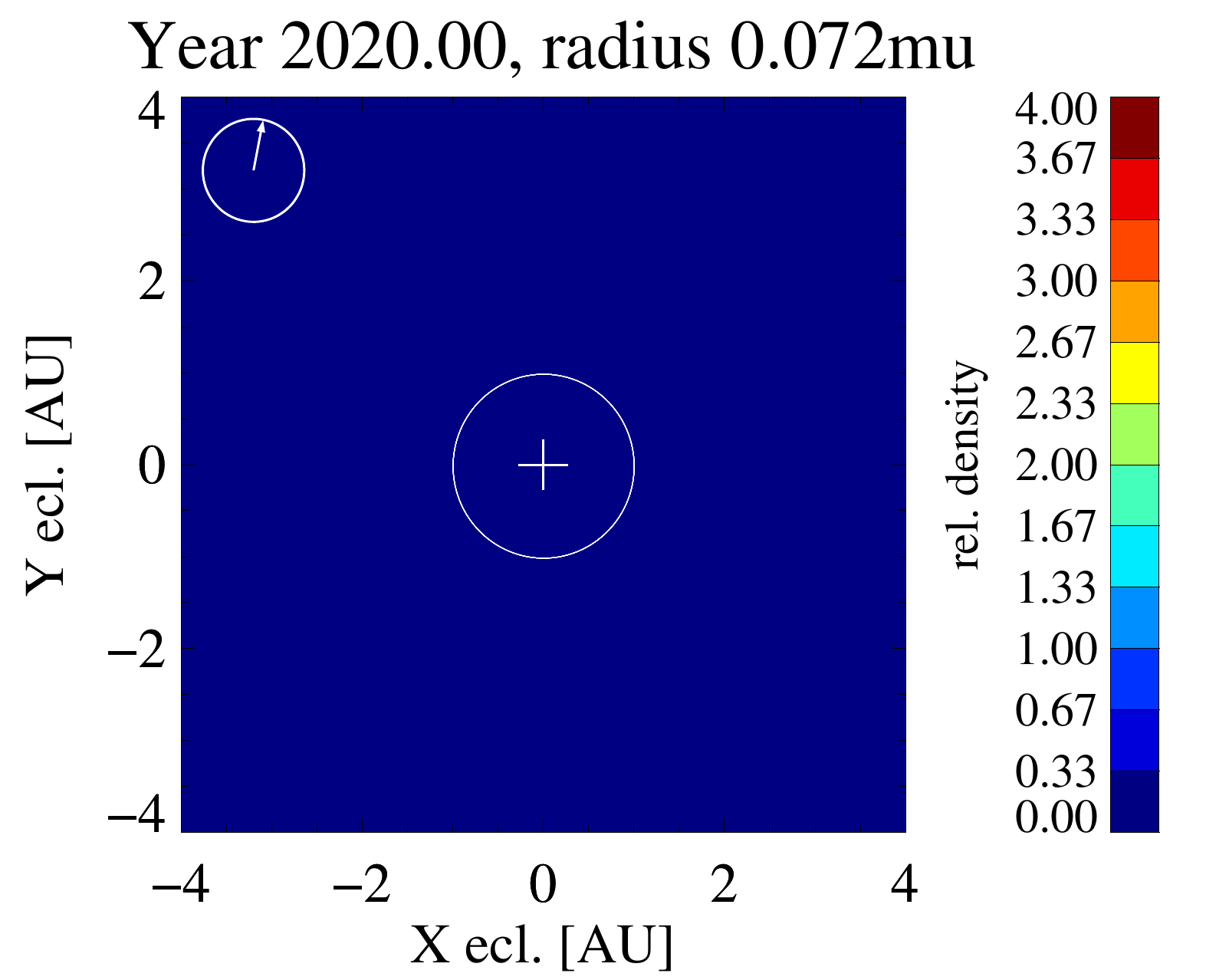} \\
\includegraphics[width=0.3\textwidth]{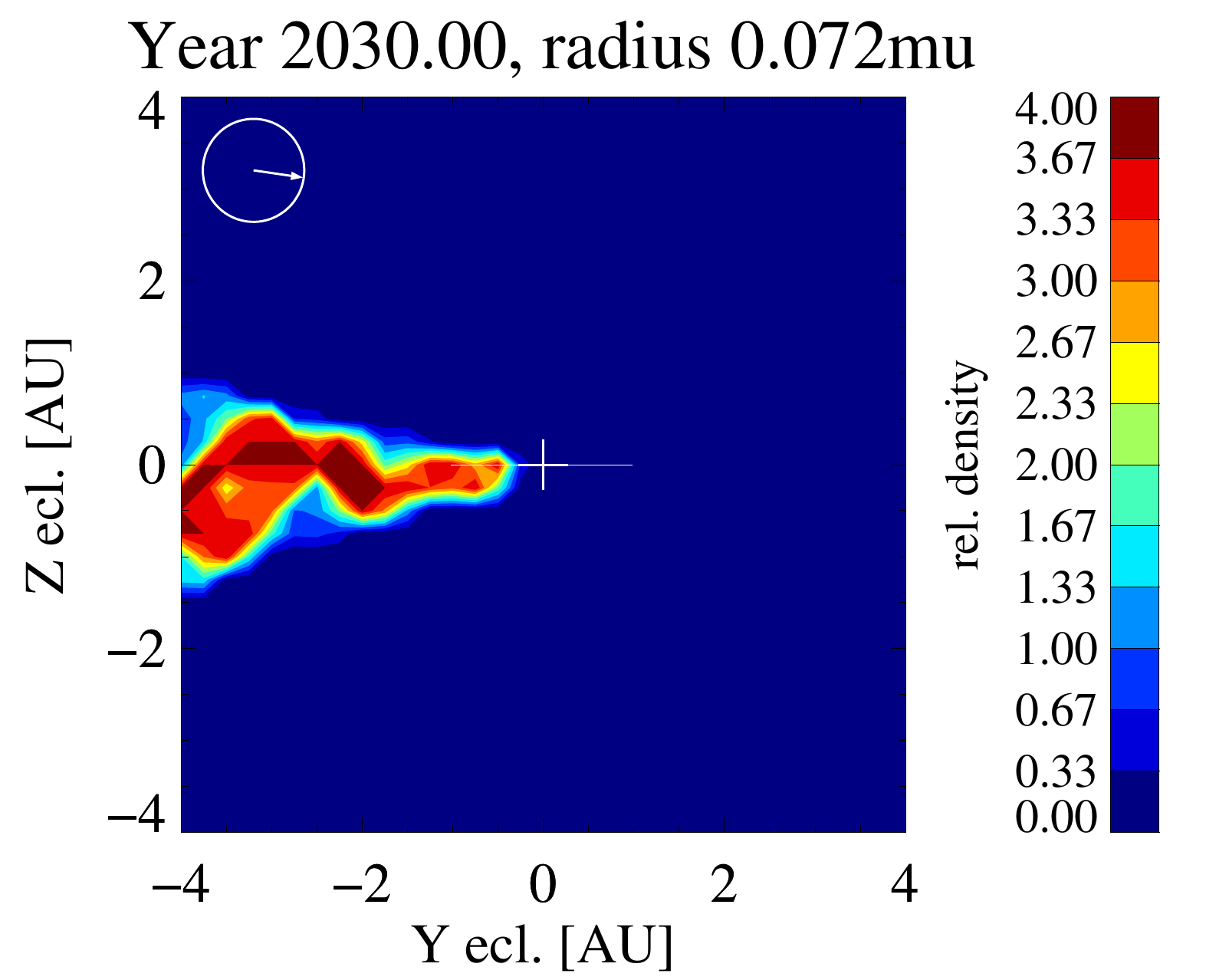} &
\includegraphics[width=0.3\textwidth]{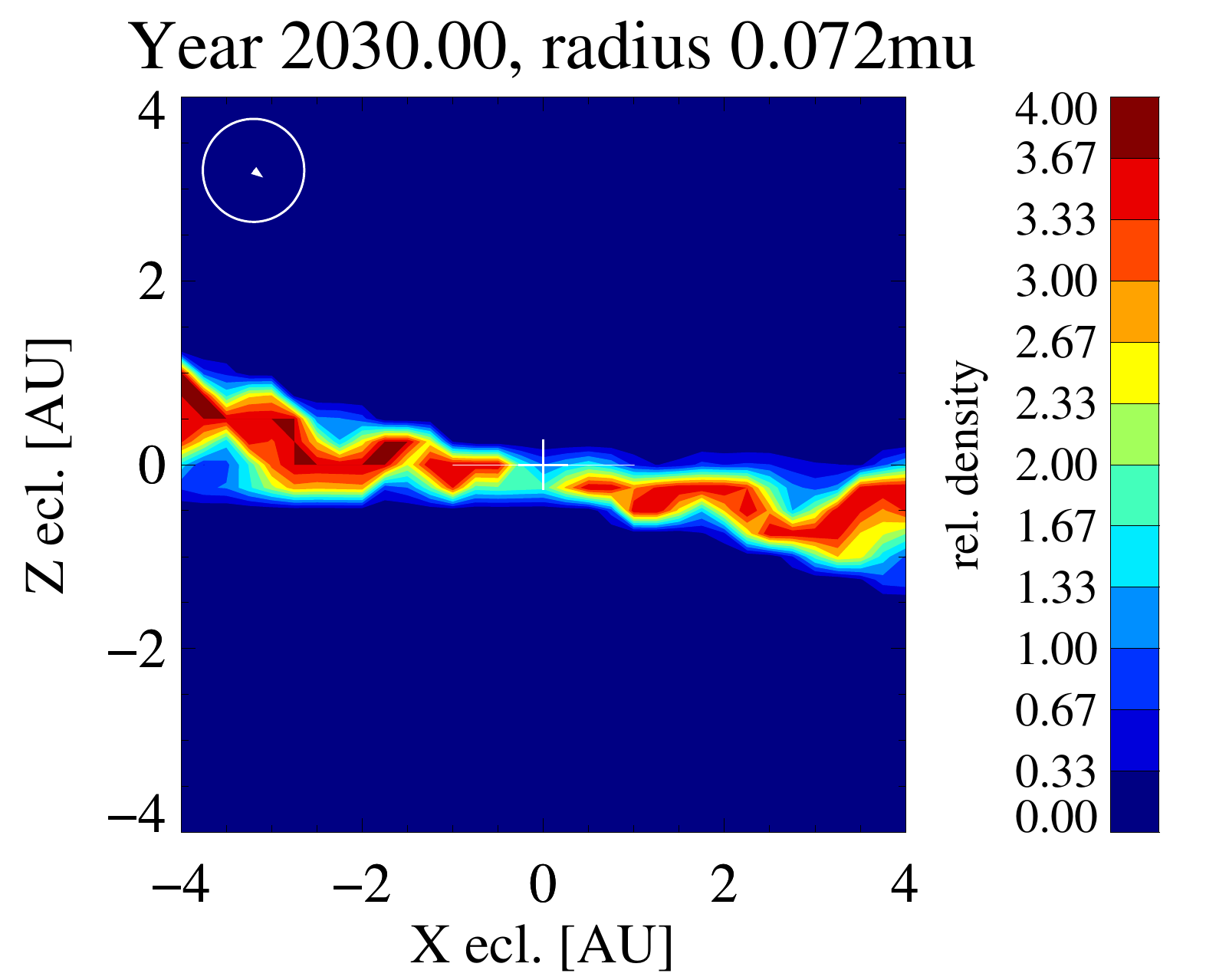} &
\includegraphics[width=0.3\textwidth]{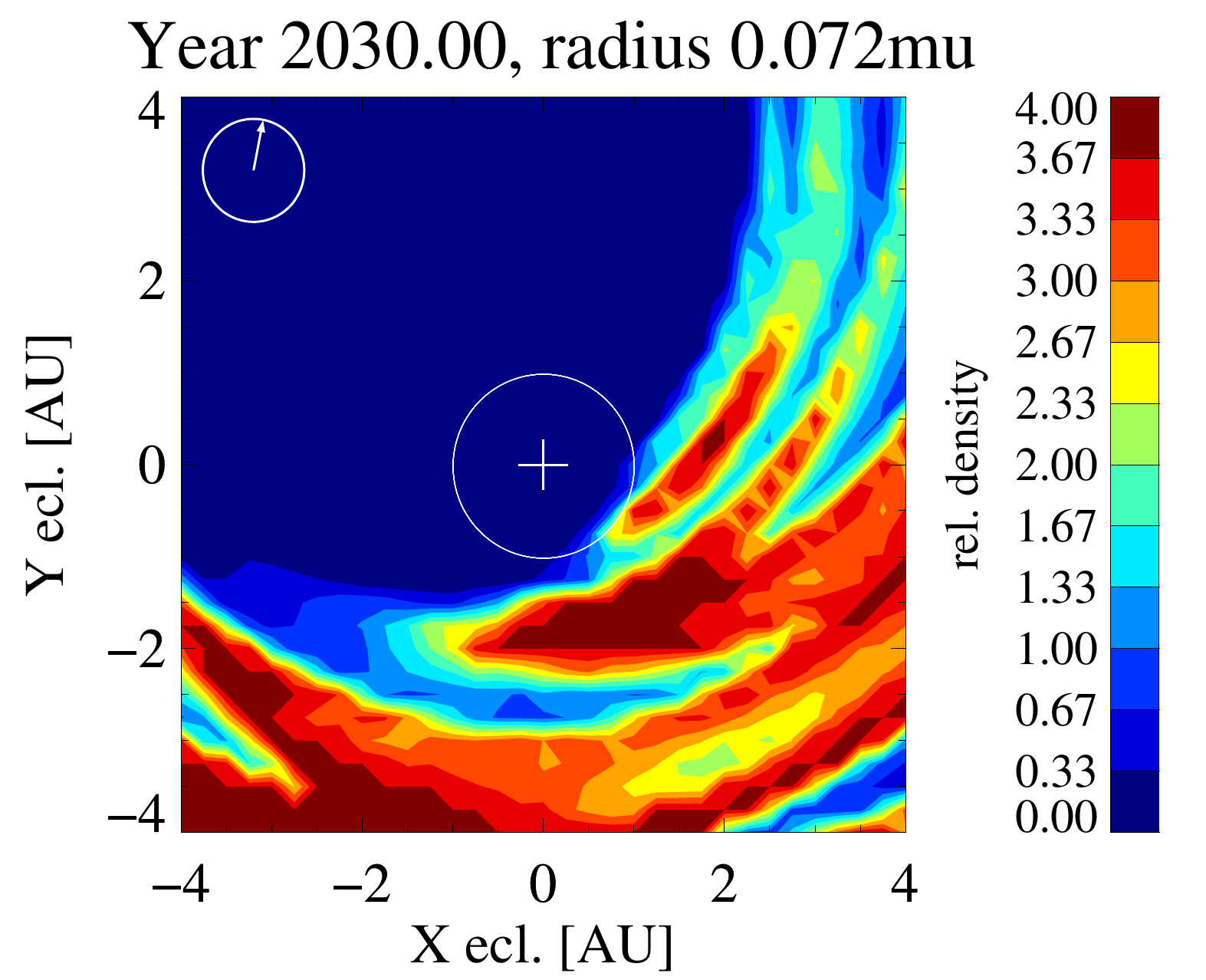} \\
\includegraphics[width=0.3\textwidth]{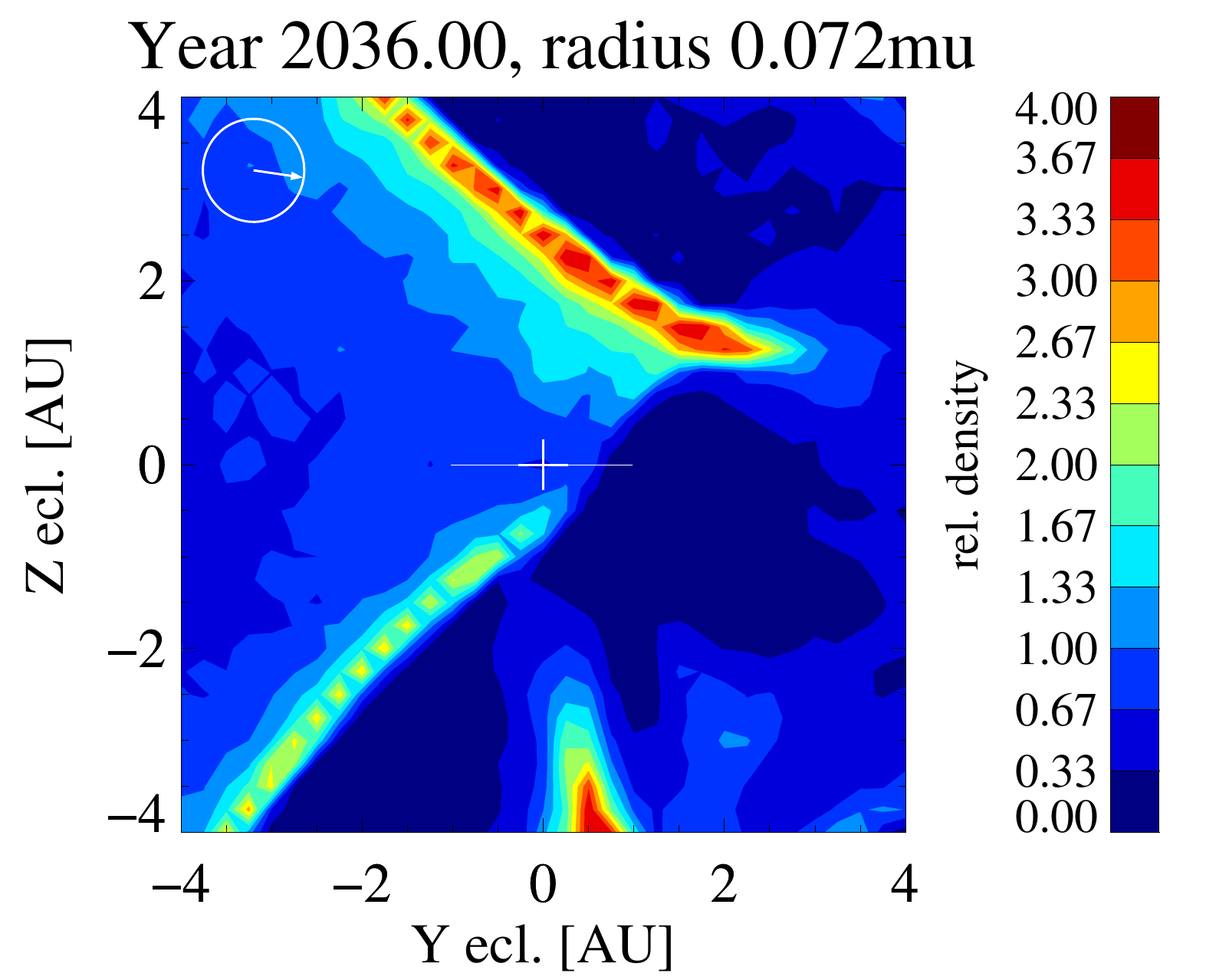} &
\includegraphics[width=0.3\textwidth]{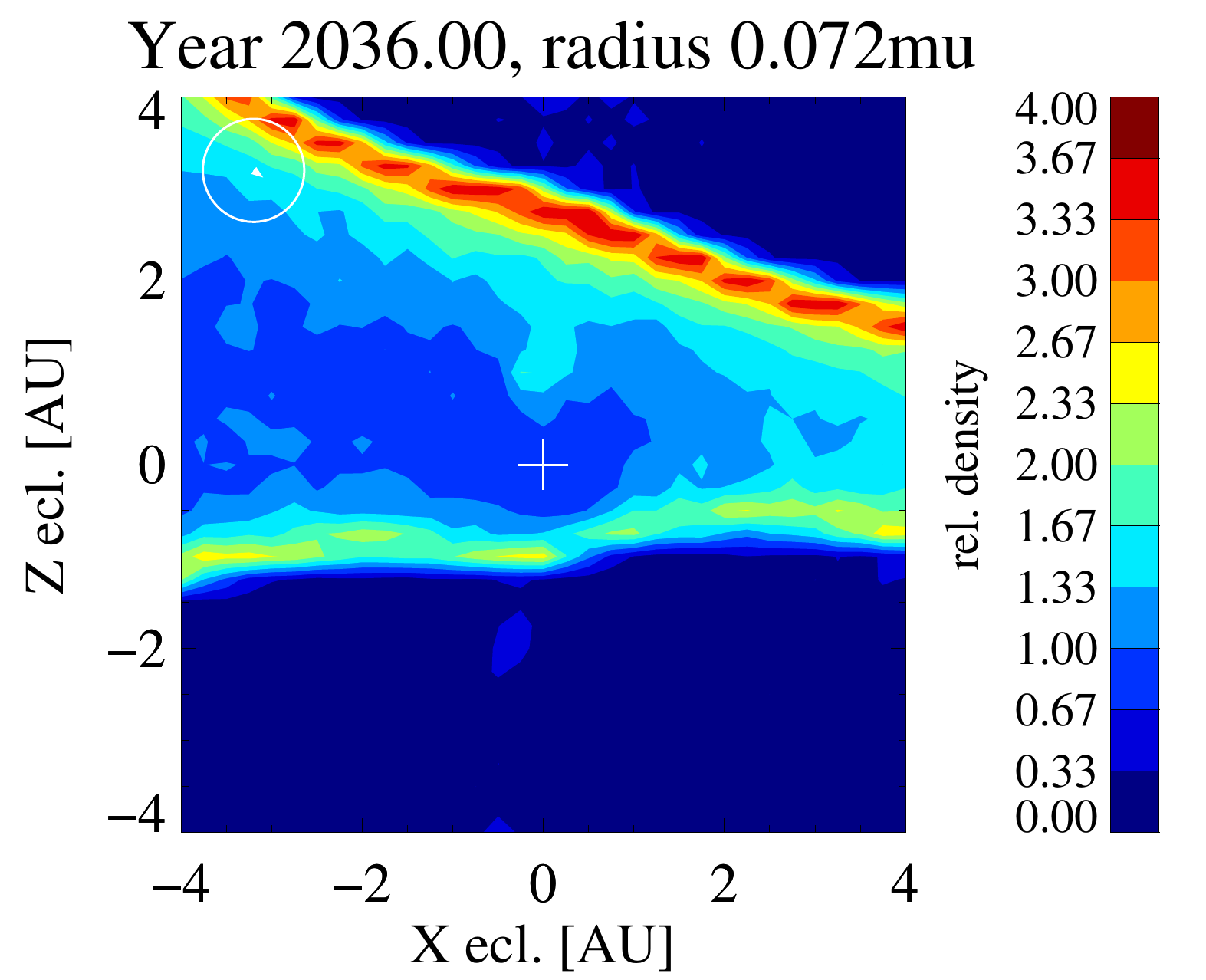} &
\includegraphics[width=0.3\textwidth]{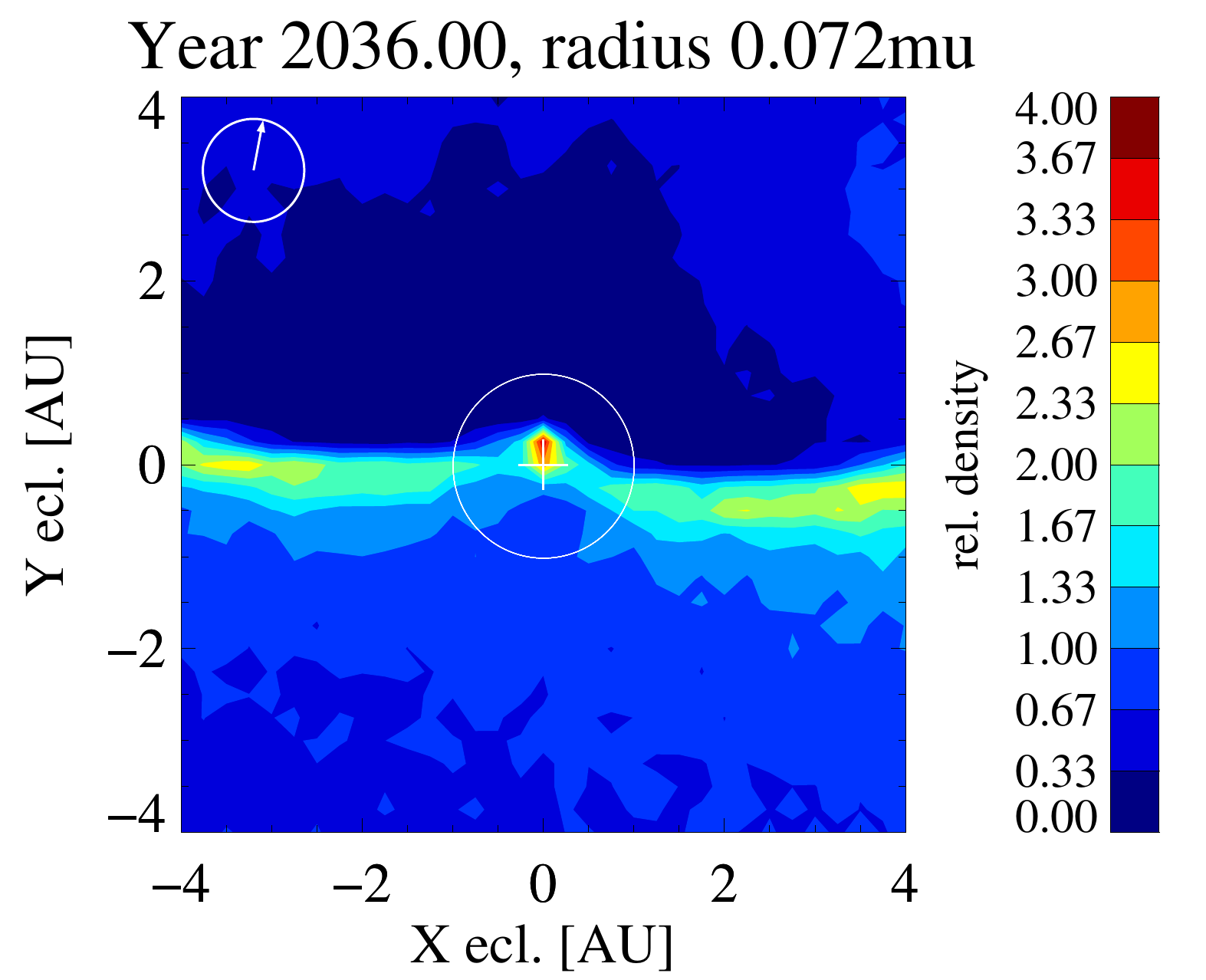} \\
\end{tabular}
\caption{Cuts along the ecliptic coordinate planes through the simulated density cubes for particles of size 
$0.072\,\mathrm{\mu m}$ at different epochs. The years have been chosen to select a representative case in the defocussing phase (2020), the focussing phase (2030), and a transitional phase (2036). The Sun is in the centre. The ISD density is colour coded: dark blue: no ISD particles reach this region of space; green, yellow, and red colours represent density enhancements with respect to the initial density at 50 AU. The projection of the original flow direction (at 50 AU) is shown in the upper left corner of each plot.}
\label{fig:denscuts072}
\end{figure*}

\begin{figure*}
\noindent
\centering
\begin{tabular}{ccc}
\includegraphics[width=0.3\textwidth]{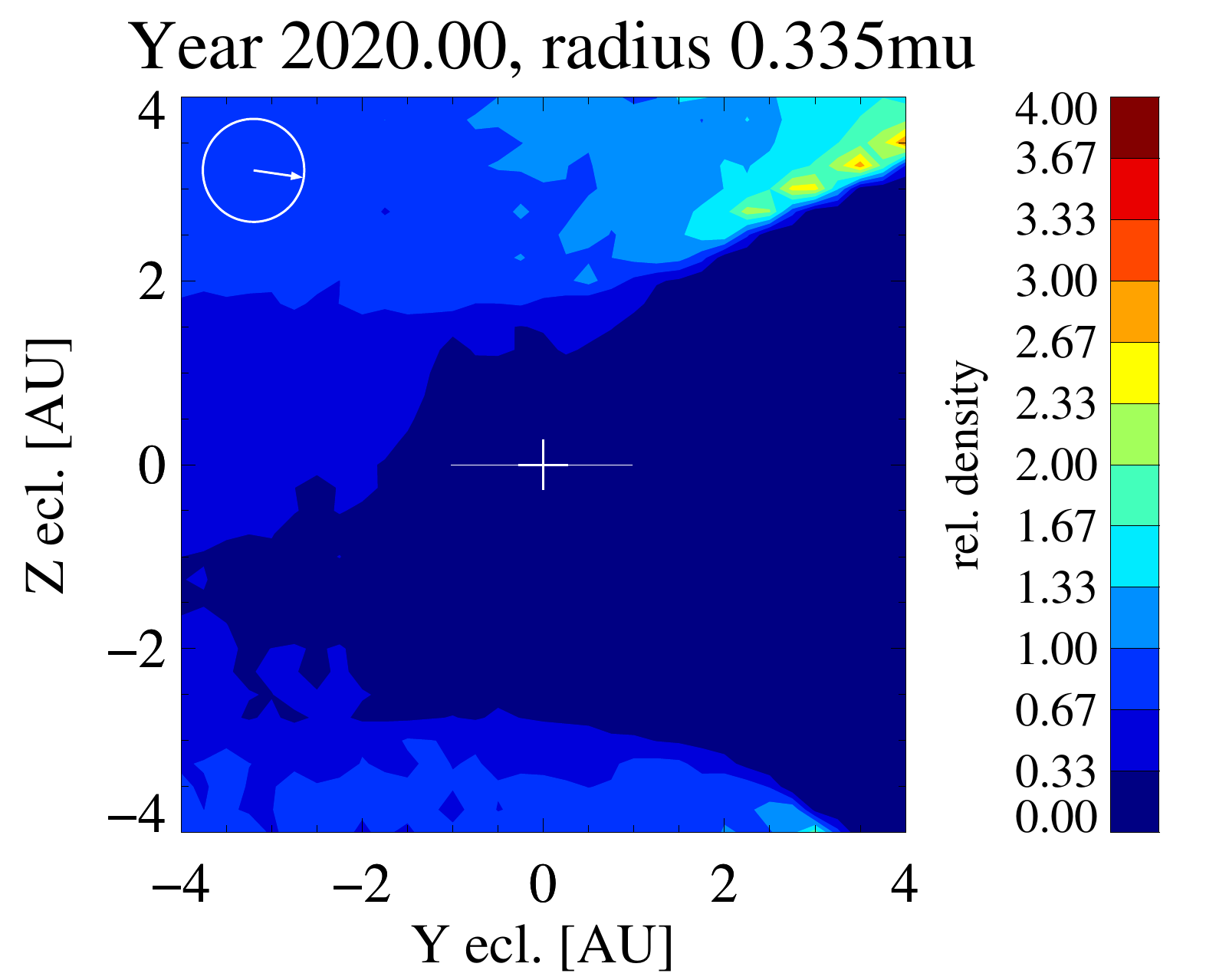} &
\includegraphics[width=0.3\textwidth]{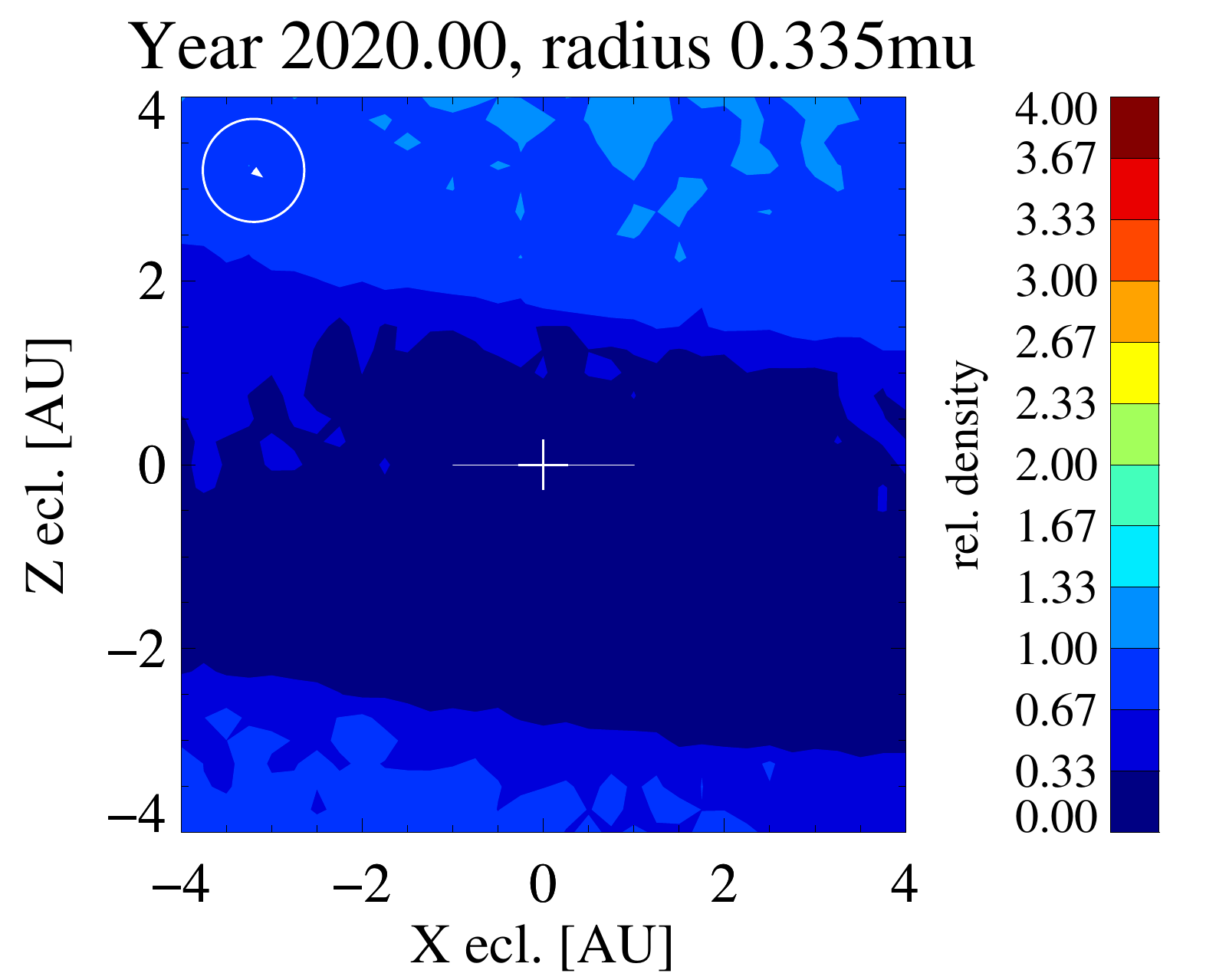} &
\includegraphics[width=0.3\textwidth]{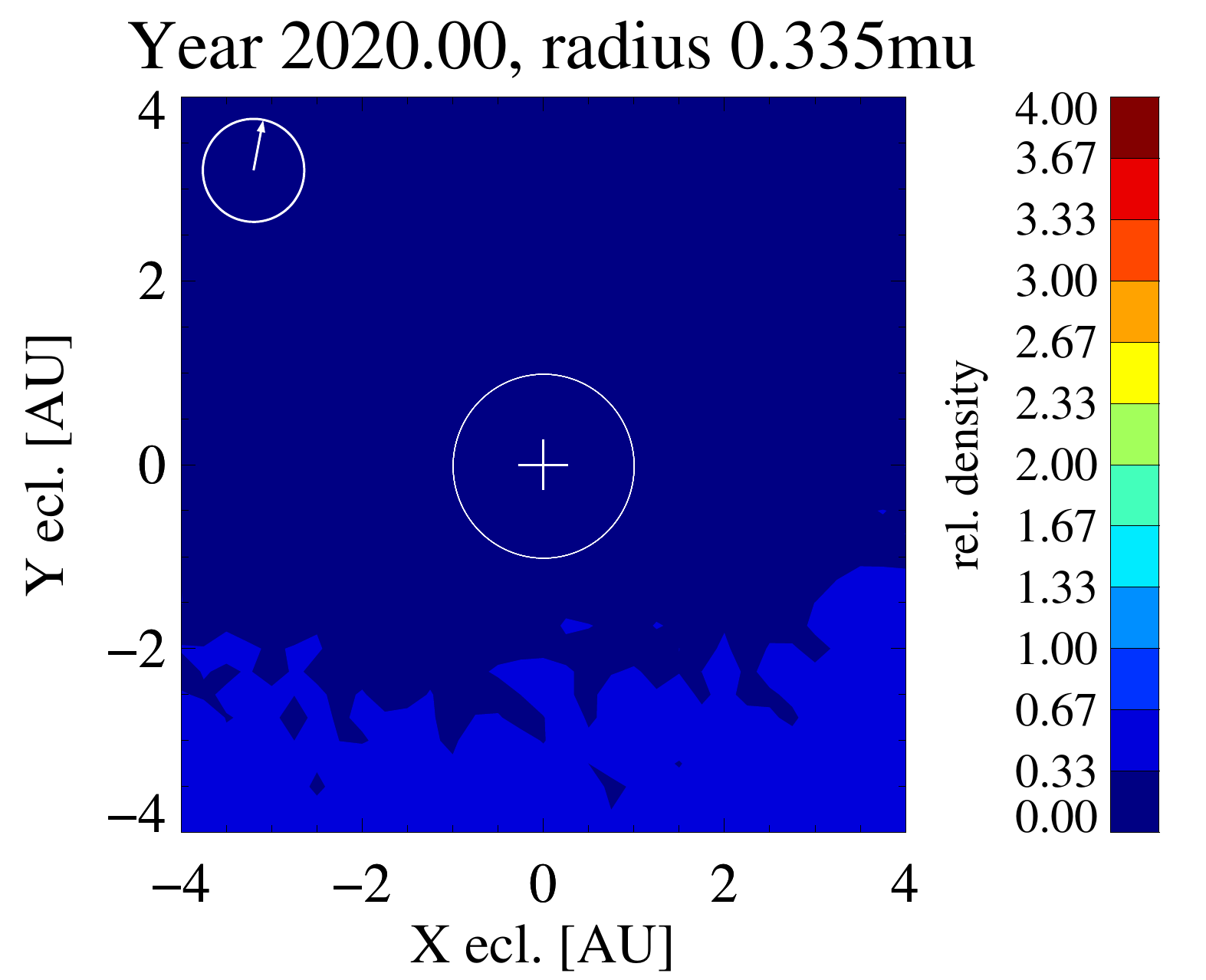} \\
\includegraphics[width=0.3\textwidth]{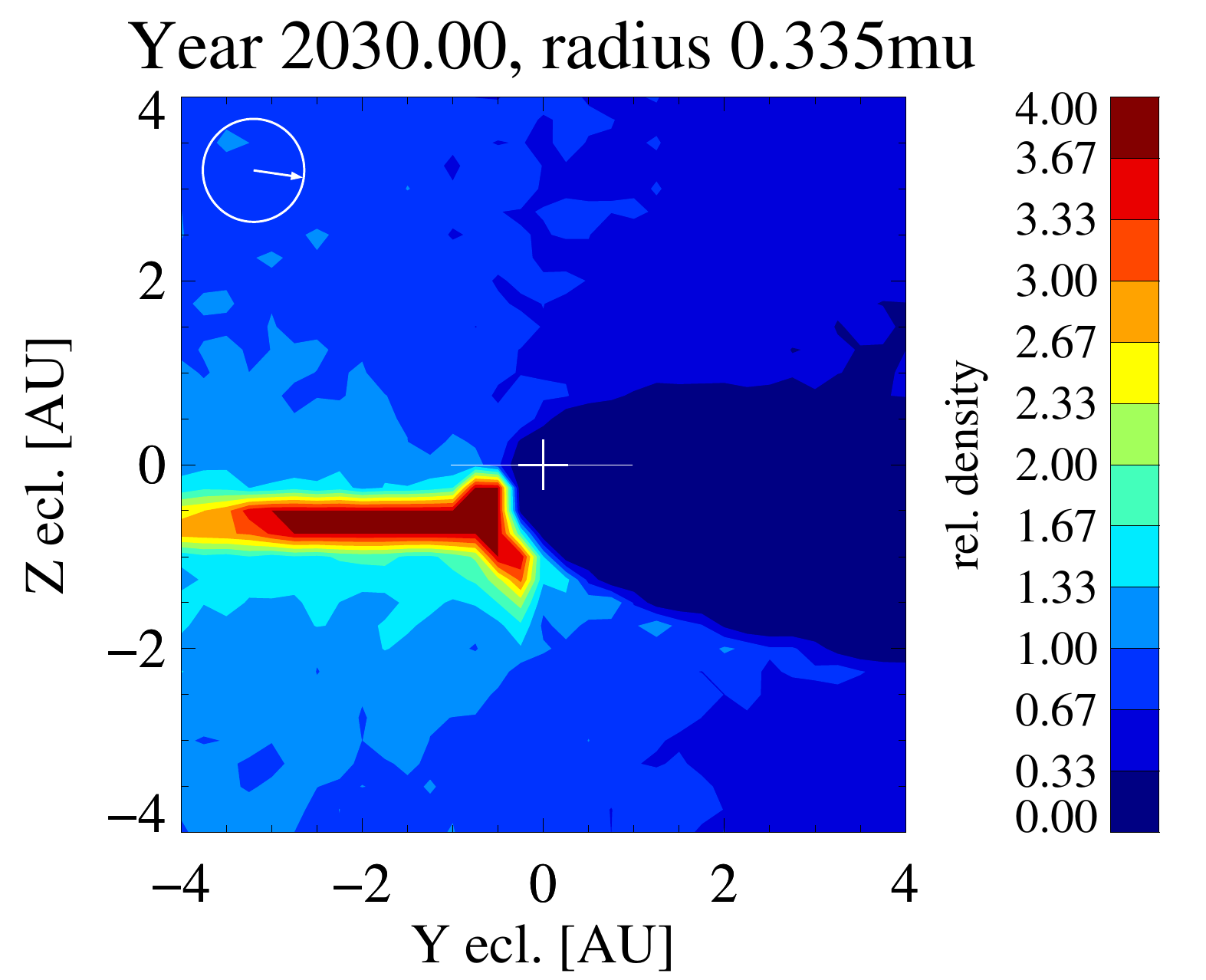} &
\includegraphics[width=0.3\textwidth]{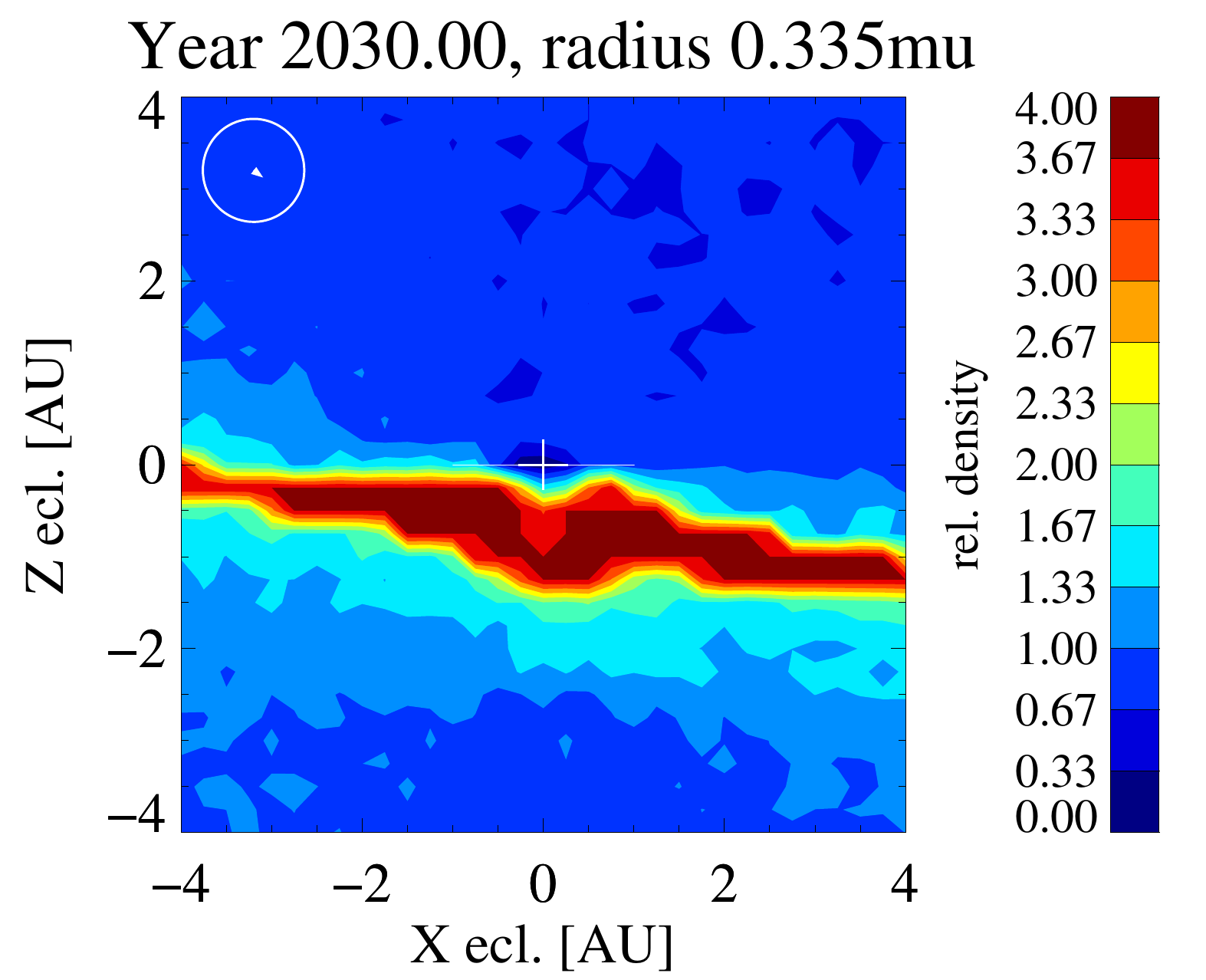} &
\includegraphics[width=0.3\textwidth]{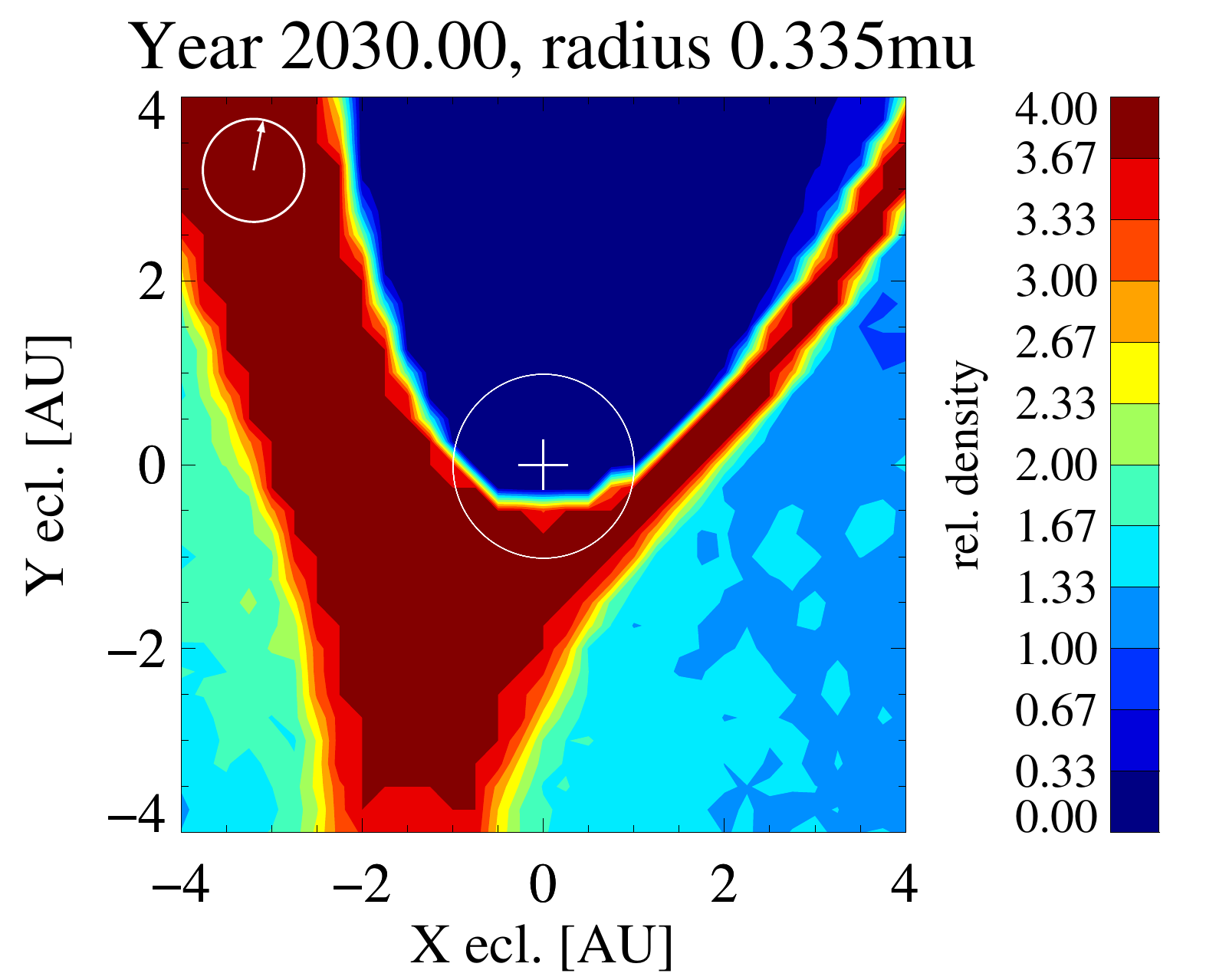} \\
\includegraphics[width=0.3\textwidth]{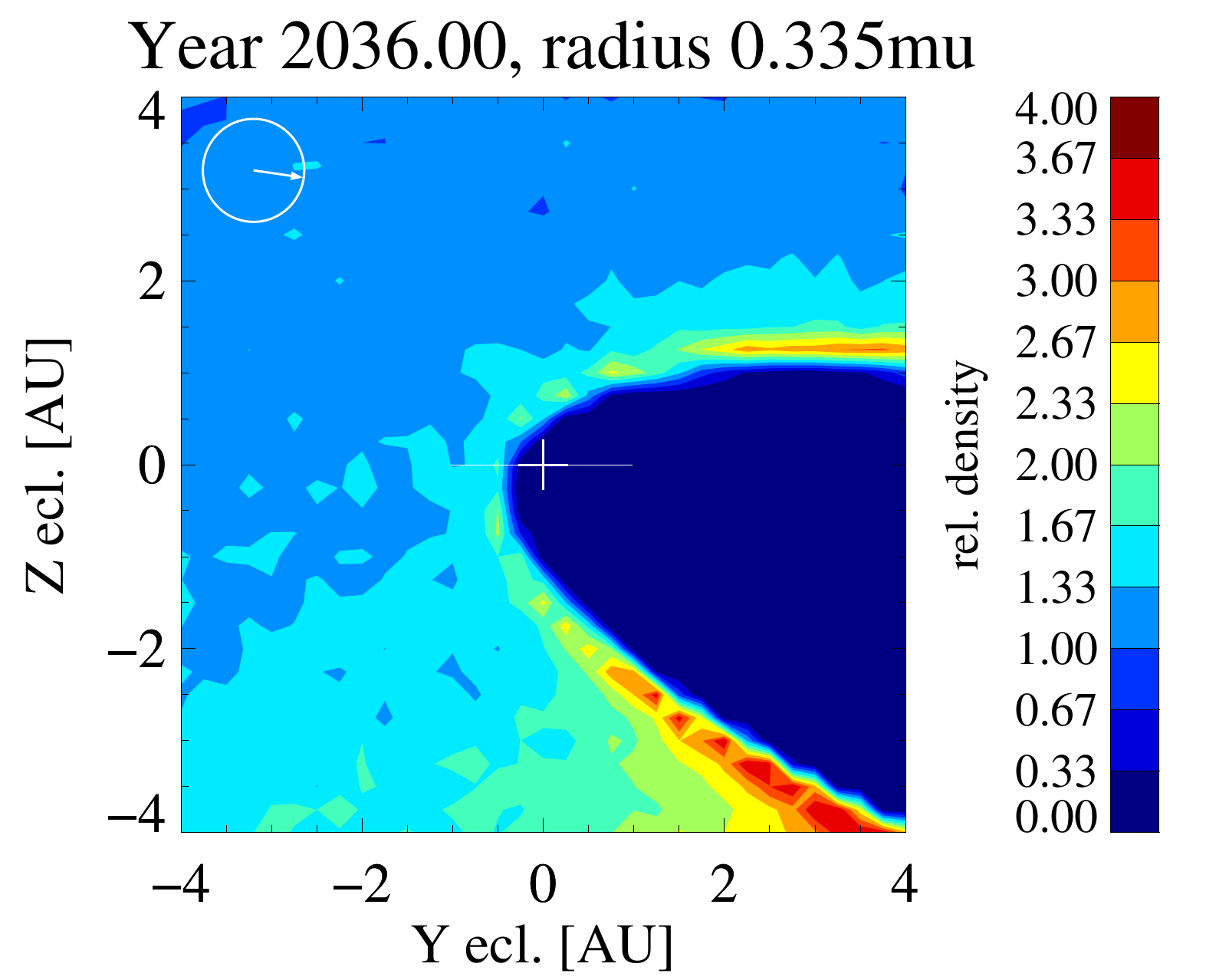} &
\includegraphics[width=0.3\textwidth]{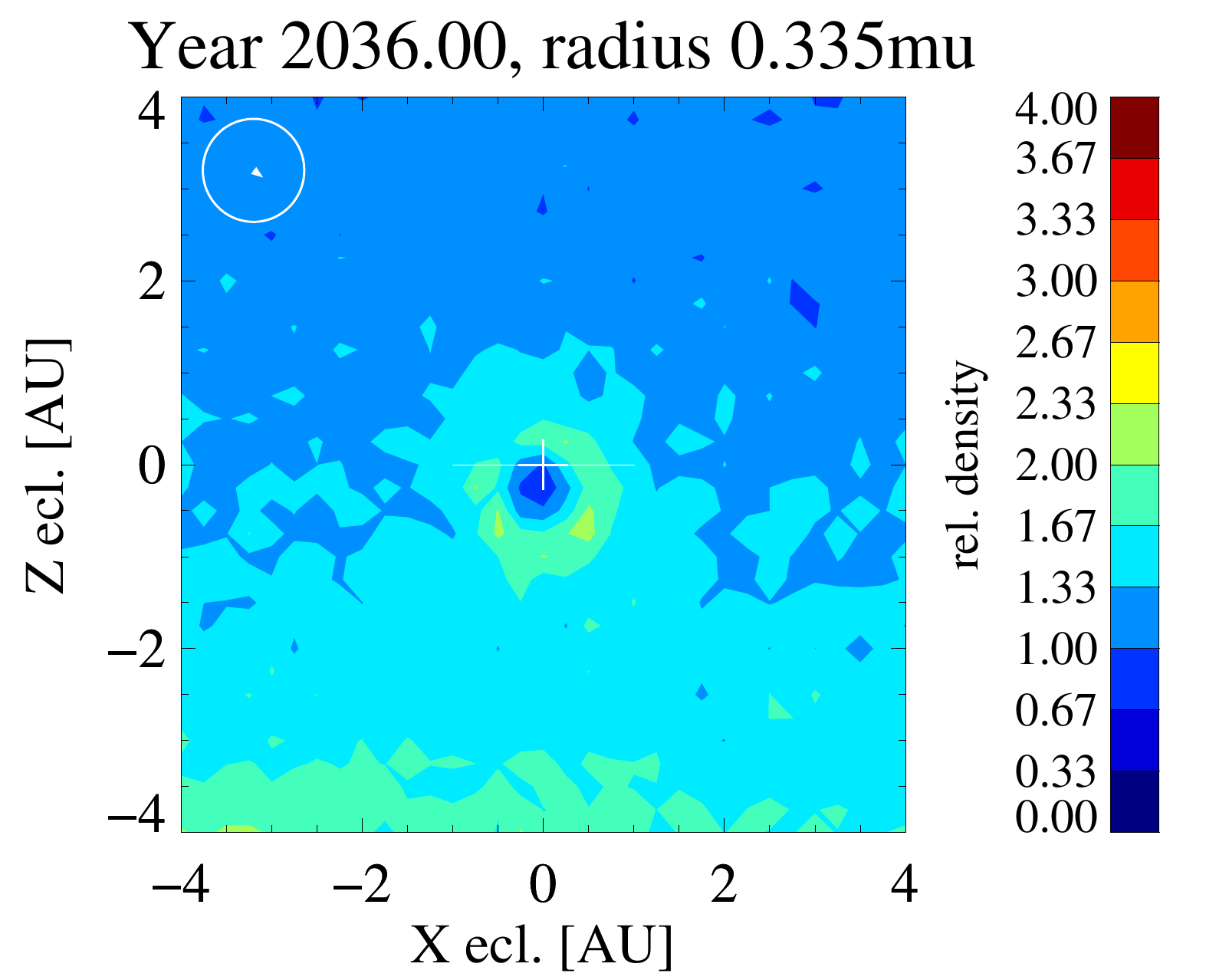} &
\includegraphics[width=0.3\textwidth]{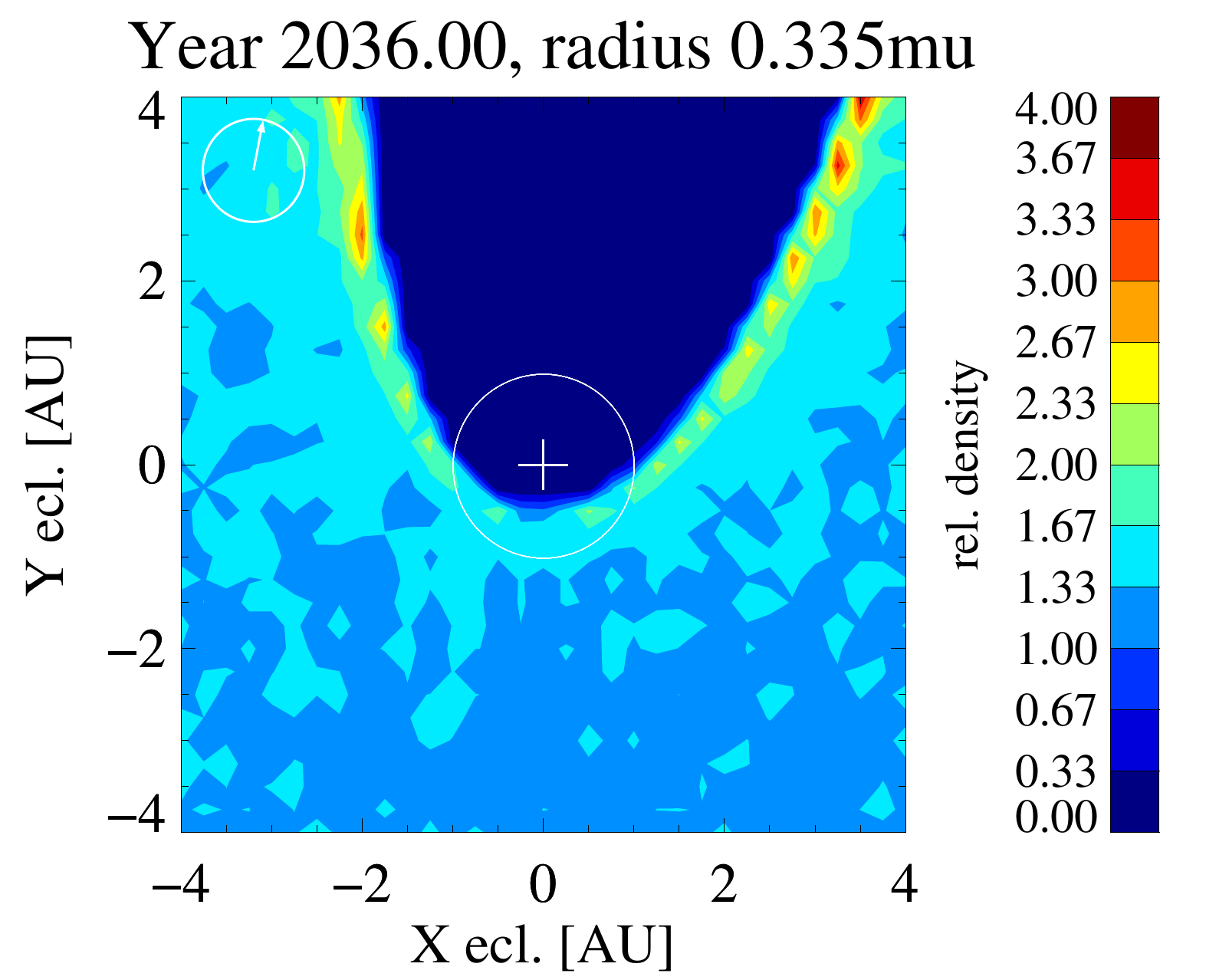} 
\end{tabular}
\caption{Cuts along the ecliptic coordinate planes through the simulated density cubes for particles of size $0.335\,\mathrm{\mu m}$ at different epochs. The years have been picked to select a representative case in the defocussing phase (2020), the focussing phase (2030), and a transitional phase (2036). The Sun is in the centre. The ISD density is colour coded: dark blue: no ISD particles reach this region of space; green, yellow, and red colours represent density enhancements with respect to the initial density at 50 AU. The projection of the original flow direction (at 50 AU) is shown in the upper left corner of each plot.}
\label{fig:denscuts335}
\end{figure*}

\begin{figure*}
\noindent
\centering
\begin{tabular}{ccc}
\includegraphics[width=0.3\textwidth]{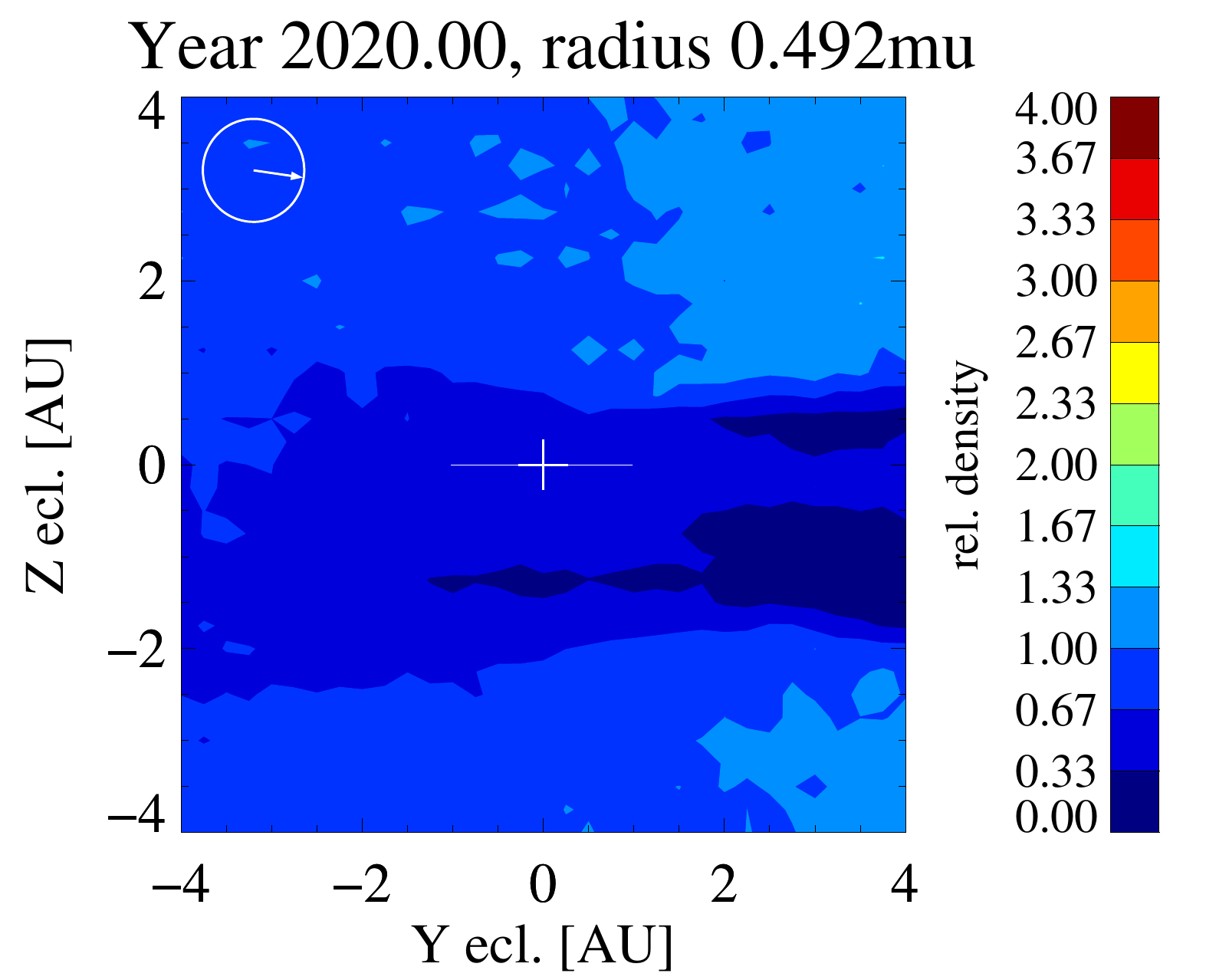} &
\includegraphics[width=0.3\textwidth]{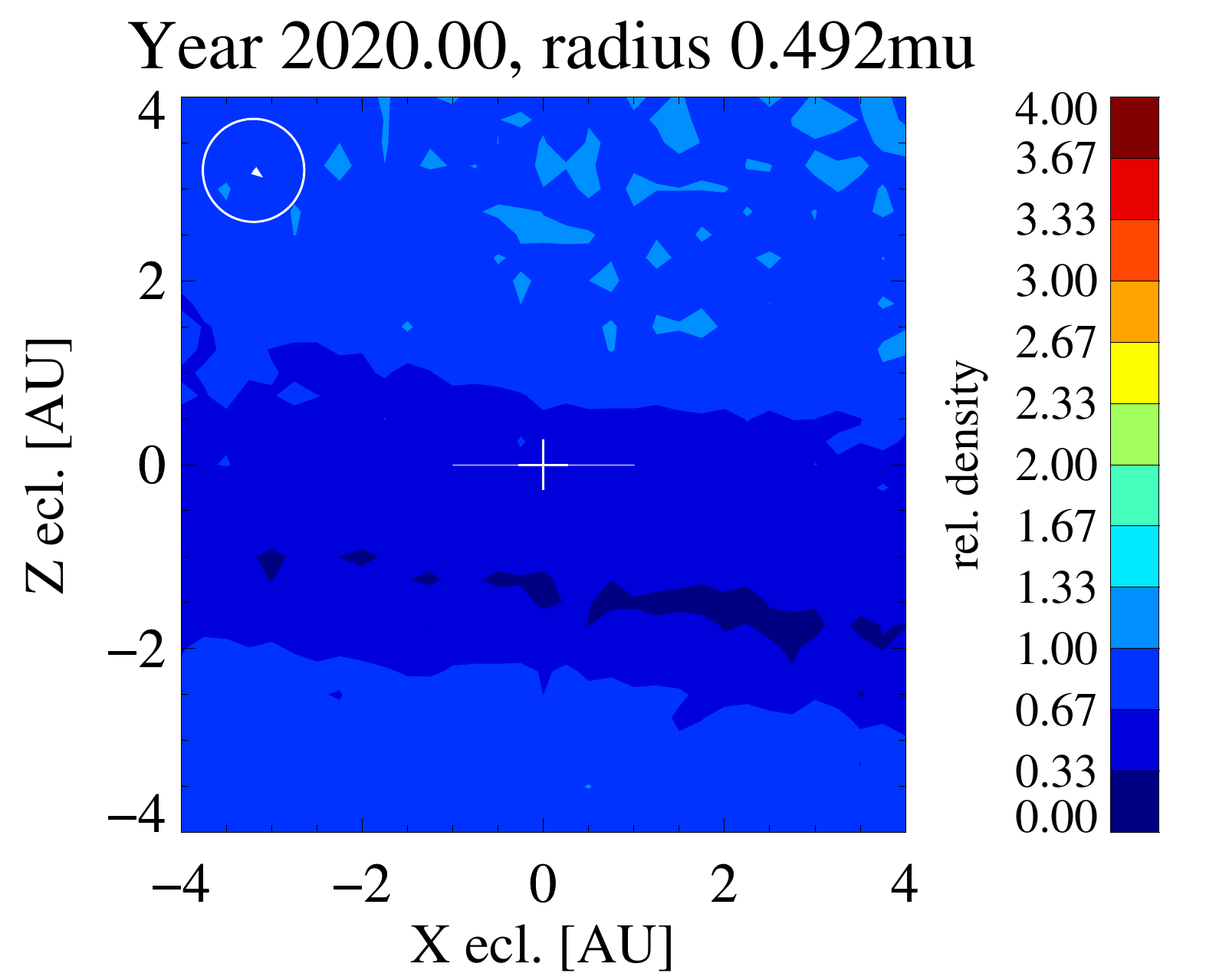} &
\includegraphics[width=0.3\textwidth]{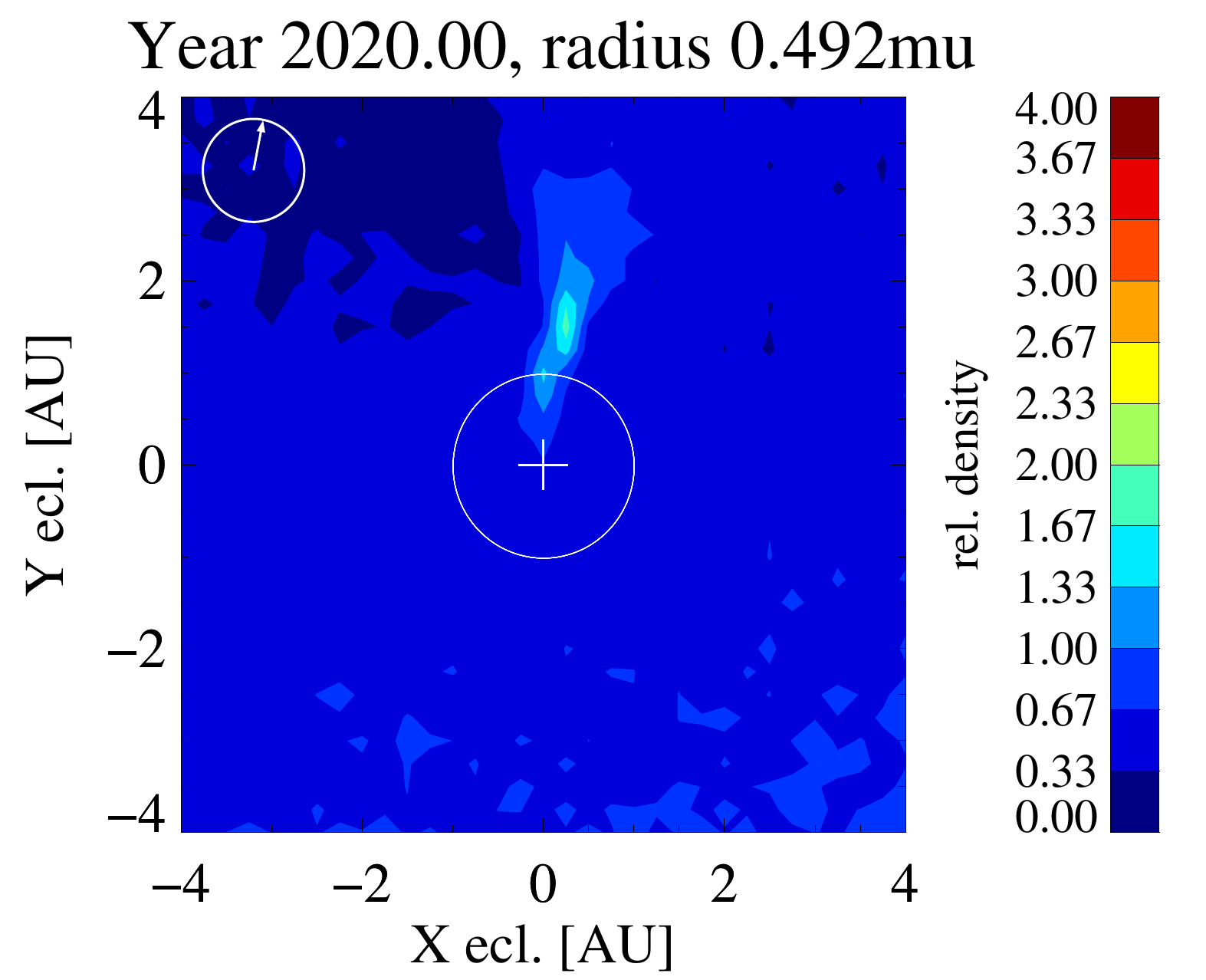} \\
\includegraphics[width=0.3\textwidth]{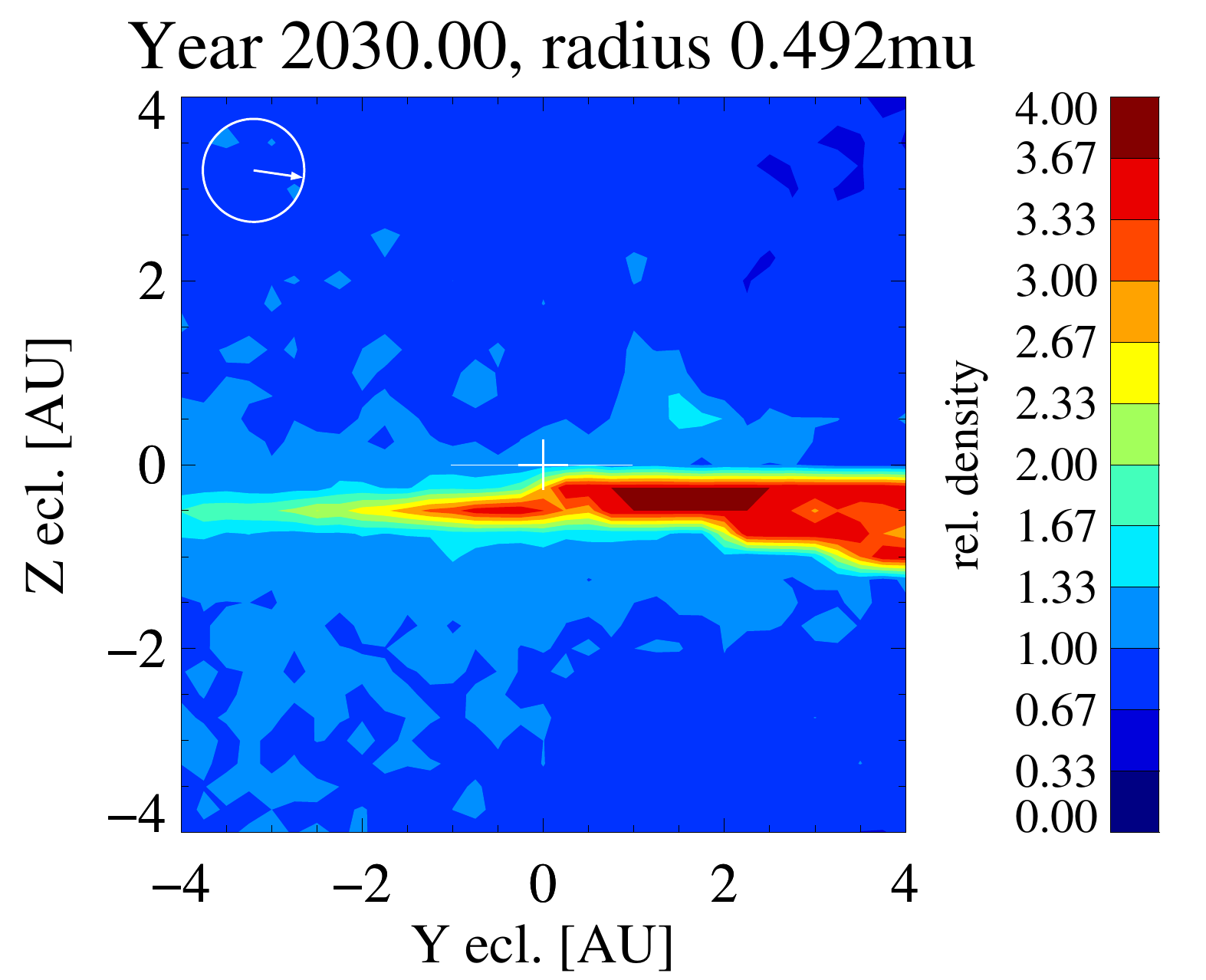} &
\includegraphics[width=0.3\textwidth]{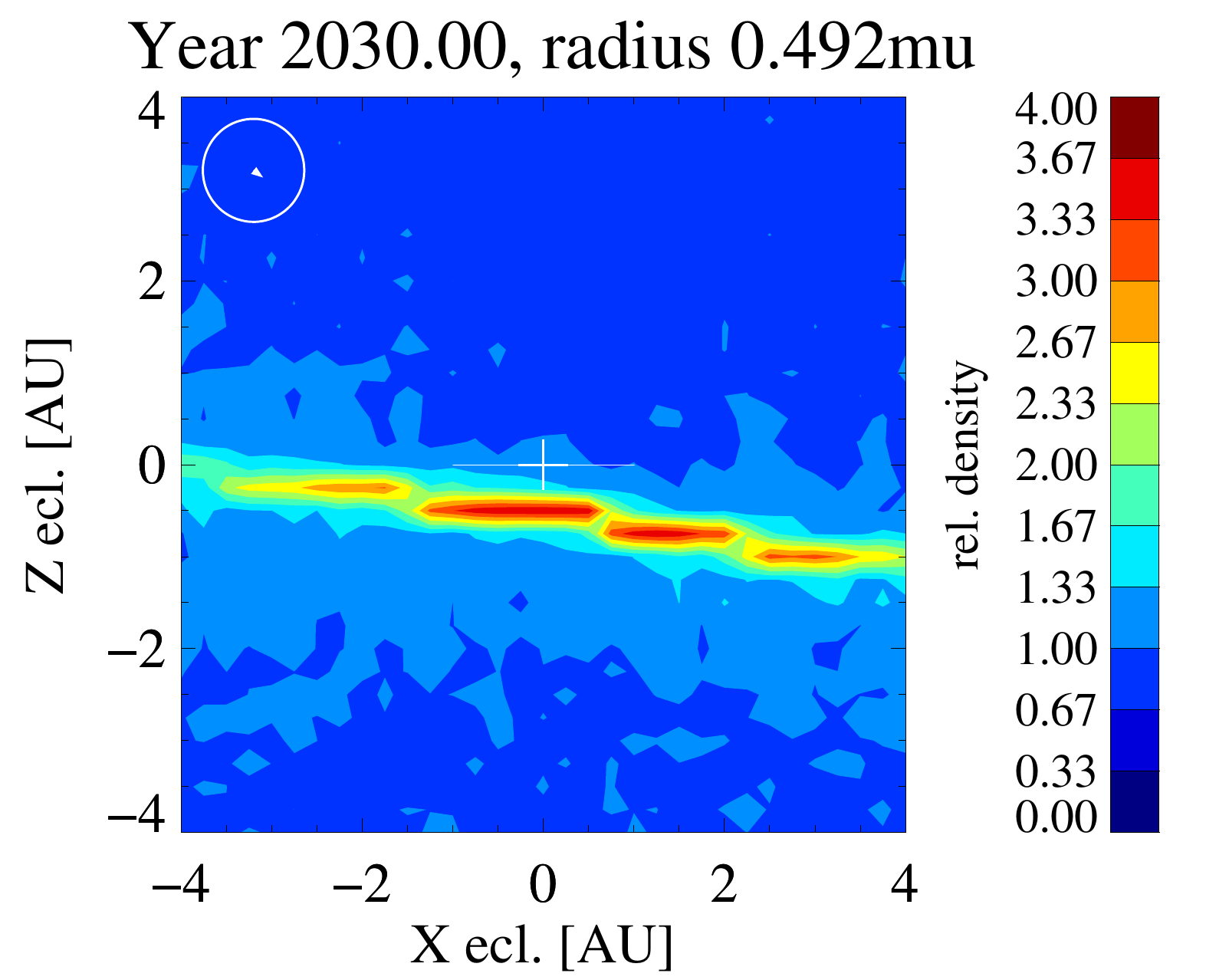} &
\includegraphics[width=0.3\textwidth]{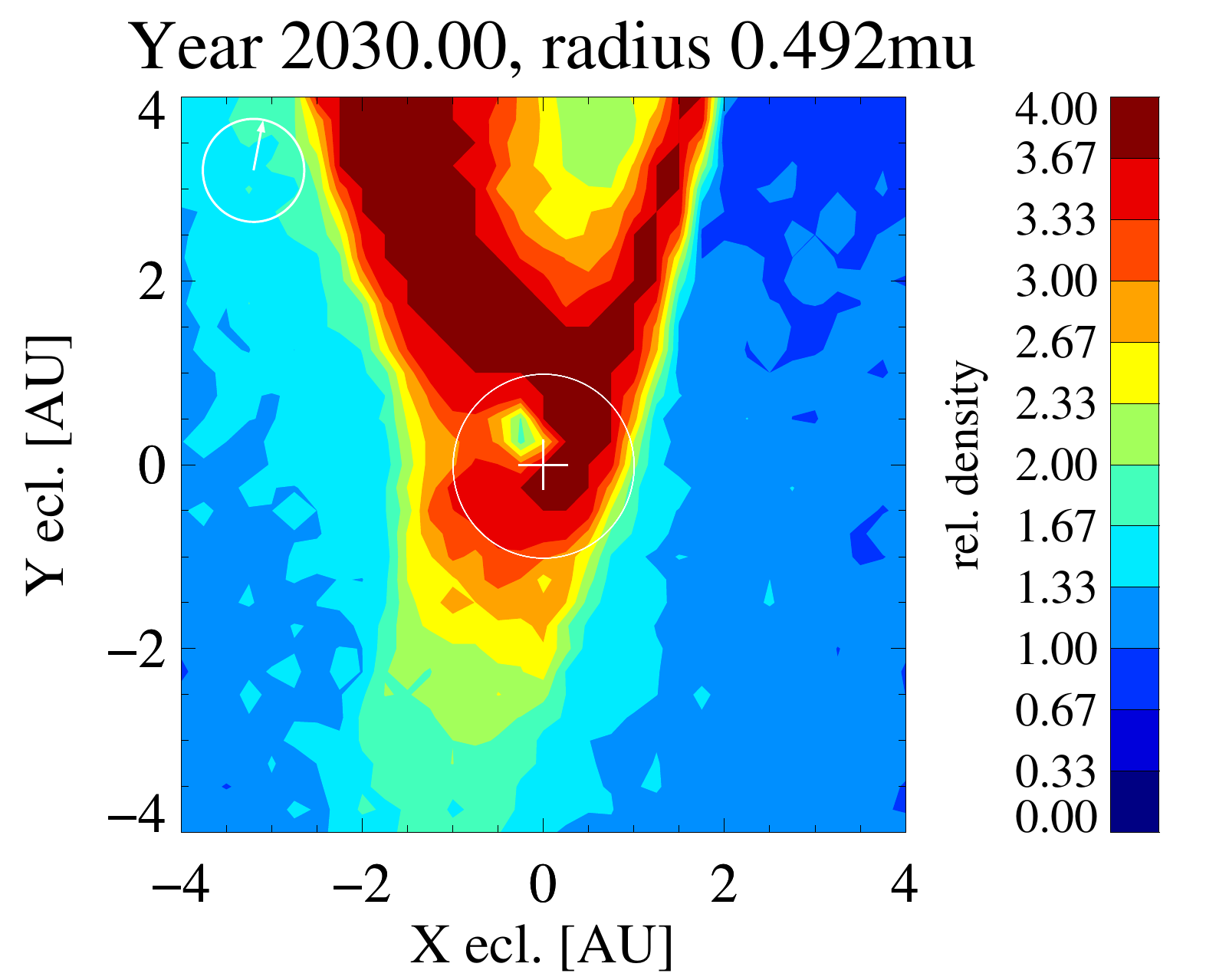} \\
\includegraphics[width=0.3\textwidth]{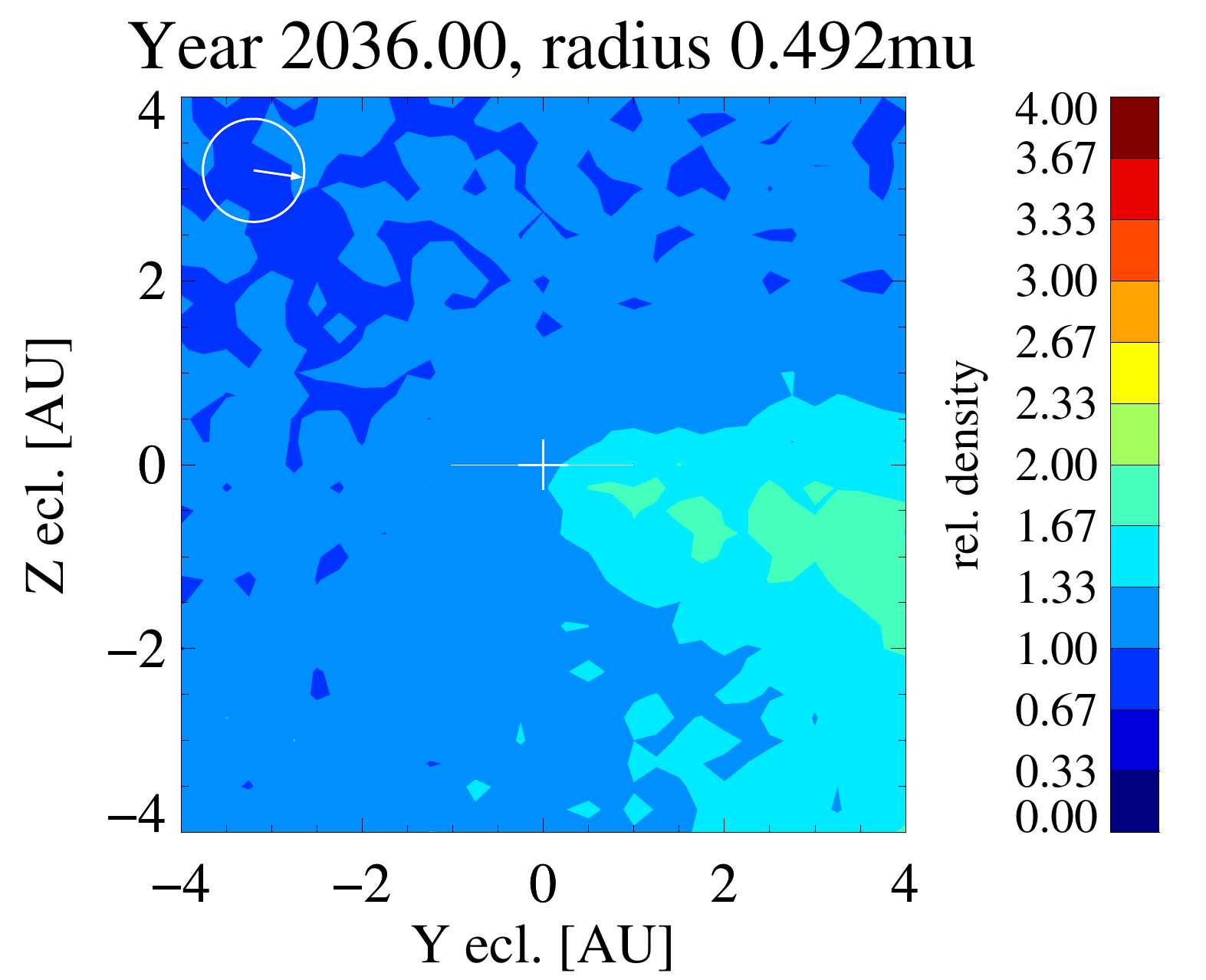} &
\includegraphics[width=0.3\textwidth]{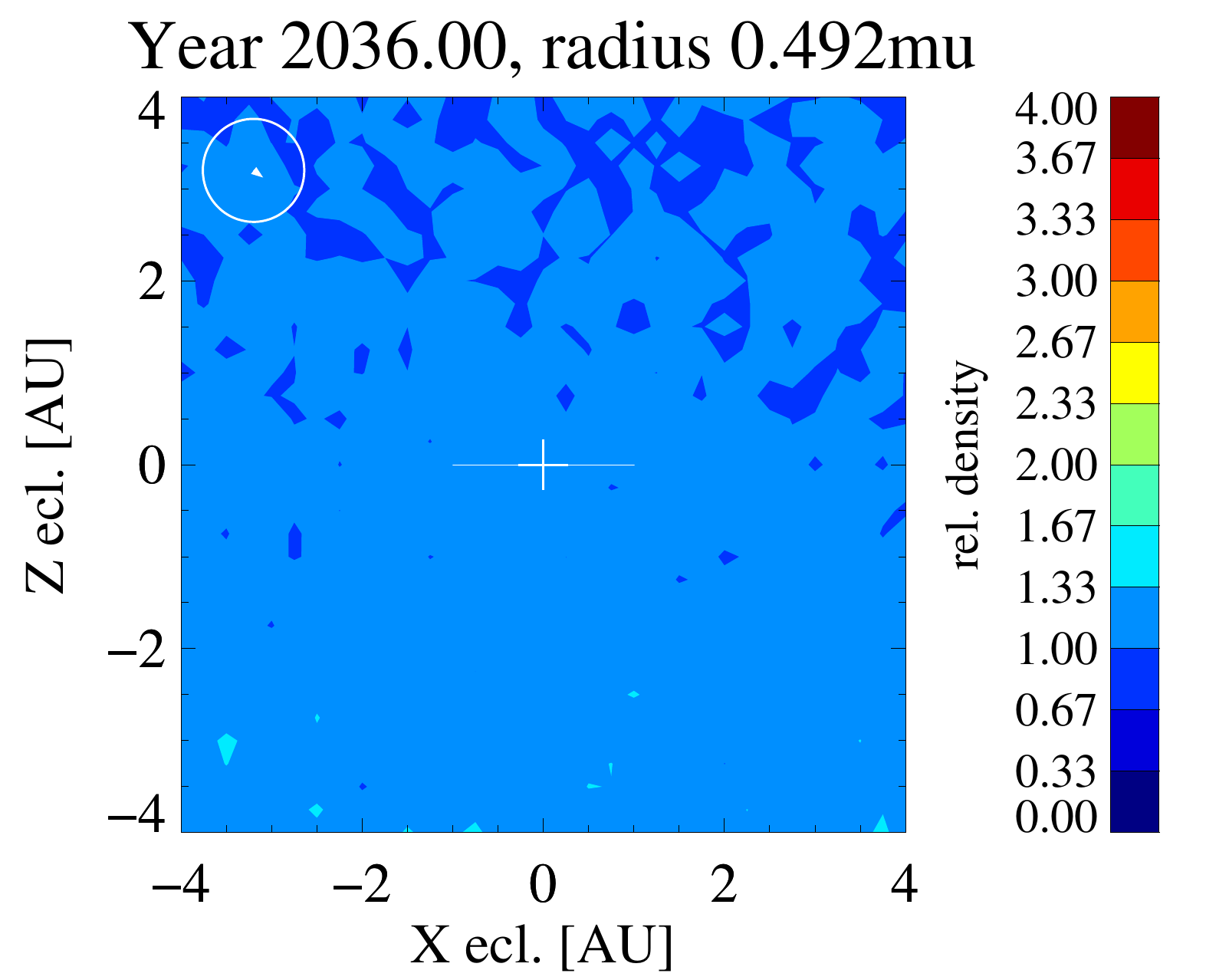} &
\includegraphics[width=0.3\textwidth]{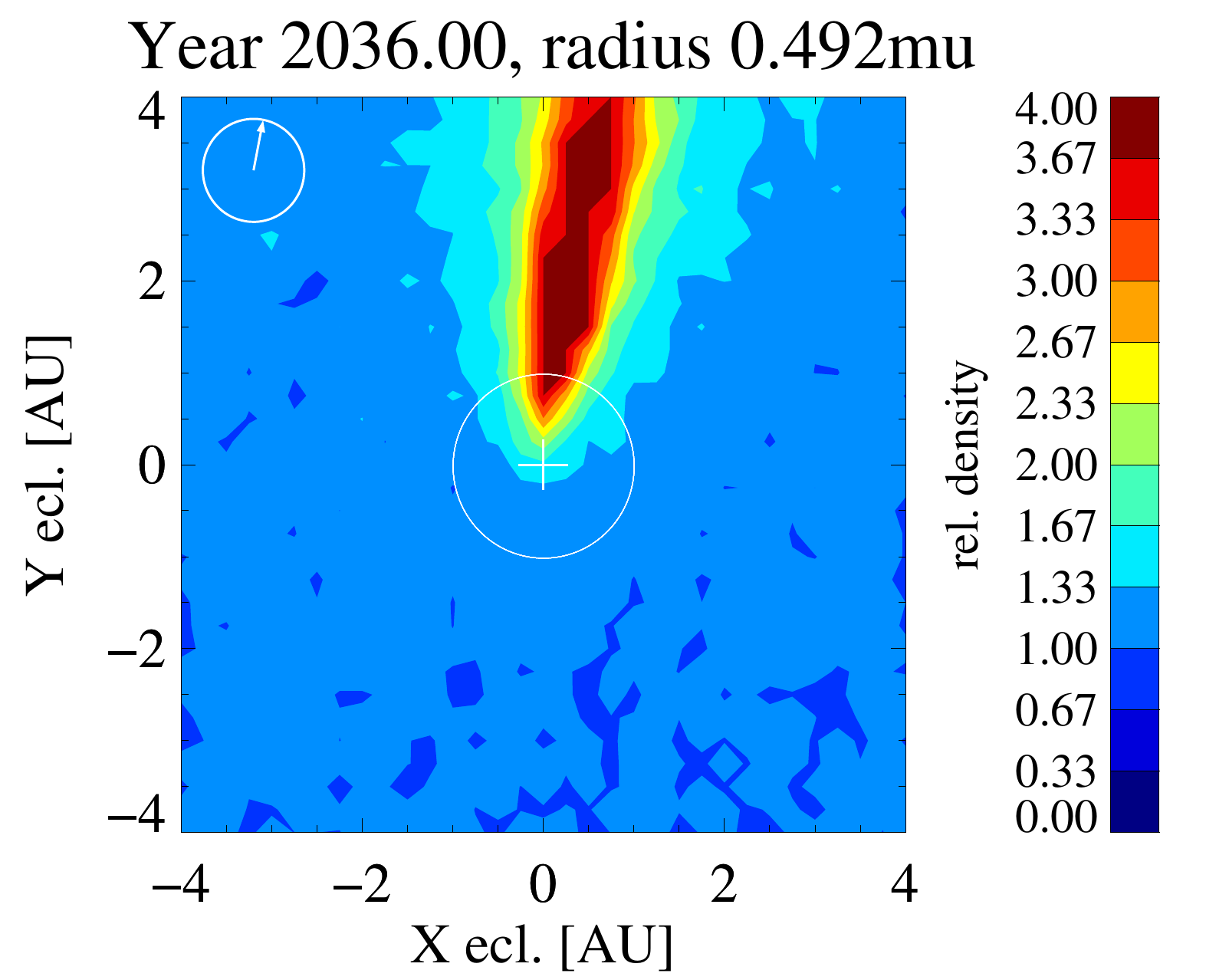} \\
\end{tabular}
\caption{Cuts along the ecliptic coordinate planes through the simulated density cubes for particles of size 
$0.492\,\mathrm{\mu m}$ at different epochs. The years have been picked to select a representative case in the defocussing phase (2020), the focussing phase (2030), and a transitional phase (2036). The Sun is in the centre. The ISD density is colour coded: dark blue: no ISD particles reach this region of space; green, yellow, and red colours represent density enhancements with respect to the initial density at 50 AU. The projection of the original flow direction (at 50 AU) is shown in the upper left corner of each plot.}
\label{fig:denscuts492}
\end{figure*}

Our simulations give a detailed insight into the flow properties of the ISD in the inner solar system within 4 AU of the Sun. The resulting flow pattern is a combination of the effects of solar gravity and radiation pressure ($\beta$) and the Lorentz force from the solar magnetic field. The relative importance of these forces depends on a particle's $\beta$ and $Q/m$ (Table~\ref{tbl:simsizes}). We characterise the results using a selection of plots for three different sized particles (0.07\,\mum, 0.34\,\mum, and 0.49\,\mum, assuming the adapted astronomical silicates curve)
representing the Lorentz-force-dominated, radiation pressure, and gravity-dominated regimes. We show cuts through the three-dimensional density cubes along the ecliptic Cartesian coordinate planes (y-z, x-z, and x-y) at different epochs. The two-dimensional cuts through the three-dimensional ISD density (Fig.~\ref{fig:denscuts072} to \ref{fig:denscuts492}) demonstrate the rich and clumpy structure of the distribution of interstellar material throughout the inner solar system, which varies markedly with the solar cycle and particle size.\footnote{Simulations using smaller cell sizes and/or a randomisation of $Q/m$ in a small interval might reduce some clumpy patterns.}

Figure~A1 shows the densities of $0.07\,\mum$ ISD particles. During electromagnetically defocussing conditions in 2020 no such small particles reach the region within 4 AU from the Sun; whereas strong concentrations of ISD particles are found during focussing conditions in 2030. During the transitional phase (e.g. 2036) ISD concentrations are found at higher (or lower) latitudes. The density distributions of 0.34\,$\mu$m ISD particles are shown in Figure~A2. For these particles the repulsion by solar radiation pressure is the dominant force ($\beta$ = 1.17); hence, an exclusion zone around the Sun ($\beta$ cone) is formed. However, the electromagnetic forces are still strong enough to prevent this type of particle from reaching the inner solar system during defocussing conditions (2020) and to strongly focus the flux towards the ecliptic plane where interstellar particles flow around a flattened $\beta$-cone (x-y plane in 2030, cf. Fig. 45-46 in \citet{sterken2012a}). 
During the transitional phases (2036), ISD particles are concentrated at the edge of
the $\beta$-cone at higher and lower latitudes. Solar gravity is the strongest force for 0.49\,\mum\ particles, hence a pronounced focussing region is observed downstream from the Sun. Nevertheless, some focussing (2030) and defocussing effects (2020) are observed.

The Earth moves around the Sun on a roughly circular orbit at 1AU radius, thereby probing different regions of the ISD flow. This causes an annual variation of apparent flow direction and its deviation (deflection angle) from the upstream inflow direction (ecliptic longitude/ latitude $l_{\mathrm{ecl}} = 259^{\circ}$, $b_{\mathrm{ecl}} = 8^{\circ}$) which is most pronounced for medium-sized and large particles ($0.34\,\mum$ and $0.49\,\mum$, respectively, Fig. A3).


The solar magnetic field changes periodically with a 22-year cycle, with a polarity flip every 11 years; its effects are therefore intrinsically time-dependent. In the years 2014-2025, and 2036 onwards, the result is an overall defocussing of the ISD flow with a depletion of ISD in the inner solar system. In contrast, a focussing towards the solar equatorial plane occurs in the years 2025-2036, leading to an increased dust density. As the force depends on $Q/m$, the overall effect is mostly seen in the small particles $a_d\lesssim0.1\,\mum$. Particles in the range $0.2\,\mum\lesssim a_d \lesssim 0.4\,\mum$ are moderately affected, whereas particles with radii $a_d\gtrsim 0.4\,\mum$ do not respond significantly to the Lorentz force.

\end{appendix}

\end{document}